\documentclass[floatfix,prc,aps,twocolumn]{revtex4}
\usepackage{amssymb}
\usepackage{graphicx,psfrag}

\begin{document}
\title{Heavy Flavor Azimuthal Correlations in Cold Nuclear Matter}

\author{R. Vogt$^{1,2}$}

\affiliation{
  $^1$Nuclear and Chemical Sciences Division,
  Lawrence Livermore National Laboratory, Livermore, CA 94551, USA\break
  $^2$Physics Department, University of California, Davis, CA 95616, USA
  }

\begin{abstract}
  {\bf Background:}  It has been proposed that the azimuthal distributions of
  heavy flavor quark-antiquark pairs may be modified in the medium of a
  heavy-ion collision.  {\bf Purpose:}  This work tests this proposition
  through next-to-leading order (NLO)
  calculations of the azimuthal distribution,
  $d\sigma/d\phi$, including transverse momentum broadening, employing
  $\langle k_T^2 \rangle$ and fragmentation
  in exclusive $Q \overline Q$ pair production.  While these studies were done
  for $p+p$, $p + \overline p$ and $p+$Pb collisions, understanding azimuthal
  angle correlations between heavy quarks in these smaller, colder systems is
  important for their interpretation in heavy-ion collisions.
  {\bf Methods:}  First, single
  inclusive $p_T$ distributions calculated with the exclusive HVQMNR code are
  compared to those calculated in the fixed-order next-to-leading logarithm
  approach.  Next the azimuthal distributions are calculated and sensitivities
  to $\langle k_T^2 \rangle$, $p_T$ cut, and rapidity are studied at
  $\sqrt{s} = 7$~TeV.  Finally, calculations are compared to 
  $Q \overline Q$ data in elementary $p+p$ and $p + \overline p$ collisions
  at $\sqrt{s} = 7$~TeV and 1.96 TeV as well as to the
  nuclear modification factor $R_{p {\rm Pb}}(p_T)$ in $p+$Pb collisions at
  $\sqrt{s_{NN}} = 5.02$~TeV measured by ALICE.  {\bf Results:}  The low $p_T$
  ($p_T < 10$~GeV) azimuthal distributions are very sensitive to the $k_T$
  broadening and rather insensitive to the fragmentation function.  The NLO
  contributions can result in an enhancement at $\phi \sim 0$ absent any other
  effects.  Agreement with the data was found to be good. {\bf Conclusions:}
  The NLO calculations, assuming collinear factorization and introducing
  $k_T$ broadening, result in significant modifications of the azimuthal
  distribution at low $p_T$ which must be taken into account in calculations
  of these distributions in heavy-ion collisions.
\end{abstract}
\maketitle
\section{Introduction}
\label{intro}

Recently there has been interest in heavy flavor correlations and how
they might be modified in heavy-ion collisions \cite{Andre,mischke,gossiaux}.
However, before drawing any conclusions, it is worth studying these correlations
in $p+p$ collisions and how they may be affected in cold nuclear matter.
This paper focuses on heavy quark pair correlations in azimuthal angle.

Heavy flavor correlations were previously studied
at high $p_T$ and high pair invariant mass where they contribute to the
background for $Z^0$ boson decays \cite{Camelia}.  They were also studied
in more elementary collisions, including the effects of final-state radiation
in event generators \cite{Vermunt} and in
the $k_T$-factorization approach \cite{Szczurek}.

Correlated
production, specifically of the azimuthal angle between heavy flavors,
$\phi$, either
by direct reconstruction of both $D$ mesons or of a $D$ meson and the decay 
product of its partner, either a light hadron or a lepton \cite{Andre}, is a
stronger test of $Q \overline Q$ production than single inclusive distributions.
Naively, at 
LO $Q \overline Q$ pairs are produced back-to-back
with a peak at $\phi = \pi$.  Higher order production will, however, 
result in a more isotropic distribution in $\phi$ due to light parton
emission in the final state.  

Single inclusive heavy flavor production is discussed and compared with
data in Sec.~\ref{sec:single}.
Because it is not possible to study pair correlations with current
approaches to single inclusive distributions such as FONLL \cite{FONLL} and
GV-VFNS \cite{GM-VFN}, the
exclusive HVQMNR NLO code \cite{MNRcode}
is employed to calculate the azimuthal
correlations.  By modifying the fragmentation function and the intrinsic
transverse momentum broadening in the HVQMNR code, it is possible to
reproduce the shape of the single heavy flavor transverse momentum
distributions from codes like FONLL.
This gives confidence in the approach to the calculation of
the pair distributions.  In this section, methods of calculating exclusive
$Q \overline Q$ pair production are briefly introduced and a comparison of the
next-to-leading order calculation with $Q \overline Q$ production in leading
event generators is discussed.

After demonstrating that the calculational approaches give reasonably
equivalent results, the sensitivity of the azimuthal distributions to the heavy
flavor fragmentation function and the intrinsic transverse momentum, $k_T$,
is explored.
Sensitivities to the heavy quark transverse momentum cut, the rapidity range
probed, and the renormalization and factorization scales are also discussed
in Sec.~\ref{sec:azi_dist}.  Section~\ref{sec:sensitivity}
describes the sensitivity of the azimuthal
distributions to the size of $\langle k_T^2 \rangle$, independent of
the fragmentation function.

The results are compared to 
$Q \overline Q$ data from $p+p$ and $p + \overline p$ collisions
in Sec.~\ref{sec:dataComp}
and to ALICE $R_{p{\rm Pb}}(p_T)$ single $D$ meson data from $p+$Pb collisions
at $\sqrt{s_{_{NN}}} = 5.02$~TeV in Sec.~\ref{sec:pPb}.  In Sec.~\ref{sec:pPb},
predictions for cold matter effects on the $\phi$ distribution are also shown.

\section{Heavy flavor production}
\label{sec:single}

\subsection{Single inclusive production approaches}

There are currently two approaches to heavy flavor production at colliders: 
collinear factorization and the $k_T$-factorization 
approach, usually employed at low $x$.  

There are two main methods of calculating the spectrum of single inclusive open 
heavy flavor production in perturbative QCD assuming collinear factorization.  
The underlying idea is similar
but the technical approach differs.  Both seek to cure the large logarithms of 
$p_T/m$ arising at all orders of the perturbative expansion which can spoil the 
convergence.  The first terms in the expansion are the leading (LL) and 
next-to-leading logarithmic (NLL) terms, $\alpha_s^2[\alpha_s \log(p_T/m)]^k$ 
and $\alpha_s^3[\alpha_s \log(p_T/m)]^k$ respectively.

It is worth noting that the single inclusive heavy flavor
$p_T$ distribution is finite at leading order (LO) as $p_T \rightarrow 0$
because of the finite quark mass scale.  This is in contrast to light hadron
and jet production where there is no mass scale to regulate the low $p_T$
cross section.

While large uncertainties can arise at low $p_T$, these are due to the choice
of factorization scale and are larger for charm than for bottom quark
production \cite{Joszo,NVF}.  A next-to-leading order (NLO)
calculation that assumes production of massive
quarks but neglects LL terms, a ``massive'' formalism, can result in large 
uncertainties at high $p_T$ \cite{FFN-NDE1,FFN-NDE2,FFN-BKvNS,FFN-BvNMSS}.  
(This massive formalism is sometimes referred to as a fixed-flavor-number
(FFN) scheme.)  If, instead, the heavy quark is treated as ``massless''
and the LL and NLL corrections are
absorbed into the fragmentation functions, the approach breaks down
as $p_T$ approaches $m$ even though it improves the result at high $p_T$.
The massless formalism is sometimes referred to as the zero-mass, 
variable-flavor-number (ZM-VFN) scheme \cite{ZM-VFN1,ZM-VFN2}.  
There are methods to
interpolate smoothly between the FO/FFN scheme at low $p_T$ and the 
massless/ZM-VFN scheme at high $p_T$.

The fixed-order next-to-leading logarithm (FONLL) approach is one such
method.  In FONLL, the fixed order
and fragmentation function approaches are merged so that the mass effects are
included in an exact calculation of the leading ($\alpha_s^2$) and 
next-to-leading ($\alpha_s^3$) order cross section while also including the
LL and NLL terms \cite{FONLL}.  
The NLO fixed order (FO) result is combined with a 
calculation of the resumed (RS) cross section in the massless limit.  The FO
and RS approaches need to be calculated in the same renormalization scheme
with the same number of light flavors.
The FONLL result is then, schematically,
\begin{eqnarray}
{\rm FONLL}\, = \, {\rm FO} \, + \, ({\rm RS} \, - \, {\rm FOM0}) G(m,p_T) 
\, \, 
\end{eqnarray}
where FOM0 is the fixed order result at zero mass.
The interpolating function $G(m,p_T) \sim p_T^2/(p_T^2 + (cm)^2)$ 
is arbitrary but must approach unity for 
$m/p_T \rightarrow 0$.  Note that the number of light flavors is thus 4 for
charm and 5 for bottom quark production in FONLL, in contrast the 3 for charm
and 4 for bottom in the fixed-order calculation.

The second interpolation scheme is the generalized-mass variable-flavor-number
(GM-VFN) scheme \cite{GM-VFN}.  The large logarithms in charm production for
$p_T \gg m$ are
absorbed in the charm parton distribution function and are thus included in
the evolution equations for the parton distributions.  The 
logarithmic terms can be incorporated into the hard cross section to achieve
better accuracy for $p_T \geq m$.
By adjusting the mass-dependent subtraction terms, no
interpolating function is required \cite{GM-VFN}.

In the $k_T$-factorization approach, off-shell leading order matrix elements 
for $g^* g^* \rightarrow c \overline c$ are used
together with unintegrated gluon 
densities that depend on the transverse momentum of the gluon, $k_T$, as well 
as the usual dependence on $x$ and $\mu_F$.  The 
motivation for choosing the $k_T$-factorized approach is that, at 
sufficiently low $x$, collinear factorization should no longer hold.

The LHC data has been compared to calculations in both approaches and
those assuming collinear factorization compare well with the LHC data.
Recent ALICE data \cite{ALICEDmesonsNew}, 
at $0 < p_T < 2$~GeV supports
collinear factorization.  Their $D^0$ results at 7 TeV for $|y| < 0.5$, were
compared to FONLL, GV-VFNS and LO $k_T$-factorization calculations.
Only the $k_T$-factorized calculation is inconsistent
with the shape of the data in the $p_T$ range where the calculation should
best apply.  The forward rapidity data of LHCb at 7 TeV \cite{LHCbDmesons}
and 13 TeV \cite{Aaij:2015bpa}, also agree well with the collinear factorization
assumption.  The recent 13 TeV data from LHCb is within the uncertainty
bands of FONLL and POWHEG (discussed below) down to $p_T \rightarrow 0$, albeit
with large uncertainty bands and near the upper limit of the calculated
band.  The GM-VFNS
calculation is given only for $p_T > 3$~GeV but agrees well with the data with
small uncertainties \cite{Aaij:2015bpa}.  Perhaps a NLO $k_T$-factorized result
could lead to improved agreement with the data but, so far, collinear
factorization appears to still work well for low $x$, moderate $p_T$
charm production.

This paper will thus employ the collinear factorization approach in the
calculations.

\subsection{Exclusive approaches to heavy flavor production}
\label{sec:exclusive}

Recall that the FONLL and GM-VFNS calculations are for single 
inclusive production only and can thus not address $Q \overline Q$ pair
observables.
There are NLO heavy flavor codes that, in addition to inclusive heavy flavor
production, also  
calculate exclusive $Q \overline Q$ pair production. 

The HVQMNR code \cite{MNRcode} uses negative weight
events to cancel divergences numerically.  Smearing the parton
momentum through the introduction of intrinsic transverse momenta, $k_T$,
reduces the importance of the negative weight events at low $p_T$.  HVQMNR
does not include any resummation.   

POWHEG-hvq \cite{POWHEG} is a positive weight generator that
includes leading-log resummation.  The entire event is available since 
PYTHIA \cite{PYTHIA} 
and HERWIG \cite{HERWIG} are employed to produce the complete event after
production of the $Q \overline Q$ pair. 

The $k_T$-factorization approach can
also be employed to calculate correlated $c \overline c$ production since the
unintegrated parton densities have a transverse component,
giving finite $p_T$ and $\phi$ distributions even at LO \cite{Szczurek}.

The HVQMNR code is employed here to focus on the effects due to the NLO
contribution alone, including $k_T$-broadening and fragmentation but
excluding the parton showers which can further randomize
the pair momenta and thus affect the angular correlations.  

\subsection{Heavy flavor production in leading order event generators compared
to NLO calculations}
\label{sec:generators}

In addition to these NLO codes, heavy flavor correlations can also be simulated
employing LO event 
generators such as PYTHIA.

\begin{figure}[htpb]\centering
  \includegraphics[width=0.5\columnwidth]{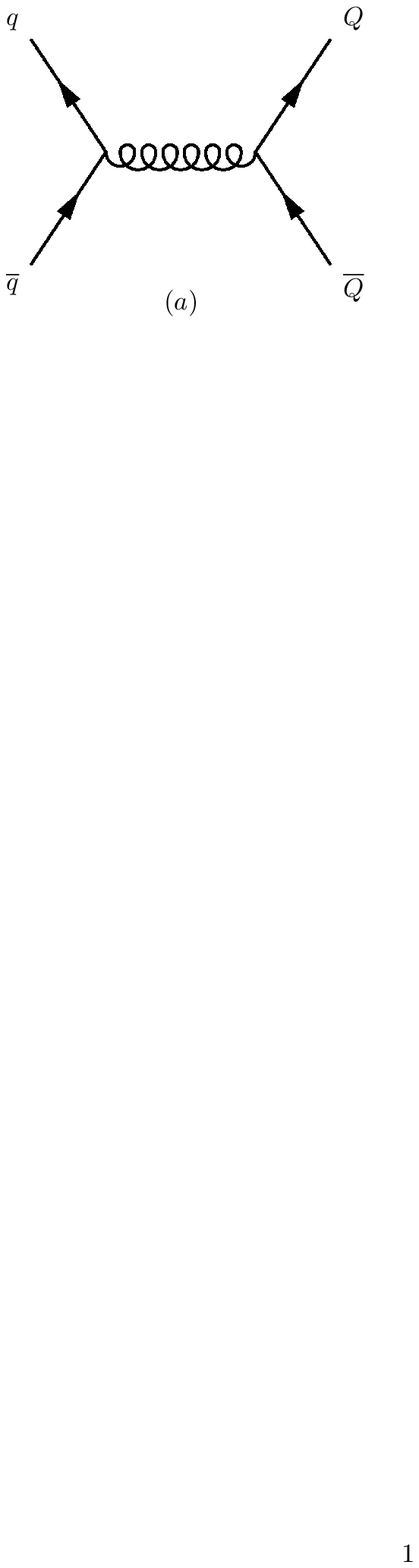}
  \includegraphics[width=0.5\columnwidth]{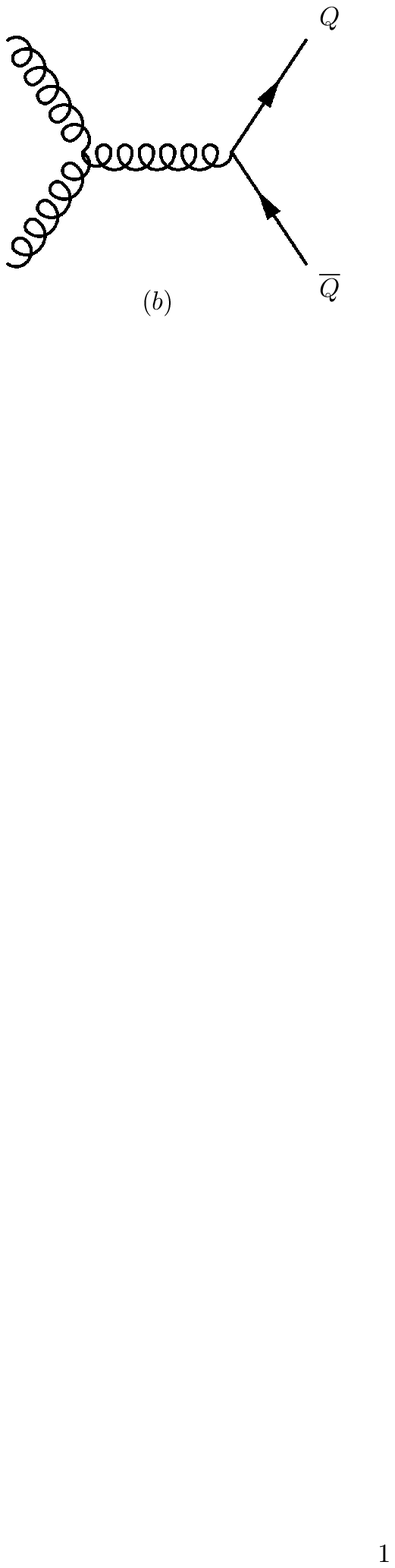} \\
  \includegraphics[width=0.5\columnwidth]{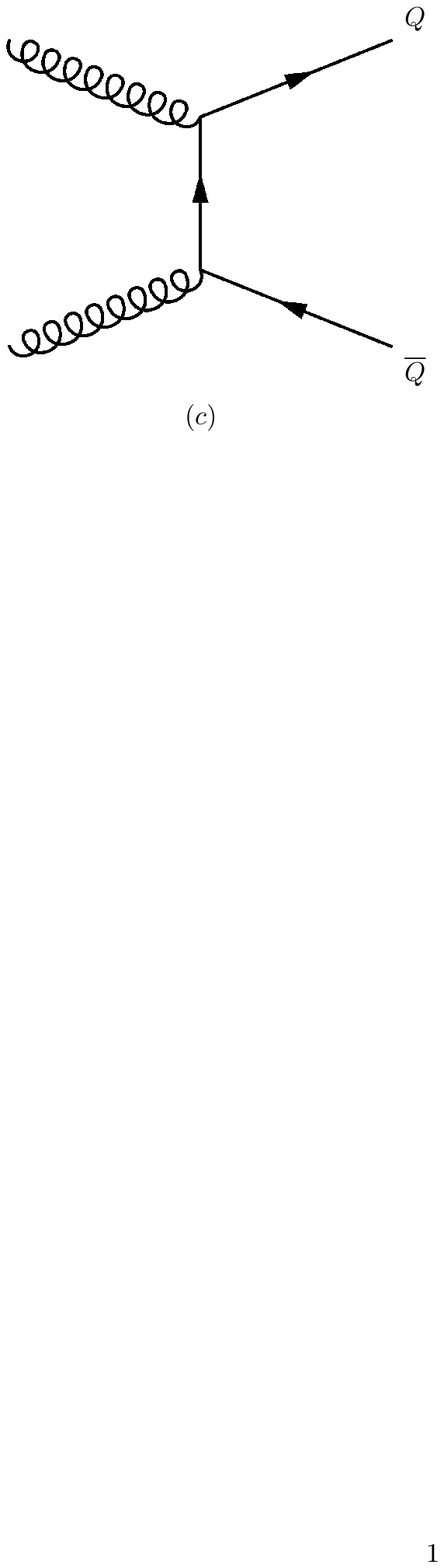}
  \includegraphics[width=0.5\columnwidth]{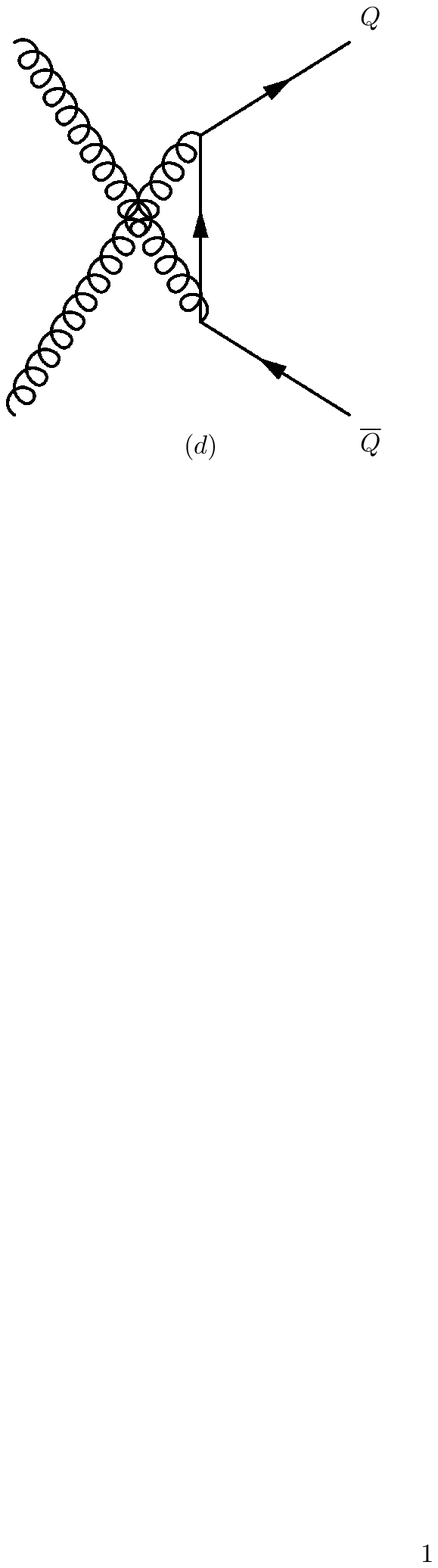}
  \caption[]{Leading order $Q \overline Q$ production diagrams with (a) the
    $q \overline q$ initial state and (b)-(d) the $gg$ contributions.}
  \label{lodias}
\end{figure}

In an event generator like PYTHIA or HERWIG, heavy flavor production is divided
into three different categories: flavor or pair creation, flavor excitation,
and gluon splitting \cite{Bedjidian:2004gd}.
This classification depends on the number of heavy quarks
in the final state of the hard process, defined as the process in the event
with the highest virtuality.  Pair creation is equivalent to the four leading
order diagrams, shown in Fig.~\ref{lodias}.  In this case the final state has
two heavy quarks, the $Q$ and $\overline Q$.  In the case of flavor excitation,
a heavy quark from the splitting $g \rightarrow Q \overline Q$ in an
initial-state parton shower is put on mass shell by scattering with a parton,
either a quark or gluon, from the other beam, $q Q \rightarrow qQ$ or
$gQ \rightarrow gQ$.  There is thus one heavy quark in the final state of the
hard scattering.  Finally, gluon splitting is defined as having no heavy
flavor in the hard scattering such as $gg \rightarrow gg$.  The $Q \overline Q$
pair is produced in an initial- or final-state parton shower via
$g \rightarrow Q \overline Q$.  Double counting of these processes is avoided
by requiring that the hard scattering should be of greater virtuality than the
parton shower \cite{Bedjidian:2004gd}.

\begin{figure}[htpb]\centering
  \includegraphics[width=0.5\columnwidth]{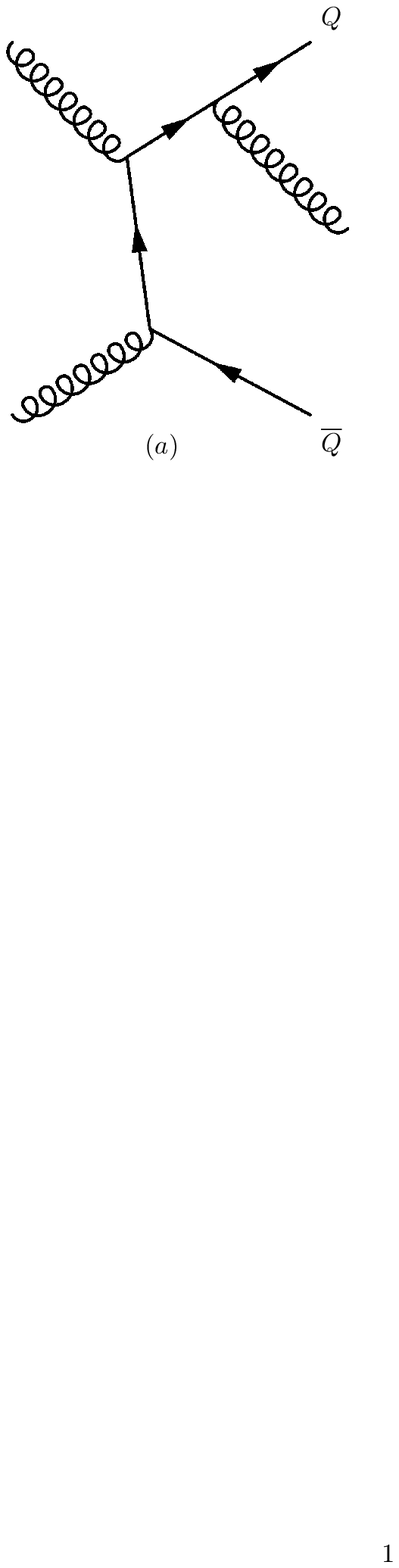}
  \includegraphics[width=0.5\columnwidth]{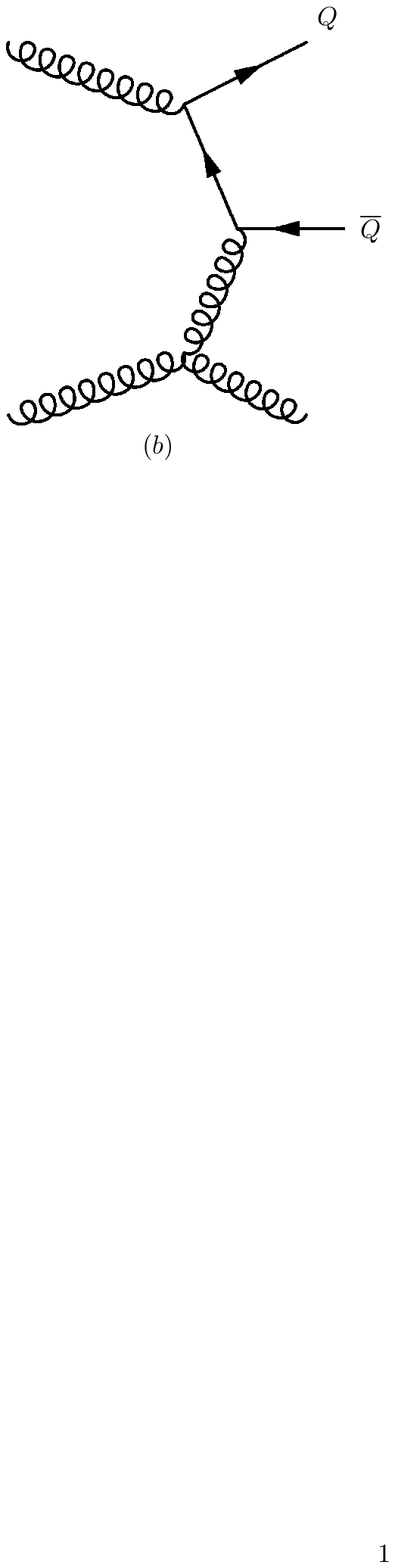} \\
  \includegraphics[width=0.5\columnwidth]{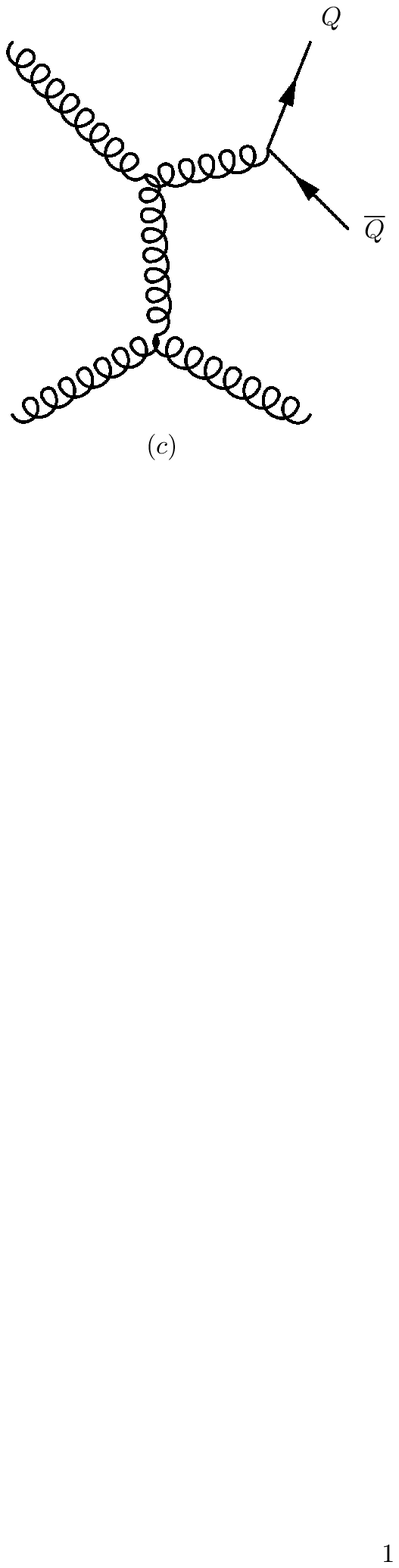}
  \includegraphics[width=0.5\columnwidth]{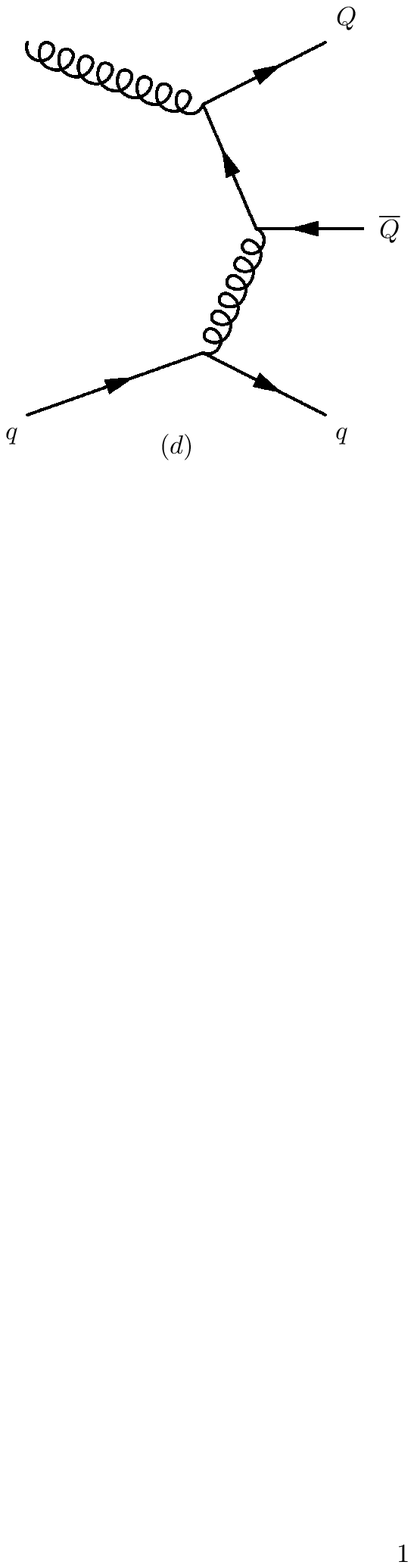} 
  \caption[]{Examples of
    real contributions to next-to-leading order $Q \overline Q$
    production.  Diagrams (a)-(c) illustrate contributions to
    $gg \rightarrow Q \overline Qg$ while (d) shows an example of
    $qg \rightarrow q Q \overline Q$ production.}
  \label{nlodias}
\end{figure}

There are parameters that can be tuned, depending on the generator employed,
that can match the distributions from the LO generator to those of a NLO
calculation.  However, that does not mean that the mix of physics processes
is identical.  For example, all three of the $Q \overline Q$ production
processes in PYTHIA contribute to perturbative QCD production at next-to-leading
order with $2 \rightarrow 3$ processes.  Pair creation is realized at NLO by
gluon emission from one of the final-state heavy quarks, as in
Fig.~\ref{nlodias}(a), the NLO version of Fig.~\ref{lodias}(c).  Flavor
excitation off a gluon from the opposite hadron is shown in
Fig.~\ref{nlodias}(b) while gluon splitting is shown in Fig.~\ref{nlodias}(c).

The diagrams in Fig.~\ref{nlodias}(a)-(c) are an illustration of the possible
NLO diagrams included in $gg$-initiated $Q \overline Q$ production,
$gg \rightarrow Q \overline Q g$.  For example, virtual corrections and crossed
diagrams are not shown in Fig.~\ref{nlodias}.  In a NLO calculation, the
amplitudes from these $gg$ contributions are summed with weights determined by
the spin and color factors of each diagram,
not parton showers and individual virtualities.
Instead, production processes are separated by the initial state.  The $gg$
initial state is dominant at collider energies over most of the rapidity range.
The next largest contribution is due to $qg$ or $\overline qg$ scattering,
$qg \rightarrow q Q \overline Q$, shown in Fig.~\ref{nlodias}(d), and
not present at leading order.  This
diagram is considered to be flavor excitation in event generators.
Finally, real and virtual corrections in the $q \overline q$ channel,
$q \overline q \rightarrow Q \overline Qg$, makes a small contribution to
$Q \overline Q$ production at high energies.

Due to the common use of event generators in simulations and analysis, there is
a great temptation to try and analyze heavy flavor production in terms of
separate production mechanisms, as characterized by creation, excitation and
splitting.  However, as described here, there is no distinction, all three are
components of NLO production.  Separating the diagrams in the initial $gg$
process in such a manner will result in improper interferences in the
calculation and an incorrect cross section.  Thus it is preferable
to do these types of analyses with an exclusive NLO code such as described in
Sec.~\ref{sec:exclusive}.

\subsection{$k_T$ broadening and fragmentation}

The transition from bare quark distributions to those of the final-state
hadrons is accomplished by including a fragmentation function  and intrinsic
transverse momentum, $k_T$, broadening.  The implementation of these two
effects are described here.

\subsubsection{Intrinsic $k_T$ broadening}

Results on open heavy flavors at fixed-target energies 
indicated that some level of
transverse momentum broadening was needed to obtain agreement with the low $p_T$
data after fragmentation was applied \cite{MLM1}.
Broadening is typically applied by including some intrinsic transverse momentum,
$k_T$, smearing to the initial-state parton densities.
The implementation of intrinsic $k_T$ in HVQMNR is not handled in the 
same way as calculations of other hard processes due to the nature of the code.
In HVQMNR, the cancellation of divergences is done numerically.  
Since adding further numerical Monte-Carlo integrations would slow the
simulation of events, as well as require multiple runs in the same
kinematics but with different intrinsic $k_T$ kicks, the kick is added in the
final, rather than the initial, state. 
The Gaussian function $g_p(k_T)$ \cite{MLM1},
\begin{eqnarray}
g_p(k_T) = \frac{1}{\pi \langle k_T^2 \rangle} \exp(-k_T^2/\langle k_T^2
\rangle) \, \, ,
\label{intkt}
\end{eqnarray}
multiplies the parton
distribution functions for both hadrons, 
assuming the $x$ and $k_T$ dependencies in the initial partons completely
factorize.  If factorization applies, 
it does not matter whether the $k_T$ dependence
appears in the initial or final state as long as the kick is not too large.  
In Ref.~\cite{MLM1}, $\langle k_T^2 \rangle = 1$ GeV$^2$ was chosen
to describe the $p_T$ dependence of fixed-target charm production.

In HVQMNR, the $Q \overline Q$ system is boosted to the rest frame
from its longitudinal center-of-mass frame.  Intrinsic transverse
momenta of the incoming partons, $\vec k_{T 1}$ and $\vec k_{T 2}$, are chosen
at random with $k_{T 1}^2$ and $k_{T 2}^2$ distributed according to
Eq.~(\ref{intkt}).   A second transverse boost out of the pair rest frame
changes the initial transverse momentum of
the $Q \overline Q$ pair, $\vec p_T$, to
$\vec p_T + \vec k_{T 1} + \vec k_{T 2}$.  The initial
$k_T$ of the partons could have alternatively been given to the entire
final-state system, as is essentially done if applied in the initial state,
instead of to the $Q \overline Q$ pair.  There is no difference if the
calculation is LO but at NLO an additional light parton can 
also appear in the final state, making the correspondence inexact.  
In Ref.~\cite{MLM1}, the difference between the two implementations is claimed 
to be small if $\langle k_T^2 \rangle \leq 2$ GeV$^2$.  The $p_T$-integrated
rapidity distribution is unaffected by the
intrinsic $k_T$.

The level of the intrinsic $k_T$ kick required in these calculations
was determined by comparison with the
shape of the quarkonium $p_T$ distributions.  In this case, the Color
Evaporation Model is employed to calculate production with a cut on the
pair invariant mass in the HVQMNR code,
giving an upper limit of $2m_H$ where $H = D$, $B$ for
charm and bottom respectively.  The CEM pair distributions require
augmentation by $k_T$ broadening to make them finite at $p_T \rightarrow 0$,
as do the $Q \overline Q$ pair $p_T$ distributions.  The
$k_T$ broadening has the most important effect on the azimuthal
correlation at low $p_T$ as well, as we will show.

The effect of the $k_T$ kick on the $p_T$ distribution can be expected to
decrease as $\sqrt{s}$ increases because the average $p_T$ also increases 
with energy.  However, the value of $\langle k_T^2 \rangle$ is assumed to
increase with $\sqrt{s}$ so that effect remains important for low $p_T$
production at higher energies.
The energy dependence of $\langle k_T^2 \rangle$ in Ref.~\cite{NVF} is
\begin{eqnarray}
  \langle k_T^2 \rangle = 1 + \frac{1}{n} \ln \left(\frac{\sqrt{s}}{20 \,
    {\rm GeV}} \right) \, \, {\rm GeV}^2 \, \, .
\label{eq:avekt}
\end{eqnarray}
Comparison with the RHIC $J/\psi$ data found that $n = 12$ gave the best 
description of the $J/\psi$
$p_T$ distribution both at central and forward rapidity
\cite{NVF}.  The value of $n$ was unchanged in the Improved Color Evaporation
Model \cite{MaVogt}.
A smaller value of $n$ and thus a larger $\langle k_T^2 \rangle$ is required for
the $\Upsilon$ $p_T$ distribution.  For $\Upsilon$, $n = 3$ is 
set by comparison to the
Tevatron results at $\sqrt{s} = 1.8$ TeV \cite{NVFinprep}.
These same values are also used in the charm and bottom pair distributions.

\subsubsection{Fragmentation}

\begin{figure}[htpb]
  \includegraphics[width=\columnwidth]{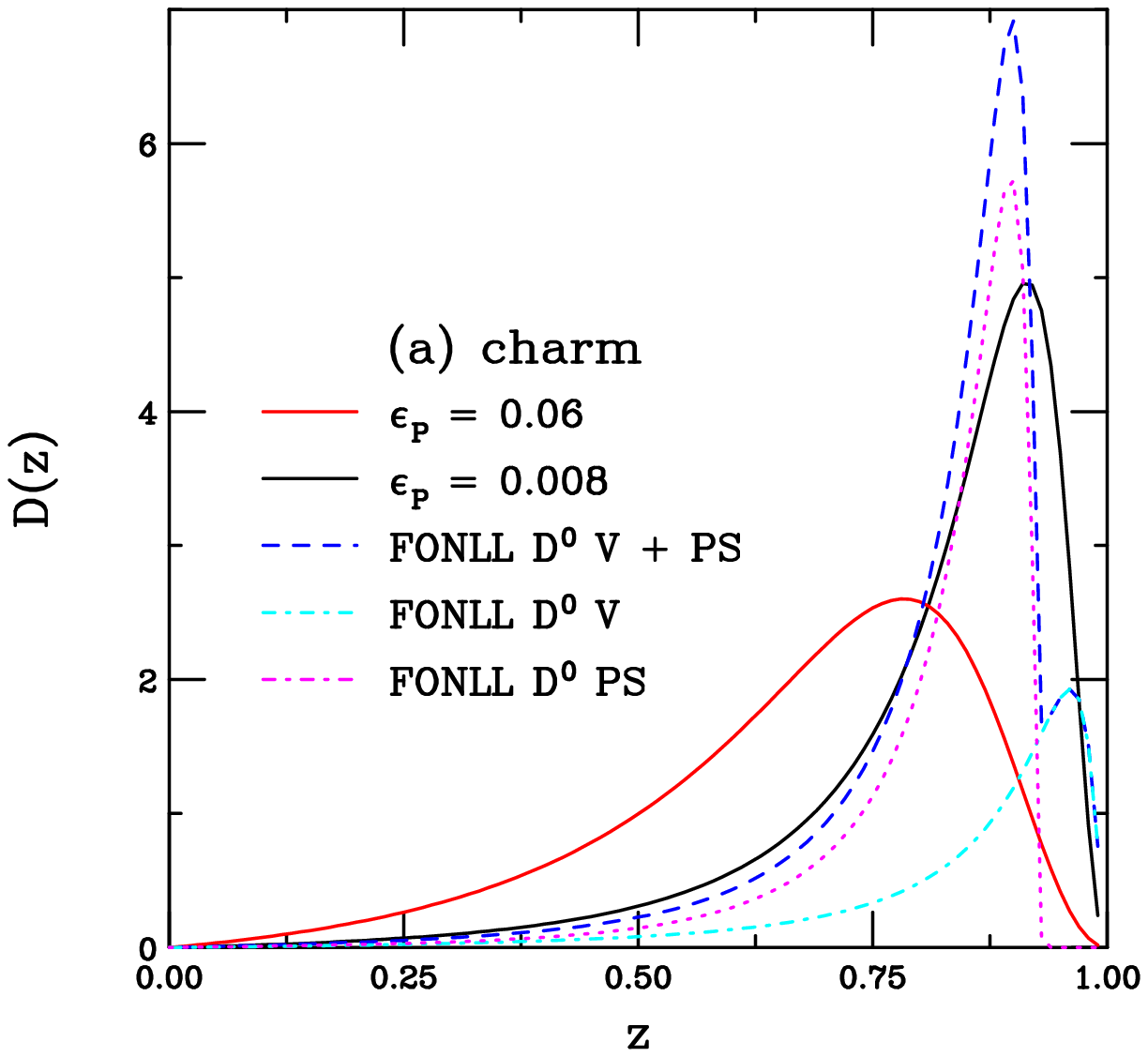}
  \includegraphics[width=\columnwidth]{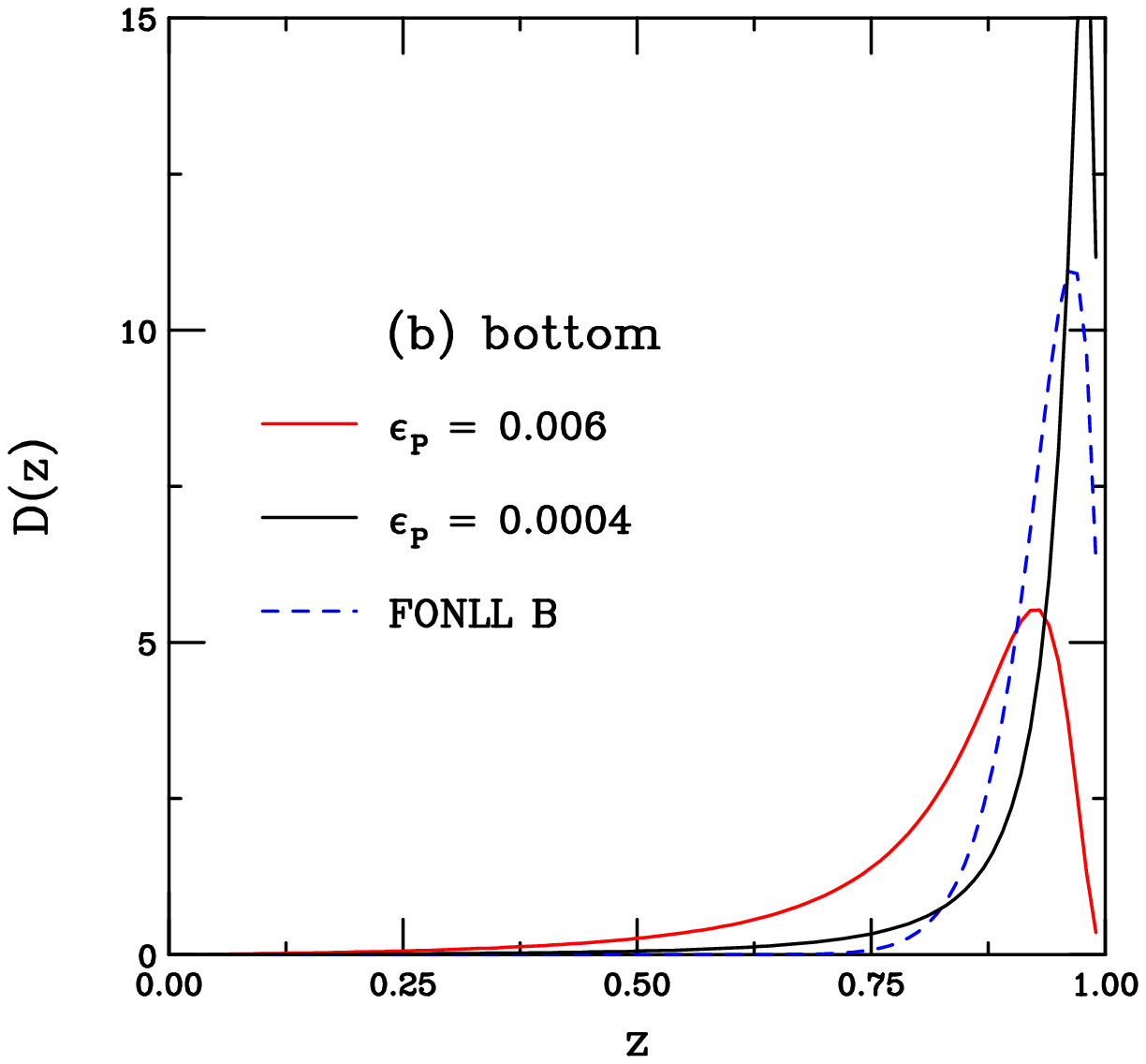}
  \caption[]{(Color online)
    The fragmentation functions used in the HVQMNR code and FONLL for (a)
    charm and (b) bottom are compared.  The red curves show the standard
    Peterson function parameter while the black curves are calculated with
    the values of $\epsilon_P$ used in this paper.  The FONLL results
    are shown in
    the dashed blue curves.  For charm quarks, the total FONLL contribution to
    $D^0$ fragmentation includes the vector (V) and pseudoscalar (PS)
    contributions, shown separately.
  }
  \label{frag_fig}
\end{figure}

The default fragmentation function in HVQMNR is
the Peterson function \cite{Pete},
\begin{eqnarray}
  D(z) = \frac{z(1-z)^2}{((1-z)^2 + z \epsilon_P)^2} \, \, ,
  \label{Eq.Pfun}
\end{eqnarray}
where $z$ represents the fraction of the parent heavy
flavor quark momentum carried by the resulting heavy flavor hadron.
In the original Peterson function, the nominal values of the fragmentation
parameter $\epsilon_P$ were 0.06 for
charm and 0.006 for bottom.  These values result in $\langle z \rangle = 0.671$
for charm and 0.828 for bottom, a reduction in the heavy quark momentum of
$\sim 33$\% and 17\% respectively, as shown by the red curves in
Fig.~\ref{frag_fig}.  Currently, the Peterson
function with $\epsilon_P = 0.06$ is considered too strong for charm
production.  The FONLL
fragmentation scheme for open heavy flavor is softer \cite{CNV}, as will be
discussed.

The value of $\epsilon_P$ in the Peterson
fragmentation function must be modified to match the FONLL $p_T$ distribution
for single $D$ and $B$ meson production since the standard $\epsilon_P$ values
for Eq.~(\ref{Eq.Pfun}) result in much softer $p_T$ distributions than those of
FONLL.  Therefore, the value of $\epsilon_P$ needs to be
reduced to match
the average ratio of the heavy quark momentum transferred to the final state
hadron, $\langle z \rangle$.  The average $z$ is calculated as
\begin{eqnarray}
  \langle z \rangle = \frac{\int_0^1 dz z D(z)}{\int_0^1 dz D(z)} \, \, .
\end{eqnarray}

Fragmentation functions for $D$ and $D^*$
mesons have been calculated in an approach consistent with an
FONLL calculation \cite{CN_CDF}.
The FONLL charm fragmentation function was determined from Mellin moments
of the distributions calculated consistently in the same framework.  The $D^0$
fragmentation function includes both pseudoscalar and vector parts that
account for the ground state, $c \rightarrow D^0$,
and excited state contributions, $c \rightarrow D^{*0} \rightarrow D^0$ and
$c \rightarrow D^{*+} \rightarrow D^0$ respectively.  The fragmentation
parameter for $D^0$ with a central charm quark mass of $m = 1.5$~GeV \cite{CNV}
was adjusted to the chosen central value of the charm quark mass, 1.27~GeV,
used here.  Assuming a linear
dependence of $\epsilon_P$ on charm mass from 1.2 to 1.7~GeV, the total FONLL
charm fragmentation function is given by the blue dashed curve in
Fig.~\ref{frag_fig}(a).  The vector and pseudoscalar contributions are shown in
the cyan and magenta curves respectively.  Individually, their average $z$
values are $\langle z \rangle = 0.812$ for the pseudoscalar channel
and 0.843 for the more massive $D^*$ in the vector channel.  Note that the
vector channel is a small overall contribution to the combined fragmentation
function, as is reflected in the average $z$ for the total FONLL fragmentation
function, 0.822.  This $\langle z \rangle$
can be compared to the result obtained with the
default value of $\epsilon_P$ in the Peterson function,
$\langle z \rangle = 0.671$.  The calculations in
this work will use the reduced value, $\epsilon_P = 0.008$, shown in the black
curve in Fig.~\ref{frag_fig}(a).  Using this $\epsilon_P$,
$\langle z \rangle = 0.820$, in good agreement with the average $z$ for
the combined FONLL charm fragmentation function.  As will be shown in 
Sec.~\ref{sec:ptdists}, the single inclusive $D$ meson $p_T$ distribution
calculated with HVQMNR for this $\epsilon_P$, combined with the value of
$\langle k_T^2 \rangle$ determined from RHIC with the energy dependence of
Eq.~(\ref{eq:avekt}), is in good agreement with the FONLL $p_T$ distribution.

The bottom quark fragmentation function in FONLL is of the form
\begin{eqnarray}
  D(z) = z(1-z)^{\epsilon_b}
\end{eqnarray}
where $\epsilon_b = 34$ for a bottom quark mass of 4.75~GeV \cite{CNV}.  Again,
assuming a linear dependence of $\epsilon_b$ on $b$ quark mass from
4.5 to 5~GeV, with
$m_b = 4.65$~GeV, $\epsilon_b = 27.5$.  The FONLL fragmentation function with
this value of $\epsilon_b$, is shown in the dashed blue curve of
Fig.~\ref{frag_fig}(b).  In this case, $\langle z \rangle = 0.934$.
In contrast, the red curve shows the default Peterson
function result for $\epsilon_P = 0.006$, with $\langle z \rangle = 0.828$.
If the same overall reduction of
the Peterson function parameter for charm is used for bottom, from 0.006 to
0.0008, $\langle z \rangle = 0.911$, insufficient to replicate the FONLL $B$
meson $p_T$ dependence.  Thus, $\epsilon_P = 0.0004$ is used here, giving
$\langle z \rangle = 0.930$, in good agreement with the average $z$ from FONLL,
as shown in the solid black curve of Fig.~\ref{frag_fig}(b).

The charm and bottom $p_T$ distributions employing these fragmentation functions
will be compared to those from FONLL in the next section.

\subsection{Single Inclusive Heavy Flavor Distributions}
\label{sec:ptdists}

In the calculations reported in this paper, the same values of the charm quark
mass and scale parameters as in Ref.~\cite{NVF} are employed here,
$(m,\mu_F/m_T, \mu_R/m_T) = (1.27 \pm 0.09 \, {\rm GeV}, 2.1^{+2.55}_{-0.85}, 1.6^{+0.11}_{-0.12})$ where $\mu_F$ is the factorization scale and $\mu_R$ is the
renormalization scale.  In the case of bottom production,   
$(m,\mu_F/m_T, \mu_R/m_T) = (4.65 \pm 0.09 \, {\rm GeV}, 1.4^{+0.77}_{-0.49},
1.1^{+0.22}_{-0.20})$ is used \cite{NVFinprep}.  The CT10 proton parton densities
\cite{CT10} are employed in the calculations.  The scale factors, $\mu_F$ and
$\mu_R$, are defined relative to the transverse mass of the pair,
$\mu_{F,R} \propto m_T = \sqrt{m^2 + p_T^2}$ where 
the $p_T$ is the $Q \overline Q$ pair $p_T$, 
$p_T^2 = 0.5(p_{T_Q}^2 + p_{T_{\overline Q}}^2)$.  At LO in the total cross section,
the $Q \overline Q$ pair
$p_T$ is zero.  Thus, while the calculation is
NLO in the total cross section, it is LO in the pair distributions. 
In the exclusive NLO calculation ~\cite{MNRcode}
both the $Q$ and $\overline Q$ variables
are retained to obtain the pair distributions.  Unless otherwise noted, the
calculations employ the central values of the heavy quark mass and scale
factors.

First, the relative importance of $k_T$ broadening and fragmentation
on the single inclusive heavy quark $p_T$ distribution is studied.  Undertaking
a study of the azimuthal correlations between heavy quarks requires finding
the values of $\langle k_T^2 \rangle$ and $\epsilon_P$ appropriate for
a reasonably faithful reproduction of the
FONLL single inclusive heavy quark distributions in the same kinematics
employing the HVQMNR code.
These same values of $\langle k_T^2 \rangle$ and $\epsilon_P$
are then used in the calculation of the azimuthal
distributions.  The values of $\langle k_T^2 \rangle$ obtained for $J/\psi$
and $\Upsilon$ production are assumed as a default.

\begin{figure}[htpb]
  \includegraphics[width=\columnwidth]{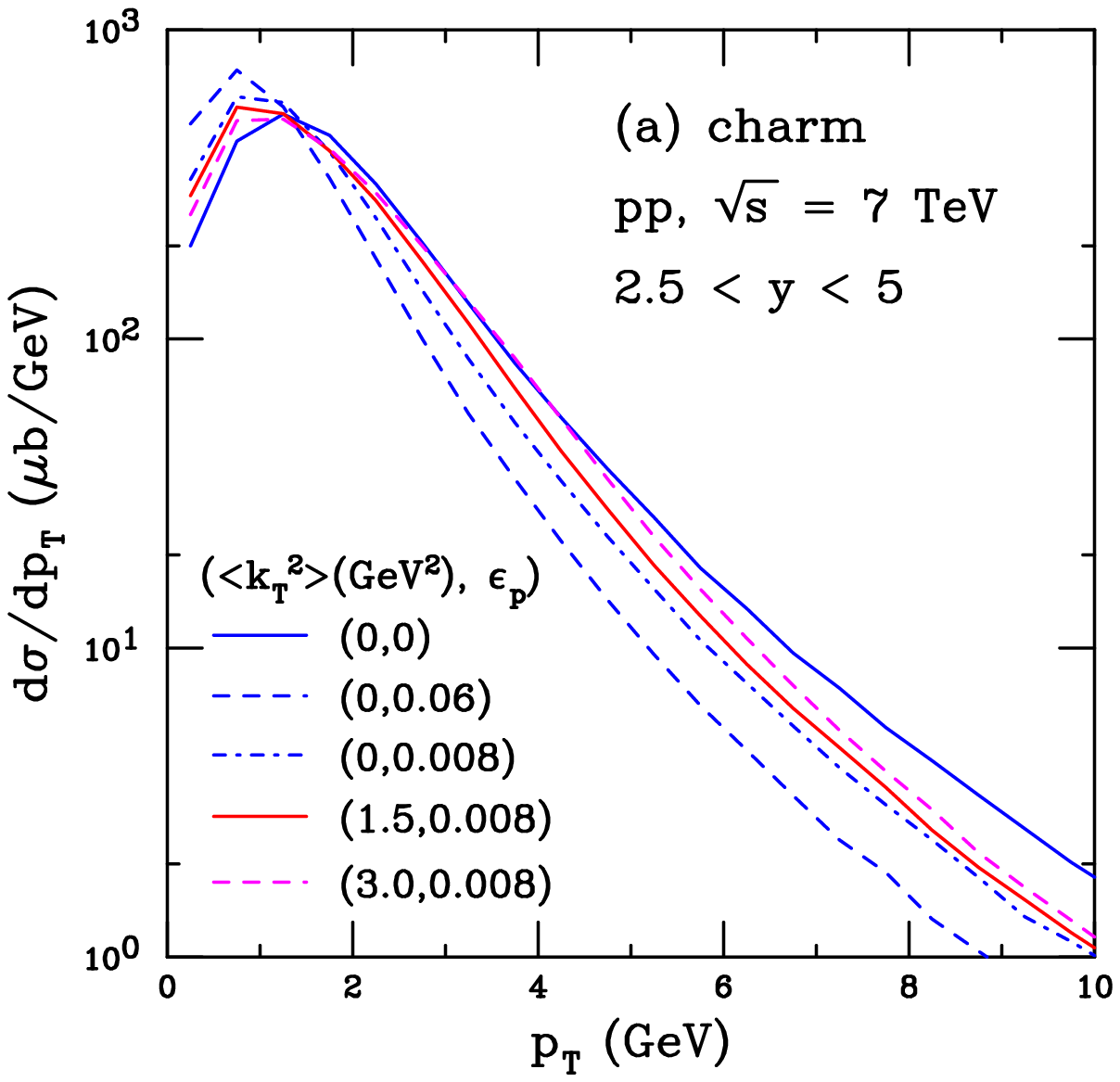}
  \includegraphics[width=\columnwidth]{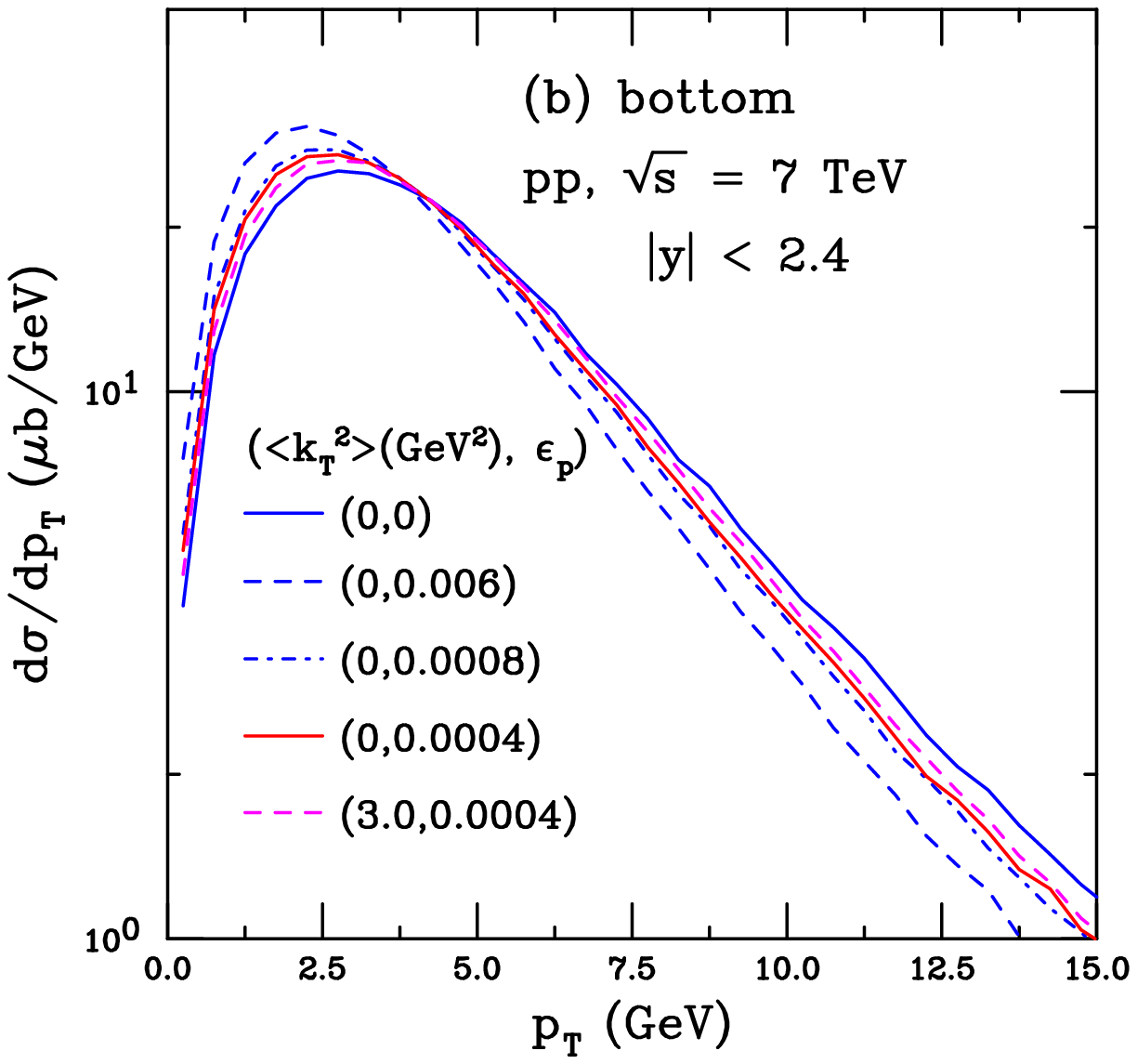}
  \caption[]{(Color online)
    The single inclusive (a) charm and (b) bottom quark distributions
    in $\sqrt{s} = 7$~TeV $p+p$ collisions at next-to-leading order using the
    HVQMNR code.  The charm
    distributions are given at forward rapidity, $2.5 < y < 5$, while the bottom
    quark distributions are given at midrapidity, $|y| < 2.4$.  Results are
    shown for various combinations of $\langle k_T^2 \rangle$ 
    (Eq.~(\protect\ref{eq:avekt})) and $\epsilon_P$
    (Eq.~(\protect\ref{Eq.Pfun})).
  }
  \label{fig1}
\end{figure}

Figure~\ref{fig1} shows how $\langle k_T^2 \rangle$ and $\epsilon_P$ affect the
low $p_T$ part of the spectrum at $\sqrt{s} = 7$~TeV.  As examples,
the charm distributions
are shown at forward rapidity, $2.5 < y < 5$, the LHCb acceptance
\cite{LHCbDmesons}, while the
bottom distributions are shown at central rapidity, $|y| < 2.4$, the rapidity
region of $b$ hadron distributions reported by ATLAS
\cite{ATLAS_bhad}.
Note that the
lower $p_T$ range is emphasized in Fig.~\ref{fig1} because this region is most
affected by $k_T$ broadening, as discussed in more detail later.  

The blue solid curves in Fig.~\ref{fig1} are the bare quark distributions, with
$\langle k_T^2 \rangle = 0$, $\epsilon_P = 0$.  This distribution is the
hardest of all the cases shown and thus has the highest average $p_T$.  (See
Table~\ref{table:avept} for the values of $\langle p_T \rangle$ with each
$\langle k_T^2 \rangle$, $\epsilon_P$ combination.)

\begin{table}
  \begin{tabular}{|c|c|c||c|c|c|}\hline
    \multicolumn{3}{|c|}{charm} & \multicolumn{3}{|c|}{bottom} \\ \hline
    $\langle k_T^2 \rangle$ & $\epsilon_p$ & $\langle p_T \rangle$
    & $\langle k_T^2 \rangle$ & $\epsilon_p$
    & $\langle p_T \rangle$ \\ 
    (GeV$^2$) & & (GeV) & (GeV$^2$) & & (GeV) \\ \hline
    0 & 0 & 1.96 & 0 & 0 & 5.38 \\
    0 & 0.06 & 1.31 & 0 & 0.006 & 4.45 \\
    0 & 0.008 & 1.58 & 0 & 0.0008 & 4.90 \\
    1.5 & 0.008 & 1.71 & 0 & 0.0004 & 5.00 \\
    3 & 0.008 & 1.84 & 3 & 0.0004 & 5.12 \\ \hline
  \end{tabular}
  \caption[]{Average $p_T$ for single inclusive charm and bottom quark
    production for the chosen
    values of $\langle k_T^2 \rangle$ and $\epsilon_P$.}
\label{table:avept}
\end{table}

The dashed curves, with the lowest average $p_T$, include no broadening but
employ the default
Peterson fragmentation function parameter from $e^+ e^-$ data
\cite{Chirin}, $\epsilon_P = 0.06$ for charm and 0.006 for bottom, corresponding
to the red curves in Fig.~\ref{frag_fig}.  The
effect on the charm quark distributions is particularly strong because this
value of $\epsilon_P$ results in a $\sim 33$\% decrease in $p_T$ relative to
the bare quark, as already discussed.
At fixed-target energies, such a strong reduction in momentum
could only be made compatible with data, which agreed rather well with the
low $p_T$ bare
charm quark distribution, by setting $\langle k_T^2 \rangle = 1$~GeV$^2$
\cite{MLM1}.
For fixed-target energies, $\sqrt{s} \sim 20$~GeV, this was
sufficient because the average $p_T^2$ of the charm quark was low since
$m_c/\sqrt{s} \sim 0.065$.  Therefore,
$\langle k_T^2 \rangle \sim \langle p_T^2 \rangle$.
However, at collider energies, $m_c/\sqrt{s}$ is
small, $\sim 1.8 \times 10^{-4}$ at $\sqrt{s} = 7$~TeV.
Thus an average $\langle k_T^2 \rangle$ on the
order of $1-3$~GeV$^2$ has a rather small effect on the shape of the $p_T$
distribution, particularly at high $p_T$.
On the other hand, $\epsilon_P$ reduces the quark $p_T$ uniformly,
independent of the center-of-mass energy.

\begin{figure}[htpb]
  \includegraphics[width=\columnwidth]{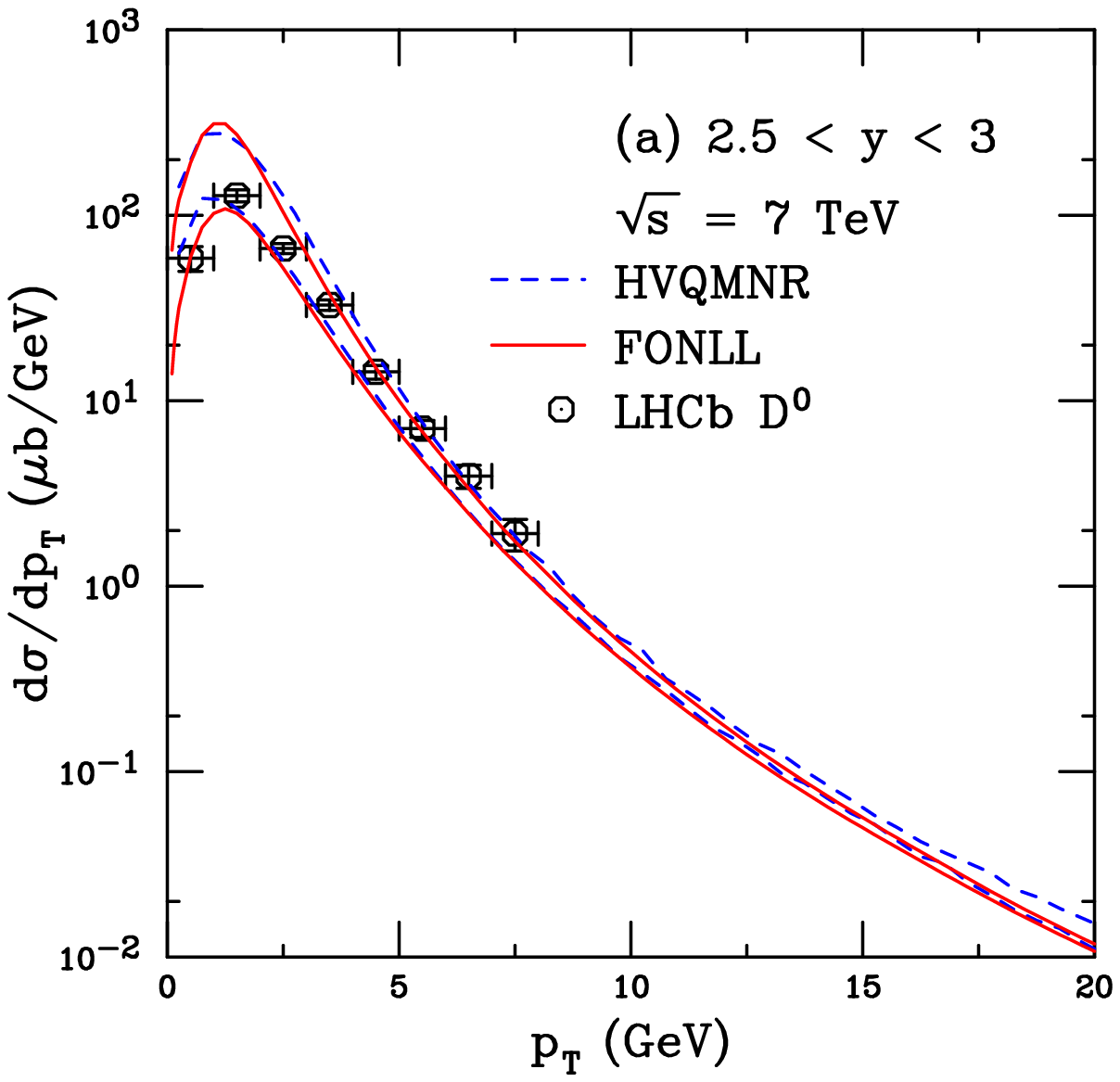}
  \includegraphics[width=\columnwidth]{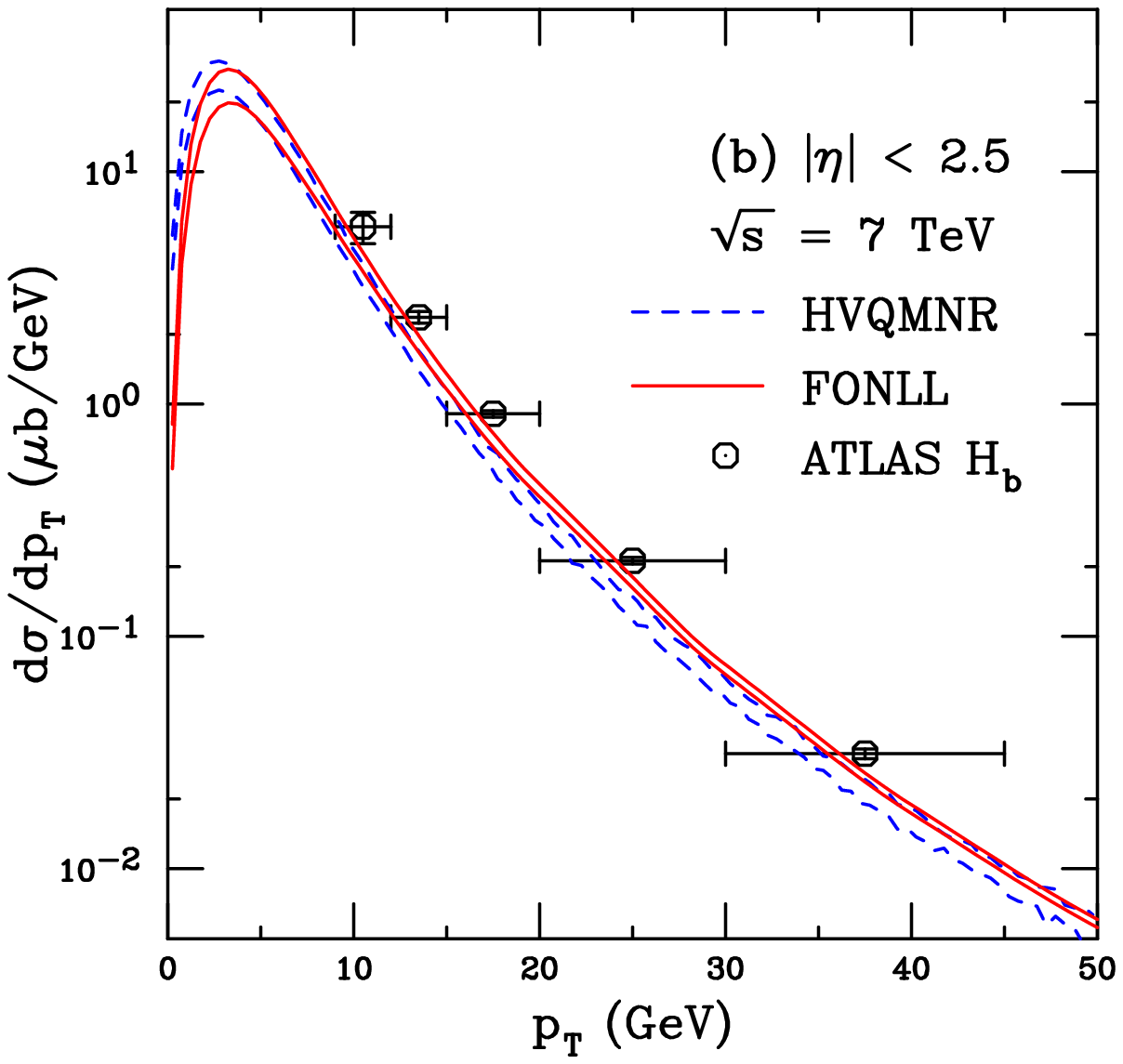}
  \caption[]{(Color online)
    The single inclusive (a) $D^0$ and (b) $b$-quark hadron, $H_b$,
    distributions in $\sqrt{s} = 7$~TeV $p+p$ collisions are compared to data
    from LHCb \protect\cite{LHCbDmesons} at $2.5 < y < 3$ and
    ATLAS \protect\cite{ATLAS_bhad} at $|\eta| < 2.5$
    respectively.  The curves show the extent of the uncertainty
    bands calculated employing Eqs.~(\protect\ref{sigmax}) and
    (\protect\ref{sigmin}).  The HVQMNR code (blue dashed curves) utilizes
    $(\langle k_T^2 \rangle ({\rm GeV}^2), \epsilon_P) = (1.5,0.008)$ for
    charm and (3,0.0004) for bottom.  The corresponding FONLL uncertainty band
    (red curves) is also shown.  The same quark mass and scale parameters are
    used in both calculations.
    }
  \label{fig2}
\end{figure}

The effect of fragmentation on the bottom quark distribution is less striking
because $\epsilon_P$ is an order of magnitude smaller.  Nonetheless,
as will be shown in Fig.~\ref{fig2}, it is clear that even this milder effect
may be too strong to be compatible with the $b$-hadron data.

While $\langle k_T^2 \rangle \sim 1$~GeV$^2$
was sufficient to mitigate the effect of fragmentation
on charm production at $\sqrt{s} = 20$~GeV \cite{MLM1},
it is clear that any value of
$\langle k_T^2 \rangle$ large enough to do so at $\sqrt{s} = 7$~TeV would be
too large to be physical.  Thus, a reduction of $\epsilon_P$ by a factor of
$\sim 7.5$, to 0.008 for charm and 0.0008 for bottom, was checked (dot-dashed
curves in Fig.~\ref{fig1}).
This reduction is adequate for charm production, giving a result intermediate
to the free quark distribution and that with the standard Peterson
fragmentation parameter $\epsilon_P$.  However, it was found
necessary to reduce $\epsilon_P$ by an additional factor of two,
to 0.0004, for bottom
(red solid curve in Fig.~\ref{fig1}(b)).  This additional reduction
is sufficient
to produce a $b$-hadron distribution similar to that of FONLL when broadening
is included.

Next, the effect of $\langle k_T^2 \rangle$ broadening is introduced in addition
to fragmentation.  As mentioned previously, the
default values of $\langle k_T^2 \rangle$ for
charm and bottom are assumed to be those found to agree with low $p_T$
quarkonium production in the Improved Color Evaporation Model \cite{MaVogt}.
The energy dependence of $\langle k_T^2 \rangle$ is given in
Eq.~(\ref{eq:avekt}).  At $\sqrt{s} = 7$~TeV, Eq.~(\ref{eq:avekt}) results in
$\langle k_T^2 \rangle \sim 1.5$~GeV$^2$ for charm and $\sim 3$~GeV$^2$ for
bottom.  It is clear that even these relatively large values, although still
less than $\langle p_T^2 \rangle$ at $\sqrt{s} = 7$~TeV, have a rather small
effect on the single inclusive heavy quark $p_T$ distribution.  Indeed, doubling
$\langle k_T^2 \rangle$ for charm (magenta curve in Fig.~\ref{fig1}(a)) results
in only a small change in the $p_T$ distribution in the range shown.
No further increase of $\langle k_T^2 \rangle$ is shown for bottom quarks
because such values, $\langle k_T^2 \rangle > 3$~GeV$^2$,
would be too large for the assumption of the equivalence of adding
$\langle k_T^2 \rangle$ in the initial or final state
\cite{MLM1}.  

In Fig.~\ref{fig2}, the final values of $\langle k_T^2 \rangle$ and
$\epsilon_P$, $\sim 1.5$~GeV$^2$, 0.008 for charm and 3~GeV$^2$, 0.0004 for
bottom, are used to calculate the uncertainty bands on the HVQMNR results 
and compare them to those of FONLL, with its default fragmentation functions.
The same mass and scale parameters are employed in the HVQMNR and FONLL
calculations.  The results are compared
to LHCb $D^0$ meson data in Fig.~\ref{fig2}(a), and ATLAS $b$-hadron data
in Fig.~\ref{fig2}(b).  In the HVQMNR calculations, the values of
$\langle k_T^2 \rangle$ and $\epsilon_P$ are used for all mass and scale
choices.  In the FONLL calculations, the fragmentation parameter depends on the
quark mass but is independent of scale.

The mass and scale uncertainties are 
calculated 
based on results using the one standard deviation uncertainties on
the quark mass and scale parameters.  If the central, upper and lower limits
of $\mu_{R,F}/m$ are denoted as $C$, $H$, and $L$ respectively, then the seven
sets used to determine the scale uncertainty are  $\{(\mu_F/m,\mu_F/m)\}$ =
$\{$$(C,C)$, $(H,H)$, $(L,L)$, $(C,L)$, $(L,C)$, $(C,H)$, $(H,C)$$\}$.    
The uncertainty band can be obtained for the best fit sets
\cite{NVF,NVFinprep} by
adding the uncertainties from the mass and scale variations in 
quadrature. The envelope contained by the resulting curves,
\begin{eqnarray}
\frac{d\sigma_{\rm max}}{dX} & = & \frac{d\sigma_{\rm cent}}{dX} \label{sigmax}  \\
& & \mbox{} \!\!\!\!\!\!\!\!\!\!\!\!\!\!\!\!\!\!\!\!\!\!\!\!\!\!\!\!\!\!
+ \sqrt{\left(\frac{d\sigma_{\mu ,{\rm max}}}{dX} -
  \frac{d\sigma_{\rm cent}}{dX}\right)^2
  + \left(\frac{d\sigma_{m, {\rm max}}}{dX} -
  \frac{d\sigma_{\rm cent}}{dX}\right)^2} \, \, , \nonumber
\\
\frac{d\sigma_{\rm min}}{dX} & = & \frac{d\sigma_{\rm cent}}{dX} \label{sigmin}  \\
& & \mbox{} \!\!\!\!\!\!\!\!\!\!\!\!\!\!\!\!\!\!\!\!\!\!\!\!\!\!\!\!\!\!
- \sqrt{\left(\frac{d\sigma_{\mu ,{\rm min}}}{dX} -
  \frac{d\sigma_{\rm cent}}{dX}\right)^2
  + \left(\frac{d\sigma_{m, {\rm min}}}{dX} -
  \frac{d\sigma_{\rm cent}}{dX}\right)^2} \, \, , \nonumber
\end{eqnarray}
defines the uncertainty on the cross section.
The kinematic observable $X$ can be {\it e.g.} $p_T$ or azimuthal angular
separation $\phi$. In the calculation labeled ``cent'', the central values of
$m$, $\mu_F$ and $\mu_R$ are used while in the calculations with subscript
$\mu$, the mass is fixed to the central value while the scales are varied and
in the calculations with subscript $m$, the mass is varied while the scales
are held fixed.

As an example, Fig.~\ref{fig2}(a)
compares the calculations in the rapidity interval
$2.5 < y < 3$ to the LHCb $D^0$ data \cite{LHCbDmesons}.
The LHCb data in the range $2.5 < y < 5$
was divided into five rapidity bins, each of 0.5 units of rapidity in width.
The lowest of these bins is shown as an example.  The agreement between the two
calculations is generally very good.  Some disagreement becomes apparent for
$p_T > 10$~GeV where the upper limit of the band calculated using the exclusive
HVQMNR code becomes slightly harder than the FONLL result, as can be expected
because FONLL was designed to cure large logarithms of $p_T/m$ and here
$p_T/m > 10$.  This region, as will be shown, is also relative insensitive to
effects of $\langle k_T^2 \rangle$ on the azimuthal distributions.

There is a somewhat larger difference in the calculations of the $b$-hadron
distributions in Fig.~\ref{fig2}(b).  In this case, the results are shown,
along with the ATLAS data \cite{ATLAS_bhad},
at midrapidity and up to considerably higher $p_T$.  It is,
however, worth noting that, at $p_T \sim 50$~GeV, $p_T/m$ for bottom is similar
to that for charm at $p_T \sim 10$~GeV.  Here the one standard deviation fit
to the $b \overline b$ cross section gives a narrower range on $m_b$ and
$\mu_F/m_T$, $\mu_R/m_T$.
To improve the agreement between the calculations still
further it would be necessary to employ $\epsilon_P = 0$ in the HVQMNR
calculation since the FONLL $b$-hadron $p_T$ distribution is equivalent to the
HVQMNR bare $b$-quark distribution.

\section{$Q \overline Q$ Pair Azimuthal Distributions}
\label{sec:azi_dist}

\begin{figure}[htpb]
\includegraphics[width=\columnwidth]{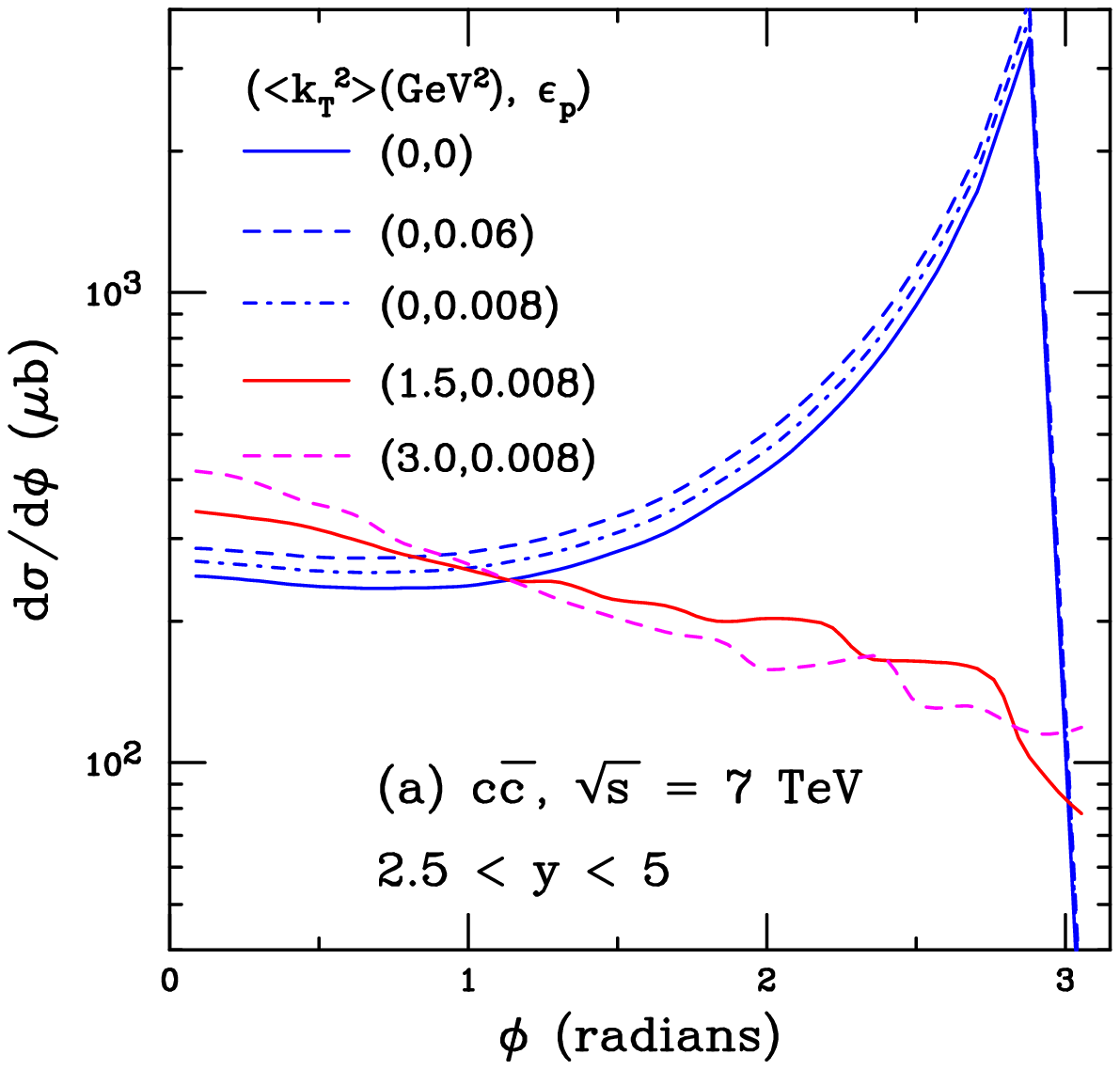}
\includegraphics[width=\columnwidth]{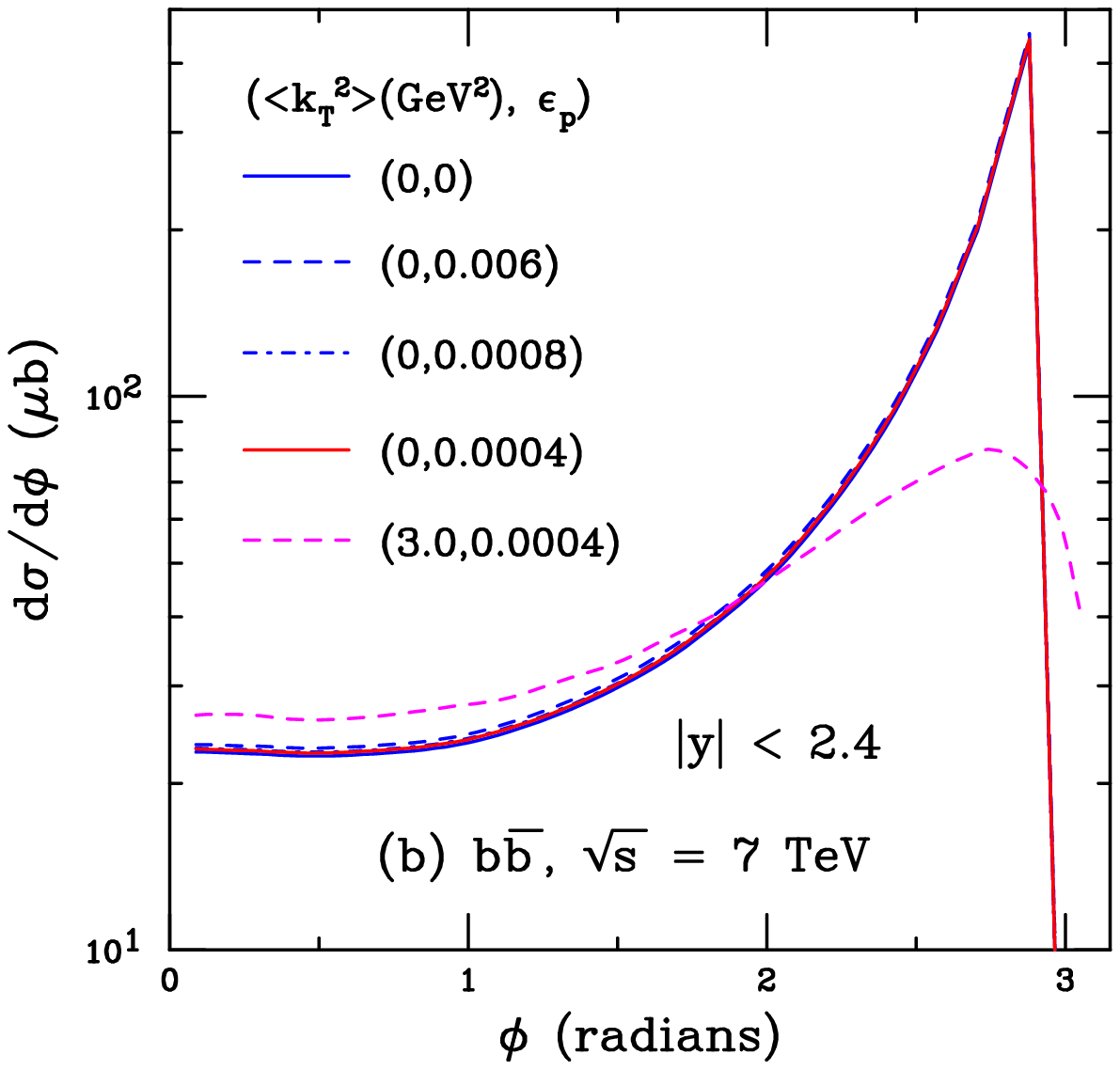}
\caption[]{(Color online)
  The NLO azimuthal distribution between two heavy quarks, $d\sigma/d\phi$
  in $p+p$ collisions at $\sqrt{s} = 7$~TeV using the HVQMNR
  code for (a) $c \overline c$ pairs at forward rapidity, $2.5 < y < 5$, and
  (b) $b \overline b$ pairs at midrapidity, $|y| < 2.4$.  The results are shown
  for the same combinations of $\langle k_T^2 \rangle$ in
  Eq.~(\protect\ref{eq:avekt}) and $\epsilon_P$ in Eq.~(\protect\ref{Eq.Pfun})
  as in Fig.~\protect\ref{fig1}.
  }
\label{fig3}
\end{figure}

The single inclusive heavy quark distributions have been calculated
both in the exclusive $Q \overline Q$ code, HVQMNR, and the inclusive-only
FONLL approach and have been shown to agree well with each other.  The next
step is to calculate the azimuthal angular separation between the two heavy
quarks.  In Fig.~\ref{fig3}, the distributions, $d\sigma/d\phi$, are calculated
to next-to-leading order for the same choices of
$\langle k_T^2 \rangle$ and $\epsilon_P$, with the central mass and scale
values, as in Fig.~\ref{fig1}, using HVQMNR only
since now an exclusive calculation is required to obtain $d\sigma/d\phi$.
First, some topological details relevant to the azimuthal distributions of
$Q \overline Q$ pair production are discussed.

The leading order perturbative QCD $2 \rightarrow 2$ diagrams for heavy
flavor production, shown in Fig.~\ref{lodias}, produce back-to-back
$Q \overline Q$ pairs with an azimuthal separation of
$\phi = \pi$.  At leading order, the $p_T$ of the pair of
heavy quarks is zero and $d\sigma/d\phi$ is represented by a delta function,
$\delta(\phi - \pi)$.
Note also that, already at leading order, the single inclusive
heavy quark $p_T$ distribution is maximally hard so that higher order
corrections do not significantly change the shape of the $p_T$ distribution
\cite{RVKfact1,RVKfact2}.

As discussed in Sec.~\ref{sec:generators}, 
at next-to-leading order, both virtual and real conntributions arise.  The
virtual corrections are typically the exchange of soft gluons at the vertices
while real corrections give rise to $2 \rightarrow 3$ processes,
see Fig.~\ref{nlodias}.
These corrections
smear out the azimuthal separation in the leading order contribution so that
while there is still a peak at $\phi = \pi$, the pairs are no longer strictly
back-to-back.  Instead, there is a tail
toward $\phi \sim 0$, even with $\langle k_T^2 \rangle = 0$
and $\epsilon_P = 0$.  

The NLO contribution can also lead to different azimuthal topologies, as
in Fig.~\ref{nlodias}(b)-(d), due to `flavor excitation' and `gluon splitting'.
These
diagrams do not harden the single inclusive heavy flavor $p_T$ distribution.
However, they do tend to make the azimuthal correlation more isotropic
\cite{Bedjidian:2004gd}.  The $2 \rightarrow 3$ processes 
produce a non-singular azimuthal
distribution at $\phi = \pi$ before any broadening or fragmentation is
taken into account.
A final-state light parton emitted
from one of the outgoing heavy quarks is likely
to be collinear to the heavy quark that emitted it.  Thus the heavy quark and
anti-quark remain close to a back-to-back configuration with a peak at
$\phi \sim \pi$.  However, if the light parton is from the initial state,
as in $qg$ scattering, this parton can be hard and its momentum can be
balanced against that of the $Q \overline Q$ pair so that $\phi \sim 0$.
In addition, the `gluon splitting' diagram of Fig.~\ref{nlodias}(c), will
result in a pair with $\phi \sim 0$ against a hard parton.
Thus, the low $p_T$ heavy quarks azimuthal distribution
is likely to be more isotropic, while high $p_T$ heavy quarks will more
likely result in a doubly-peaked $\phi$ distribution, with peaks at
$\phi \sim 0$ and $\pi$.

Fragmentation results in a hadron
with momentum nearly collinear to the original heavy quark.  Thus fragmentation
does not change the general direction of the parent heavy parton.  The resulting
heavy hadron will, however, be somewhat out of alignment with the
initial heavy quark. The larger the value of $\epsilon_P$, then, the greater
the energy lost to hadronization
and the more likely that the $Q$ and $\overline Q$ are out of
alignment, resulting in a slight enhancement at $\phi \sim 0$ relative to
$\epsilon_P = 0$.  Lower values
of $\epsilon_P$ reduce any enhancement at $\phi \sim 0$, as is evident in the
dashed and dot-dashed curves in Fig.~\ref{fig3}(a) for $\epsilon_P = 0.06$ and
0.008 respectively.

On the other hand, including the randomization of the final-state parton
momentum by $k_T$ broadening has a substantial efect on $d\sigma/d\phi$.
Indeed, the level of $k_T$ broadening required to describe $J/\psi$ production
at low $p_T$ completely changes the shape of $d\sigma/d\phi$ for
$c \overline c$, resulting in a
peak at $\phi \sim 0$ and none at $\phi \sim \pi$.  The larger value of
$\langle k_T^2 \rangle$, $\sim 3$~GeV$^2$,
increases the enhancement at $\phi \sim 0$, as shown in
Fig.~\ref{fig3}(a).

In the case of $b \overline b$ pairs, introducing $k_T$
broadening still gives a peak near $\phi \sim \pi$, even though the
$\langle k_T^2 \rangle$ for bottom shown in Fig.~\ref{fig3}(b) is as large
as the upper value of $\langle k_T^2 \rangle$ for charm in Fig.~\ref{fig3}(a).
This is because $\langle k_T^2 \rangle \sim 1.5$~GeV$^2$ is on the
order of $m_c^2$, $\sim 1.7$~GeV$^2$, while $\langle k_T^2 \rangle = 3$~GeV$^2$
for botttom is considerably less than $m_b^2 \sim 21.6$~GeV$^2$.  Thus
$\langle k_T^2 \rangle/m_c^2 \sim 1$ results in a much more isotropic $\phi$
distribution while $\langle k_T^2 \rangle/m_b^2 \sim 0.14$ results in a lower
momentum kick imparted to a bottom quark.

\begin{figure}[htpb]\centering
  \includegraphics[width=\columnwidth]{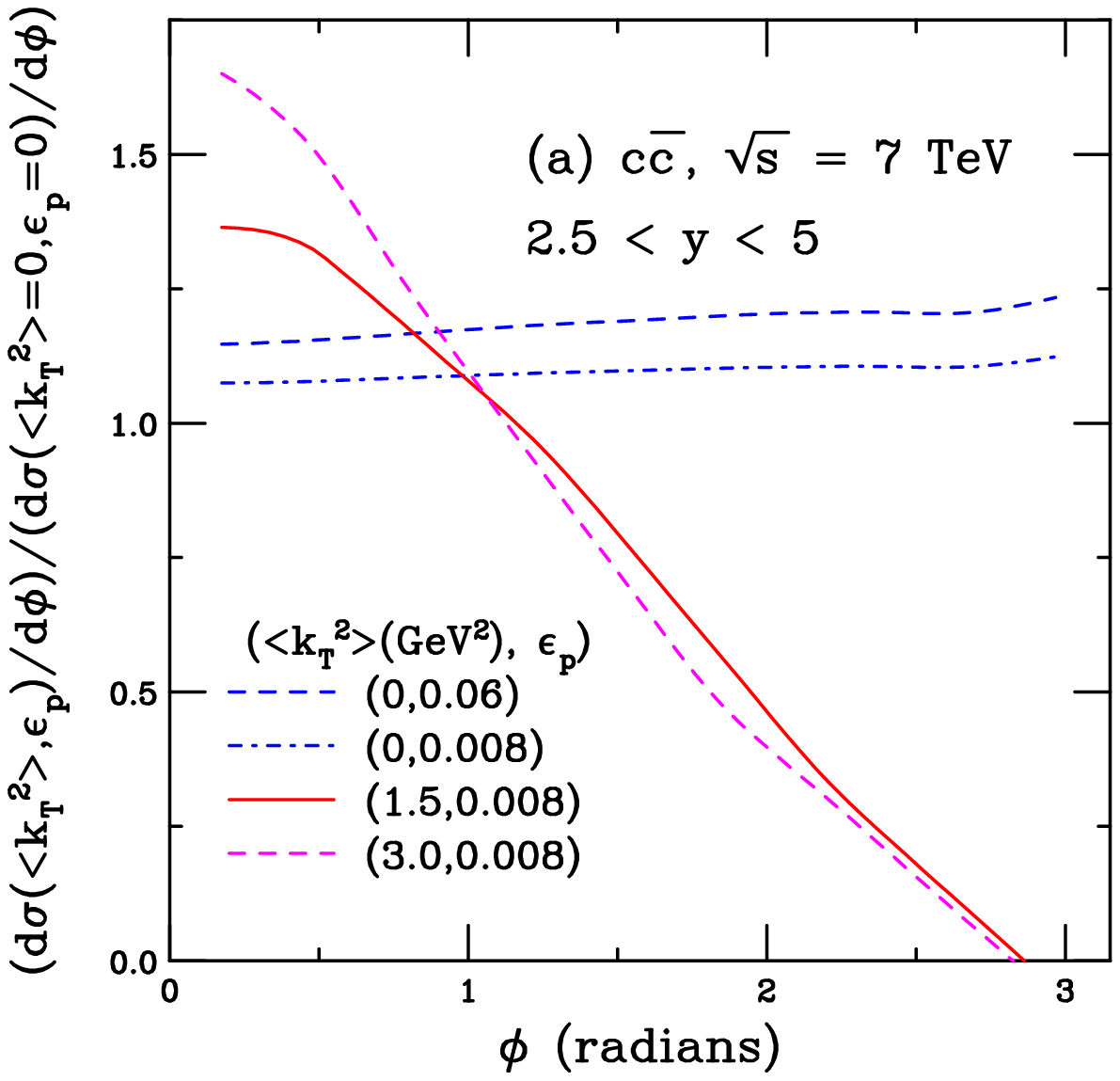}
  \includegraphics[width=\columnwidth]{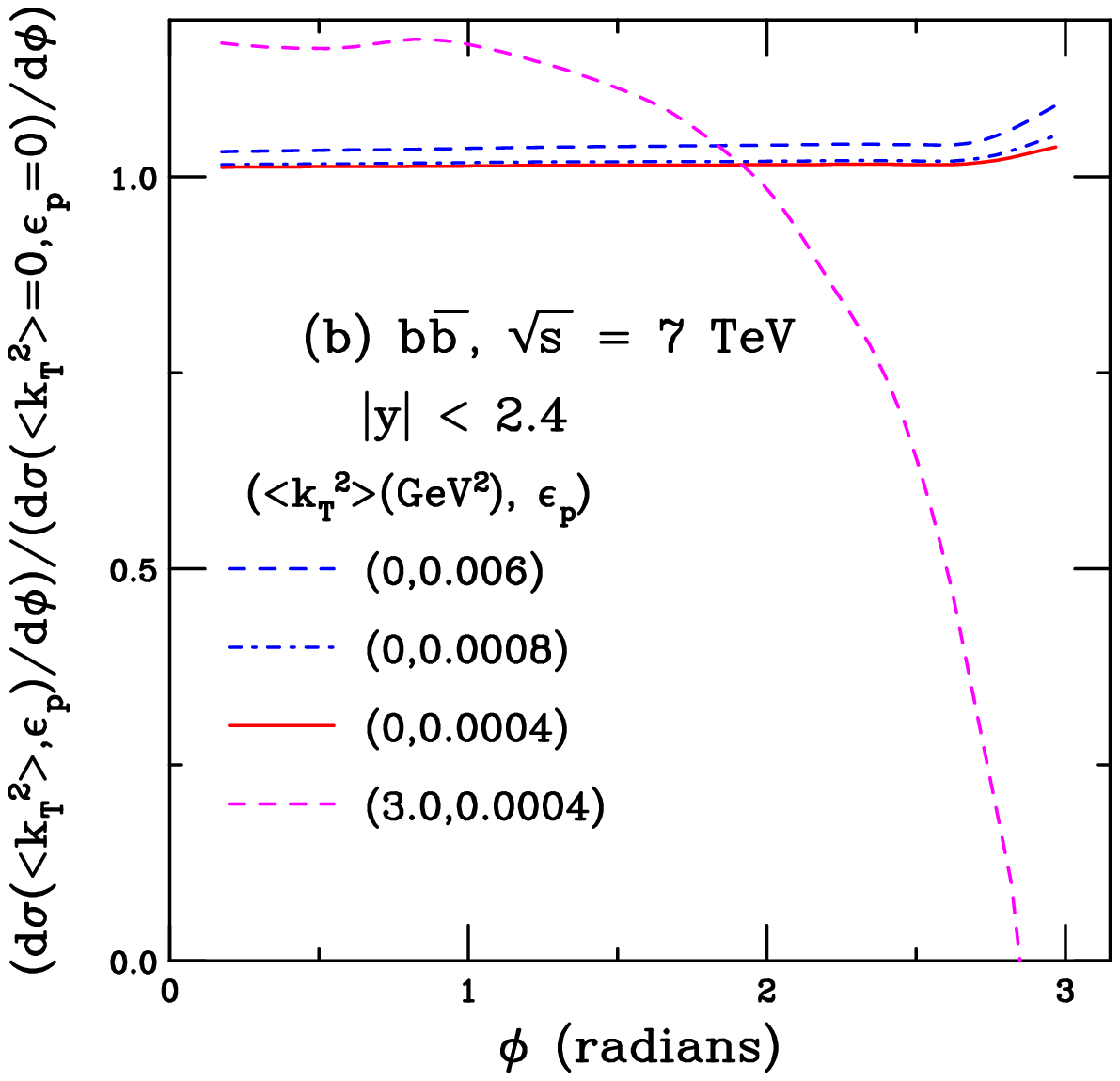}
  \caption[]{(Color online)
    The ratio of azimuthal angle distributions for the same
    $\langle k_T^2 \rangle$ and $\epsilon_P$ combinations as in
    Fig.~\protect\ref{fig3} relative to $\langle k_T^2 \rangle = 0$,
    $\epsilon_P = 0$ for (a) $c \overline c$ pairs at $2.5 < y < 5$ and
    (b) $b \overline b$ pairs at $|y| < 2.4$.  
  }
  \label{fig4}
\end{figure}

To illustrate the results more clearly, the ratio of all
calculations are shown relative to the results for bare quarks, with
$\langle k_T^2 \rangle = 0$, $\epsilon_P = 0$ in Fig.~\ref{fig4}.  In both
cases, the results with only finite $\epsilon_P$ and $\langle k_T^2 \rangle = 0$
are relatively independent of $\phi$ with the largest values of
$(d\sigma(\langle k_T^2\rangle, \epsilon_P)/d\phi)/
(d\sigma(\langle k_T^2\rangle=0, \epsilon_P =0)/d\phi)$ for the default values
of $\epsilon_P$, 0.06 for charm and 0.006 for bottom.  For charm production,
the ratio is $\sim 1.05-1.15$ at $\phi \sim  0$, increasing to $\sim 1.15-1.25$
near $\phi \sim \pi$.  For bottom, with its smaller $\epsilon_P$, the ratio is
only a few percent above unity.  Once finite $\langle k_T^2 \rangle$ is
included, there is an almost linear decrease with increasing $\phi$ for charm
while, for bottom, there is an enhancement until $\phi \sim \pi/2$ with a steep
falloff as $\phi \rightarrow \pi$.  We will examine the sensitivity of the
$\phi$ distributions to $\langle k_T^2 \rangle$ in the next section.

\begin{figure}[htpb]\centering
  \includegraphics[width=\columnwidth]{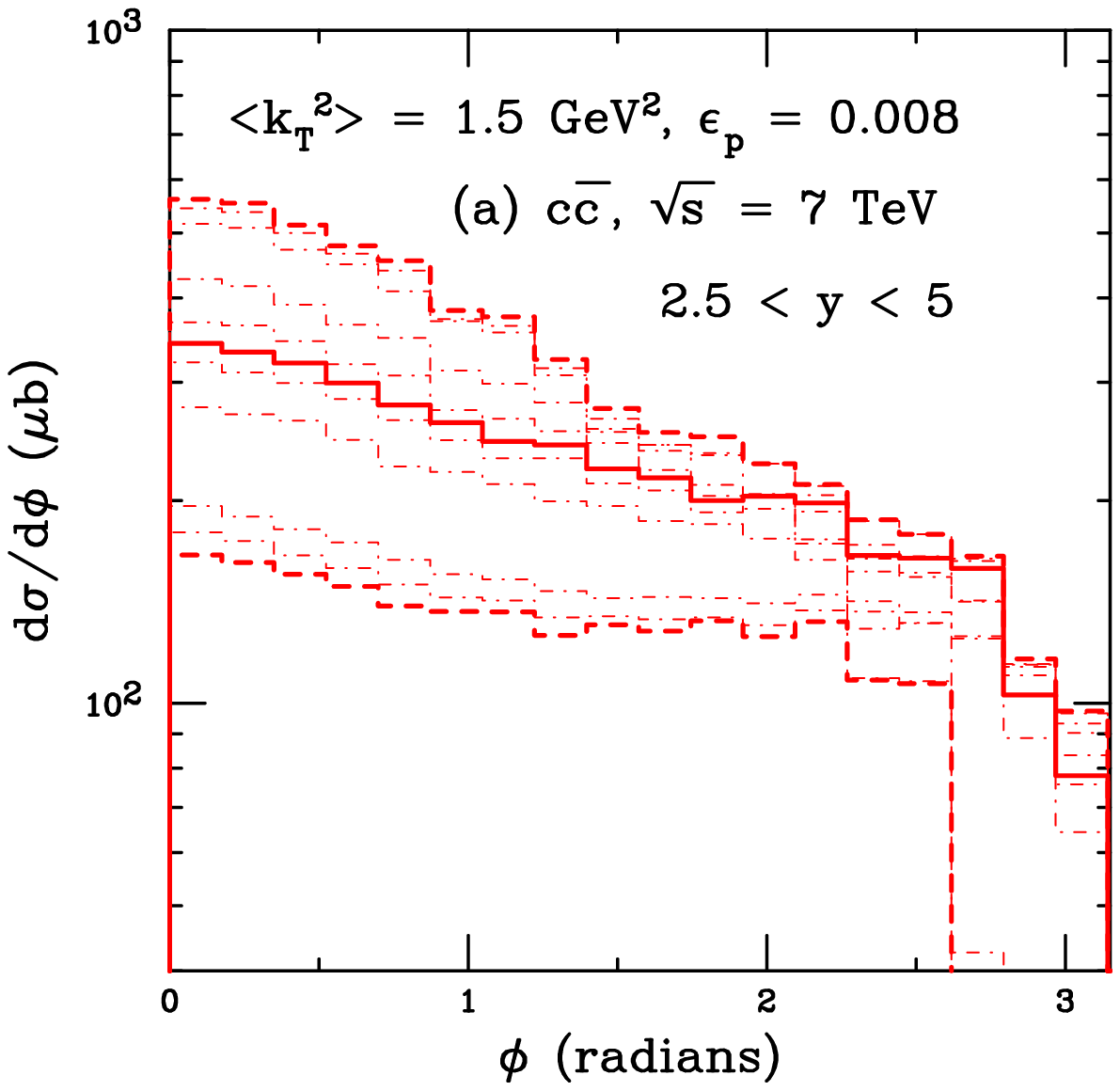}
  \includegraphics[width=\columnwidth]{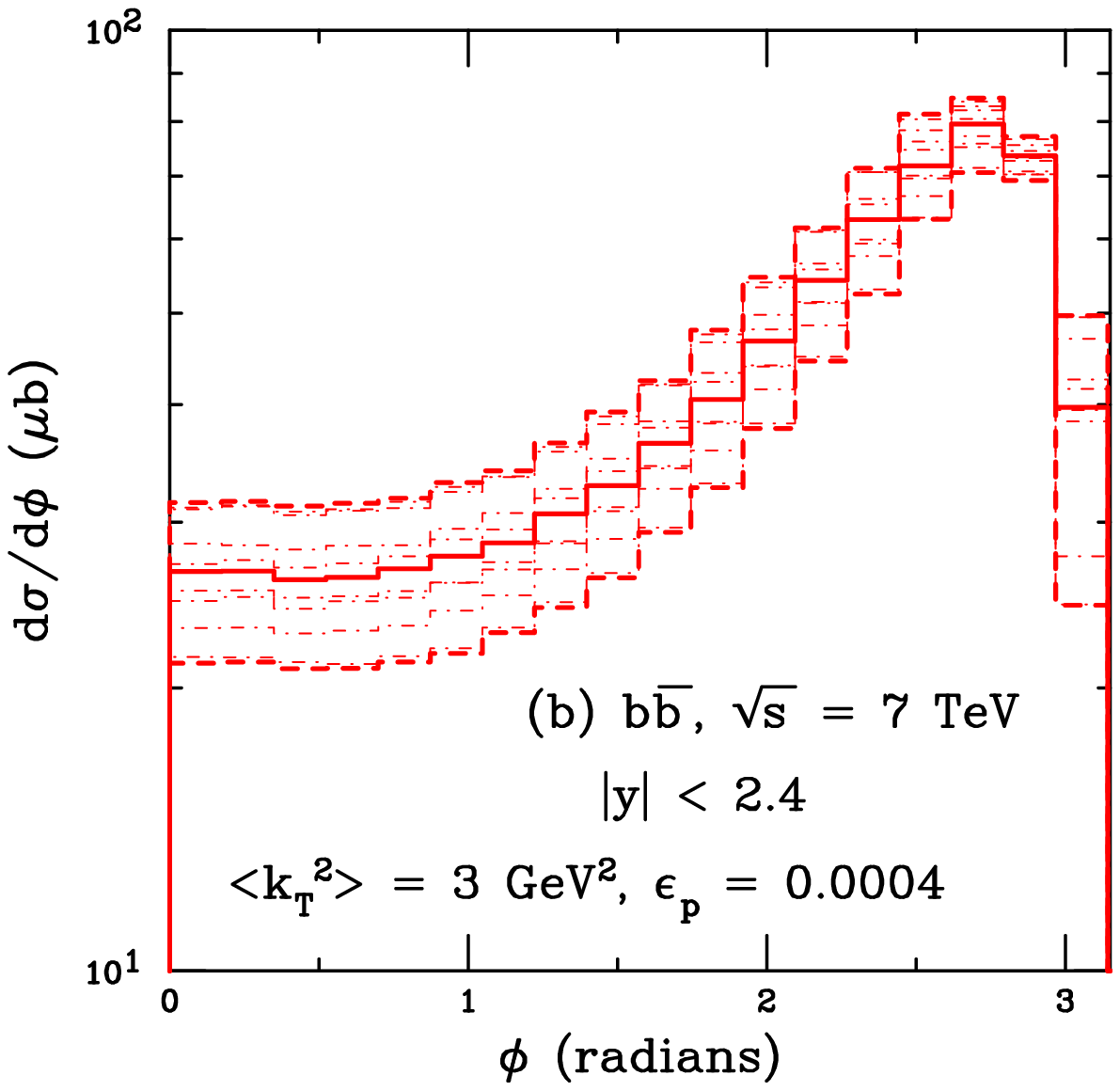}
  \caption[]{(Color online)
    The mass and scale uncertainty band on the NLO azimuthal distributions
    in $p+p$ collisions at $\sqrt{s} = 7$~TeV using the HVQMNR
    code for (a) $c \overline c$ pairs at forward rapidity, $2.5 < y < 5$, and
    (b) $b \overline b$ pairs at midrapidity, $|y| < 2.4$.  The calculations
    utilize $(\langle k_T^2 \rangle ({\rm GeV}^2), \epsilon_P) = (1.5,0.008)$
    for charm and (3,0.0004) for bottom.  The solid curves show the central
    results; the dashed, the edges of the uncertainty band; and the light
    dot-dashed curves are the individual mass and scale combinations.
  }
  \label{fig5}
\end{figure}

To determine the influence of the heavy quark mass and scale values on
$d\sigma/d\phi$ for finite $\langle k_T^2 \rangle$,
the full uncertainty bands are
shown in Fig.~\ref{fig5} for the same values of $\langle k_T^2 \rangle$ and
$\epsilon_P$ employed
in Fig.~\ref{fig2}.  In this case, in addition to the central
value and the upper and lower limits of the band, each mass and scale
combination is also shown in the light dot-dashed curves.  The results are given
as histograms in both cases even though $d\sigma/d\phi$ is relatively smooth,
especially for $b \overline b$ pairs.  The differences in the shapes of the
distributions are larger for charm.  Thus the level of the
enhancement at $\phi \sim 0$
seen for the central value varies considerably in this case.

The upper and lower limits of
the band are obtained according to Eqs.~(\ref{sigmax}) and (\ref{sigmin}) with
$X = \phi$.  On a bin-by-bin basis, the maximum and minimum mass and scale
combination may vary.  For example, changing the factorization scale changes
the slope of the $p_T$ distribution so that, at some value of $p_T$, the
maximum of $d\sigma/dp_T$ may come from a different scale choice.
The mass and scale choices affect the shape of $d\sigma/d\phi$ most strongly for
charm.  As $\phi \rightarrow \pi$, the maximum of $d\sigma/d\phi$ approaches
the central value while the minimum of the band is essentially determined by a
single mass and scale combination.

\begin{figure}[htpb]\centering
\includegraphics[width=\columnwidth]{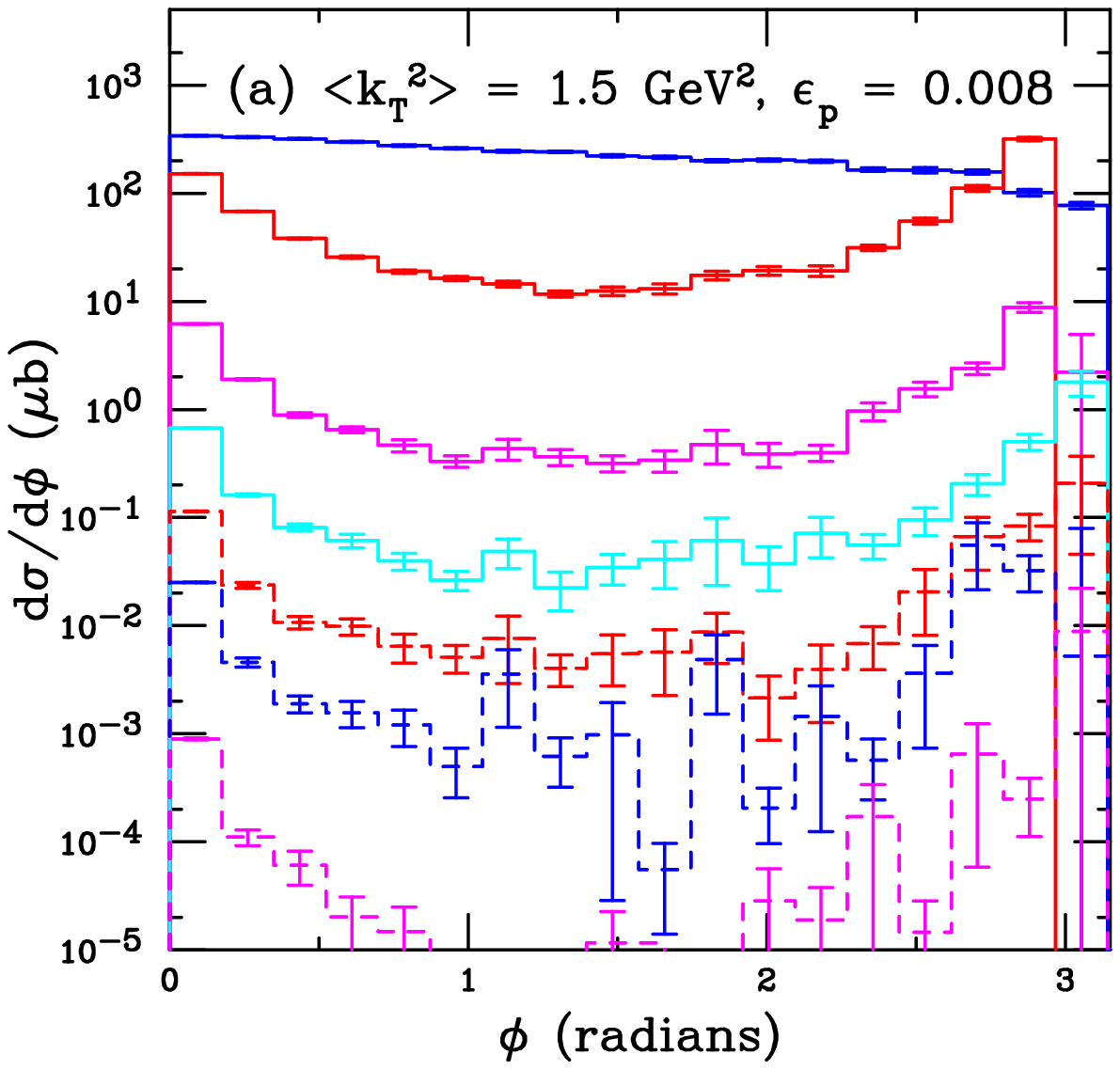}
\includegraphics[width=\columnwidth]{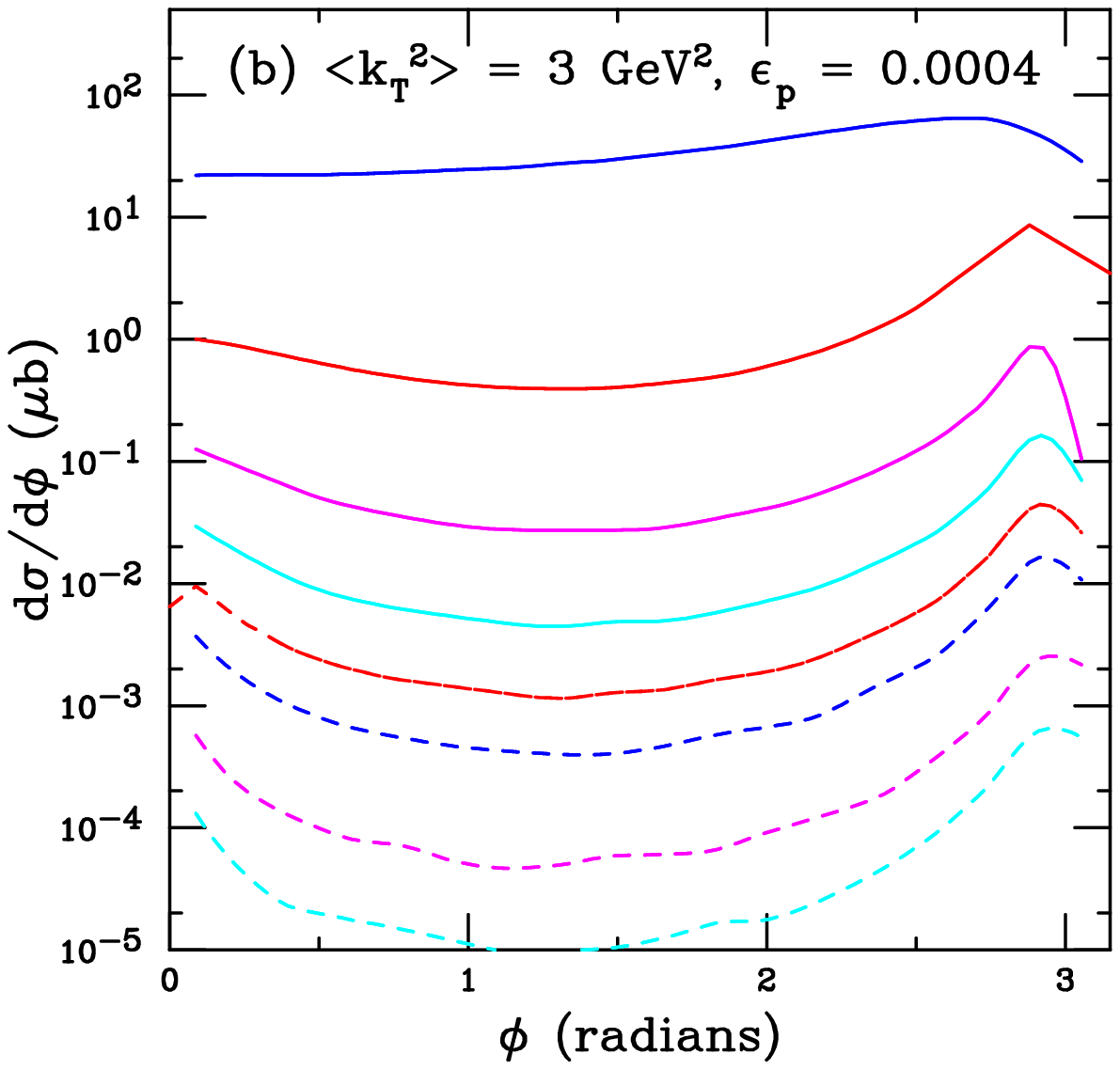}
\caption[]{(Color online)
  The effect of changing the $p_T$ cut on the azimuthat angle distribution
  between the two heavy quarks.  The calculations
  utilize $(\langle k_T^2 \rangle ({\rm GeV}^2), \epsilon_P) = (1.5,0.008)$
  for charm (a) and (3,0.0004) for bottom (b).  The $c \overline c$
  distributions are shown at forward rapidity, $2.5 < y < 5$, while the
  $b \overline b$ distributions are at midrapidity, $|y| < 2.4$.
  From top to bottom the results are: $p_T < 10$~GeV; $p_T > 10$~GeV;
  $p_T > 20$~GeV; $p_T > 30$~GeV; $p_T > 40$~GeV; $p_T > 50$~GeV;
  $p_T > 75$~GeV.
  The $c \overline c$ curves for all but the lowest $p_T$ cut are scaled up by
  $10^3$. The $b \overline b$ curves, which do not include a scale factor, also
  include a calculation with $p_T > 100$~GeV.
  }
\label{fig6}
\end{figure}

To end this section, the effect of a $p_T$ cut on $d\sigma/d\phi$ for the
values of $\langle k_T^2 \rangle$ and $\epsilon_P$ used in the calculations
of the uncertainty band in Figs.~\ref{fig2} and \ref{fig5} is shown
in Fig.~\ref{fig6}.  Results are given starting from low $p_T$,
$p_T < 10$~GeV, and increasing with $p_T$ cuts above
10, 20, 30, 40,
50 and 75~GeV for charm with an additional cut of $p_T > 100$~GeV for
bottom.  (Note that a minimum $p_T$ cut is chosen instead of a $p_T$ range
because the azimuthal distribution is dominated by the behavior at the
lowest $p_T$.)  
The low $p_T$ results, with $p_T < 10$~GeV,
are equivalent to those integrated over all $p_T$,
shown in Fig.~\ref{fig3}.  Notably, already at
$p_T > 10$~GeV, the shape of $d\sigma/d\phi$ changes with peaks at both
$\phi = 0$ and $\phi \sim \pi$ and a deepening dip developing between the two
peaks.   This is because that, as the minimum $p_T$ increases, the light
parton in the $2 \rightarrow 3$ process is more likely to either be low
momentum and aligned with one of the two heavy quarks and opposite the other
(emission from a final-state heavy quark, Fig.~\ref{nlodias}(a),
$\phi \sim \pi$)
or high momentum and balancing its momentum
against that of the $Q \overline Q$
pair ($\phi \sim 0$, Fig.~\ref{nlodias}(b)-(d)).

Note that no scale factors are
applied to separate the distributions, the decrease of $d\sigma/dp_T$ is
sufficient to separate them without introducing an additional scale factor.
However,  in the case of charm production, while no scale factor is applied
to the $p_T < 10$~GeV distribution, the remaining distributions are multiplied
by a factor of $10^3$ to be visible on the plot due to the steeply-falling
$p_T$ distribution of the charm quarks.

The $c \overline c$ distributions are shown as histograms with per bin
uncertainties because the charm
quark $p_T$ distributions decrease with $p_T$ faster than those for bottom
quarks so that, by $p_T > 50$~GeV the statistics for $c \overline c$ pairs
are poor.
There are few $c \overline c$ events produced for $p_T > 75$~GeV, as seen in
Fig.~\ref{fig6}(a), and these are mostly at $\phi \sim 0$. Note that the
$c \overline c$ results are shown for the forward rapidity region,
$2.5 < y < 5$, in the LHCb acceptance.  In this rapidity range, the $p_T$
distributions are more steeply falling than at central rapidity because the
edge of phase space for production is reached at lower $p_T$ at forward
rapidity.  However, as will be explained, the shape of $d\sigma/d\phi$ is not
strongly dependent on the rapidity range.

The $b \overline b$ distributions in Fig.~\ref{fig6}(b) show the same trends
but the harder $p_T$
distributions for bottom quarks result in smoother azimuthal distributions for
$b \overline b$ pairs even for $p_T > 100$~GeV.  The higher the $p_T$,
the more pronounced the peak at $\phi \sim 0$ becomes while the enhancement
at $\phi \sim \pi$ does not disappear.  (Note that there is no enhancement
at $\phi \sim 0$ for the lowest $p_T$ cut, $p_T < 10$~GeV, as is the case
for charm.)

While the
calculations shown in Fig.~\ref{fig6}
are for the central values of quark mass and scale factors, the trends would
remain the same for all the mass and scale combinations, especially for
$b \overline b$, see Fig.~\ref{fig5}.

\section{Sensitivity of $d\sigma/d\phi$ to $\langle k_T^2 \rangle$}
\label{sec:sensitivity}

In this section, the sensitivity of $d\sigma/d\phi$ to the size of the
intrinsic $\langle k_T^2 \rangle$
is explored to determine whether or not there is any obvious
onset of the change in $d\sigma/d\phi$.
Thus Eq.~(\ref{eq:avekt}) is modified to introduce a parameter
$\Delta$ that reduces the average $\langle k_T^2 \rangle$,
\begin{eqnarray}
  \langle k_T^2 \rangle = 1 + \frac{\Delta}{n}
  \ln \left(\frac{\sqrt{s}}{20 \, {\rm GeV}} \right) \, \, {\rm GeV}^2
  \, \, , \label{DeltakT}
\end{eqnarray}
starting with $\langle k_T^2 \rangle = 0$.  (Note that $\Delta = 1$ in
Eq.~(\ref{eq:avekt}).)
For charm pair production with $n = 12$, 
$\Delta = -3/2$, $-1$, $-1/2$, 0, 1/2, and 1, effectively changing
$\langle k_T^2 \rangle$ by $\sim 0.25$~GeV$^2$ at $\sqrt{s} = 7$~TeV
as $\Delta$ is increased by 1/2.
For bottom pair production, 
only $\Delta = -1/2$, 0, 1/2 and 1 result in $\langle k_T^2 \rangle > 0$
because, with $n = 3$, each change in $\Delta$ increases $\langle k_T^2 \rangle$
by $\sim 1$~GeV$^2$ at $\sqrt{s} = 7$~TeV.
Here $\Delta = -1/2$ results in
$\langle k_T^2 \rangle = 0.02$~GeV$^2$.

The broad $p_T$ cuts, $p_T < 10$~GeV and $p_T > 10$~GeV, are studied both at
midrapidity ($|y| < 2.4$) and forward rapidity ($2.5 < y < 5$).  Only the
results for midrapidity are shown here, however, since trends were found to be
independent of rapidity.  The figures in this section include the
$c \overline c$ and $b \overline b$ azimuthal distributions, $d\sigma/d\phi$,
as well as the ratios,
$(d\sigma(\langle k_T^2 \rangle)/d\phi)/(d\sigma(\langle k_T^2 \rangle = 0)/d\phi)$, to highlight the effects when differences in the distributions
themselves may be small.  Note that the choice of 10~GeV is somewhat arbitrary
and not indicative of any threshold behavior.  Comparison with data in the next
section will be more indicative of the importance of $\langle k_T^2 \rangle$
in a given $p_T$ range.

\begin{figure}[htpb]\centering
  \includegraphics[width=\columnwidth]{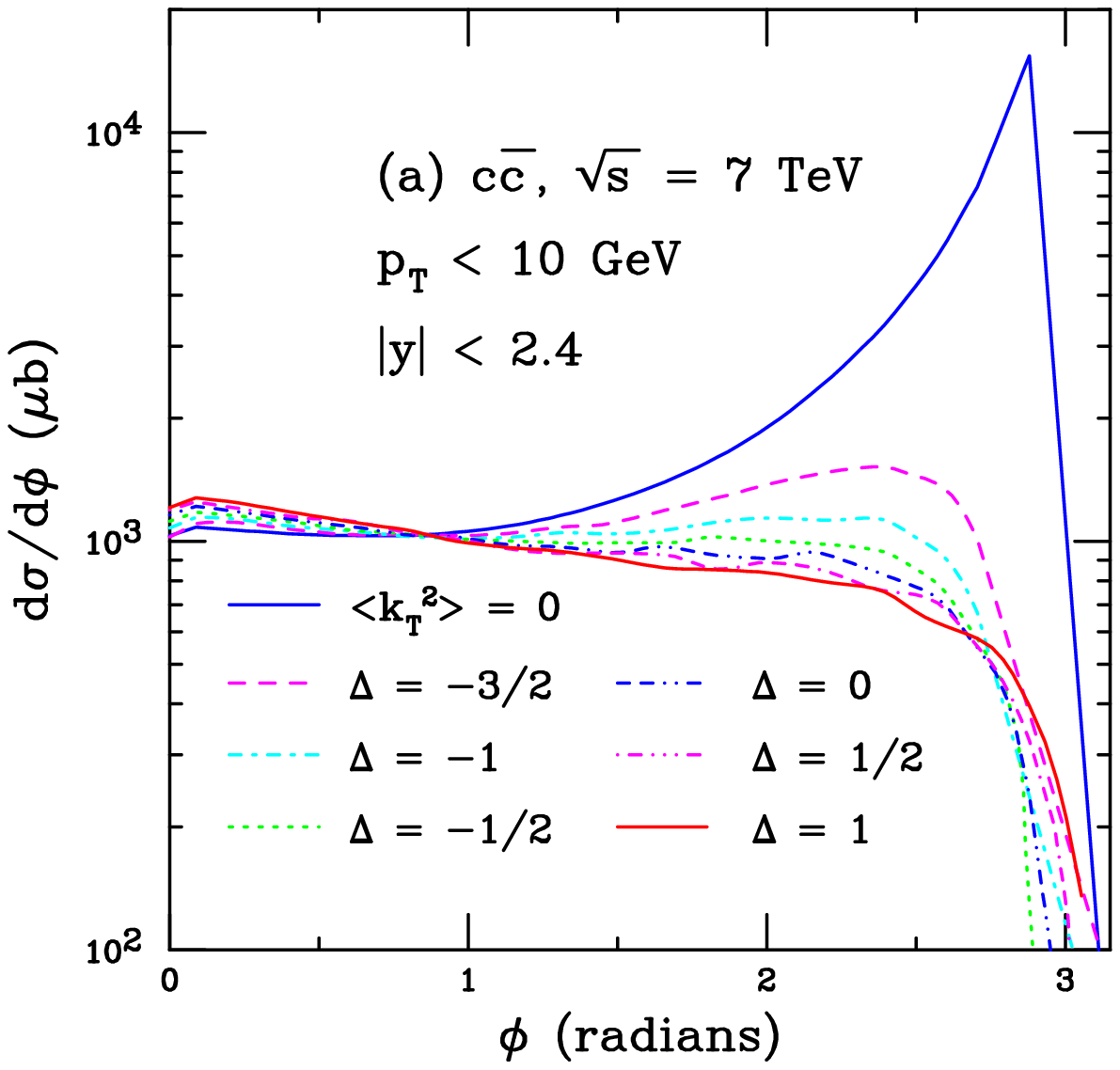}
  \includegraphics[width=\columnwidth]{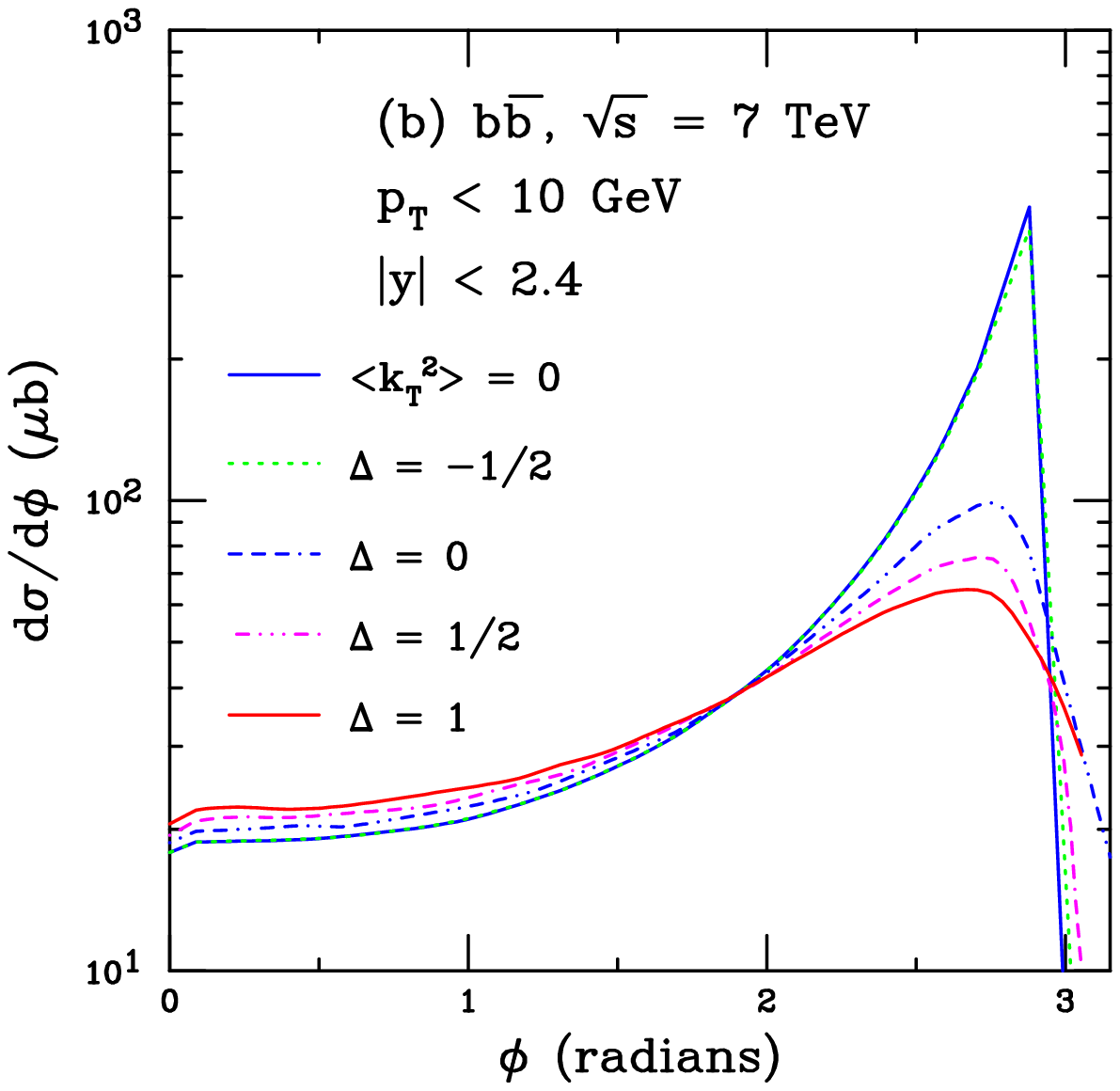}
  \caption[]{(Color online)
    The azimuthal angle distributions for (a) $c \overline c$ and
    (b) $b \overline b$ pairs in the central rapidity range $|y| < 2.4$ with
    $p_T < 10$~GeV.  Calculations are shown with $\langle k_T^2 \rangle = 0$ and
    for values of $\Delta$ from $-3/2$ to 1 for charm and $\Delta = -1/2$ to 1
    for bottom, as in Eq.~(\protect\ref{DeltakT}).
  }
  \label{fig7}
\end{figure}

Figure~\ref{fig7}(a) shows the $c \overline c$
azimuthal distributions for $p_T < 10$~GeV
at central rapidity.  Note that increasing $\langle k_T^2 \rangle$ from 0 to
0.25~GeV$^2$ already shows a significant change in
$d\sigma/d\phi$ for $c \overline c$.
The peak at $\phi \sim \pi$ is largely erased and only a weak
variation with $\phi$ can be seen on the log scale for $\phi < \pi/2$.  As
$\langle k_T^2 \rangle$ increases, the enhancement at $\phi \sim 0$ increases
while the peak near $\phi \sim \pi$ decreases and disappears.  A similar but
slower evolution is seen for $b \overline b$ in Fig.~\ref{fig7}(b).  Note that
$\langle k_T^2 \rangle = 0.02$~GeV$^2$ still largely follows the
$\langle k_T^2 \rangle = 0$ results, as may be seen in the dotted curve.
Thus sufficiently small steps in $\langle k_T^2 \rangle$ would likely reveal
a gradual change in $d\sigma/d\phi$.

\begin{figure}[htpb]\centering
  \includegraphics[width=\columnwidth]{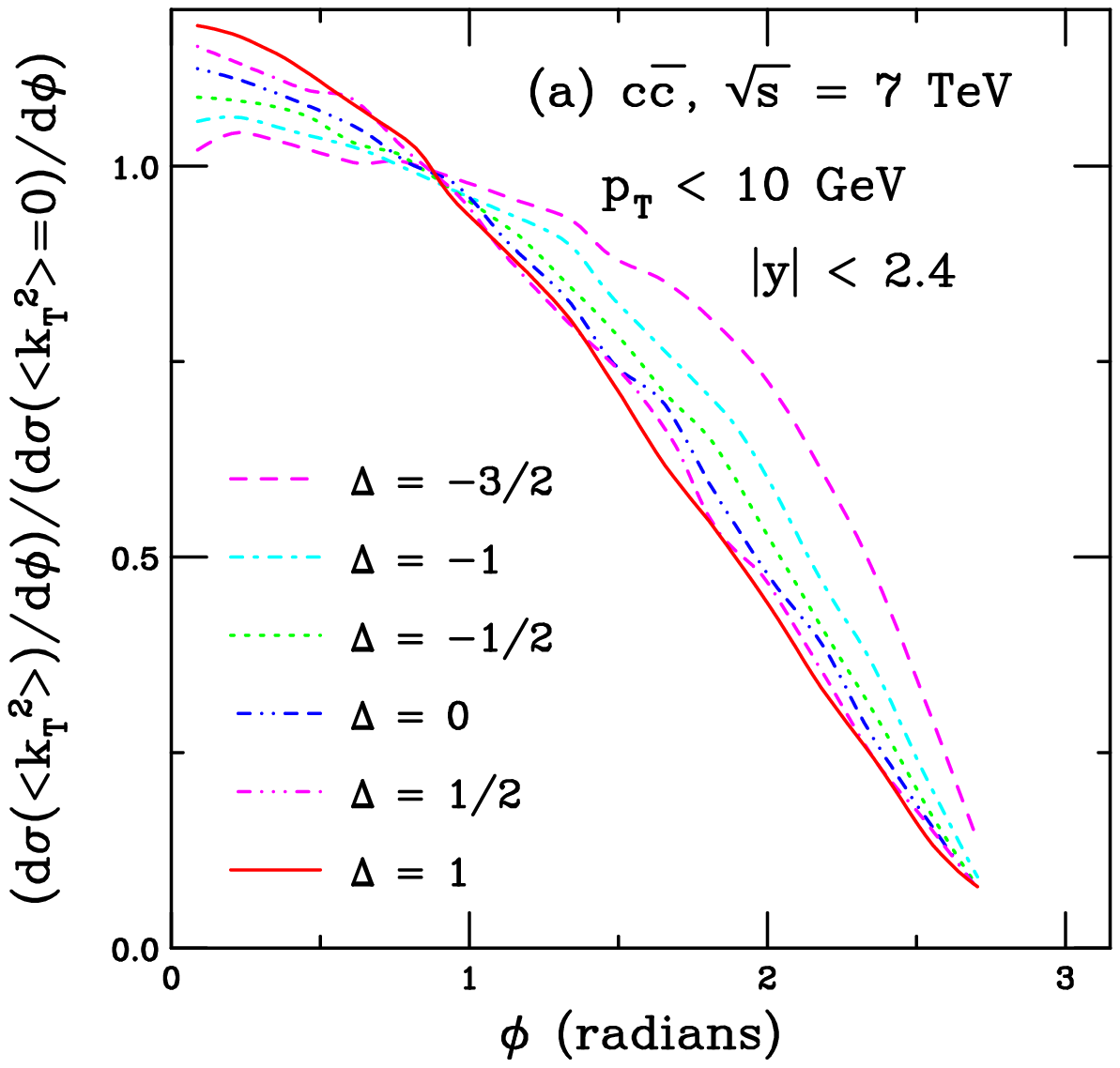}
  \includegraphics[width=\columnwidth]{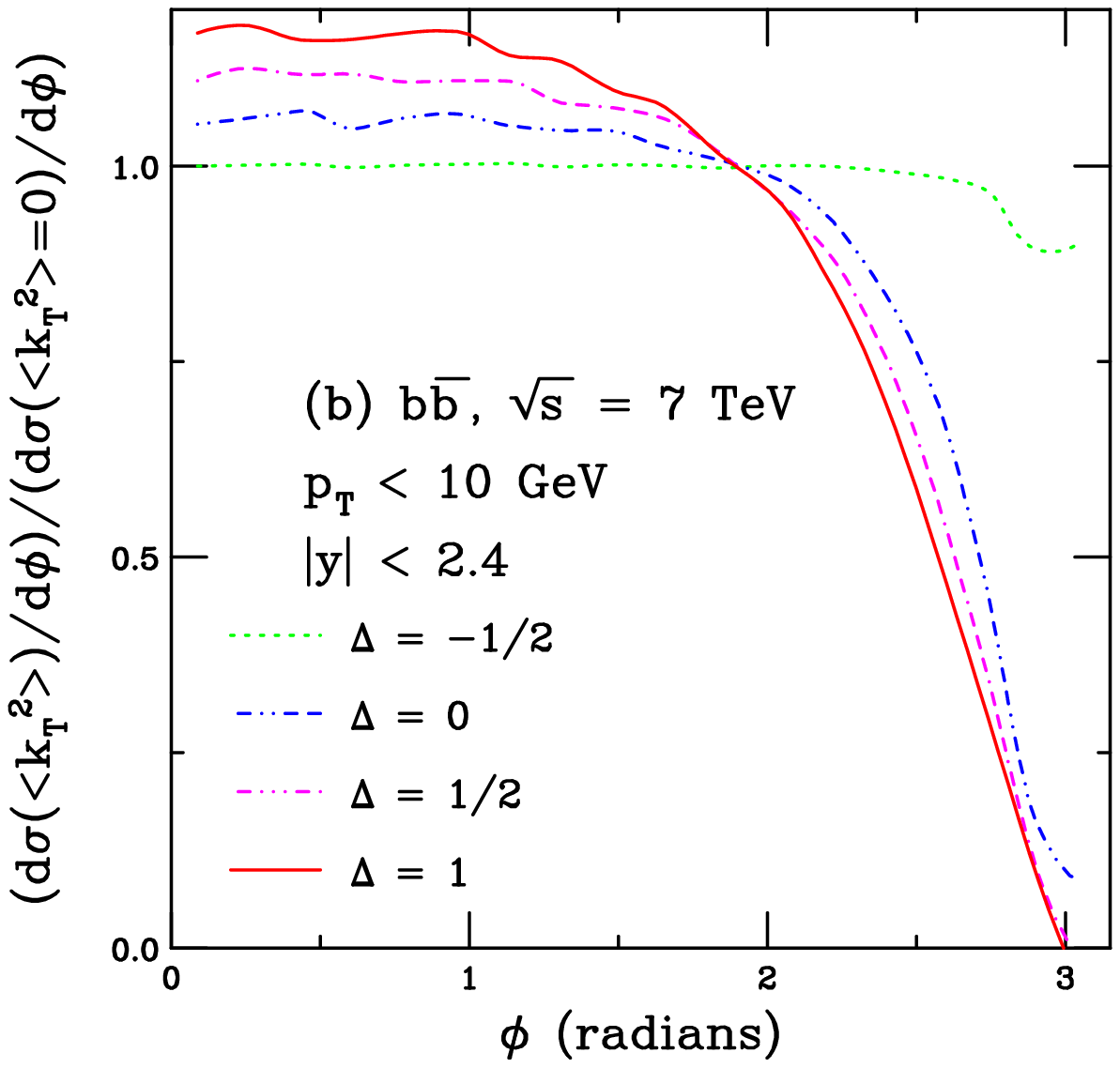}
  \caption[]{(Color online)
    The ratio of azimuthal angle distributions relative to
    that of $\langle k_T^2 \rangle = 0$ for (a) $c \overline c$ and
    (b) $b \overline b$ pairs in the central rapidity range $|y| < 2.4$ with
    $p_T < 10$~GeV.  Calculations are shown
    for values of $\Delta$ from $-3/2$ to 1 for charm and $\Delta = -1/2$ to 1
    for bottom, as in Eq.~(\protect\ref{DeltakT}).
  }
  \label{fig8}
\end{figure}

The relative changes in the distributions care seen more clearly in
Fig.~\ref{fig8} where the ratios of the azimuthal distributions relative
to that with $\langle k_T^2 \rangle = 0$ are shown.  The charm quark ratios go
from relatively flat at $\phi < \pi/2$ to an almost linear decrease with the
maximum $\langle k_T^2 \rangle$, with $\Delta = 1$ corresponding to the red
curves in Figs.~\ref{fig1}(a) and \ref{fig3}(a).  The ratio of the distributions
pivot around $\phi \sim 1$ with a smooth evolution of the ratios with $\Delta$.

On the other hand, the $b \overline b$ ratios pivot around $\phi \sim 2$, from
a slight modification at $\phi \sim \pi$ for
$\langle k_T^2 \rangle = 0.02$~GeV$^2$ to a flat ratio for $\phi < 2$ followed
by a steep descent for $\phi > 2$.

The difference in the location of the pivot point of the distributions is due to
the heavy quark mass.  The lighter mass of the charm quarks makes it more
likely that
their relative momentum will become more isotropic at low $p_T$ as
$\Delta$ increases.
Thus the point around which the slope of the ratio is changing is
closer to $\phi \sim 0$.  The bottom quarks, more than three times more massive
than the charm quarks, are less likely to have their azimuthal distributions
become completely isotropic at low $p_T$ since the $k_T$ kicks are less
effective on heavier quarks.  Thus there is a reduced peak at
$\phi \sim \pi$ for $b \overline b$
and the slopes of the ratios pivot at an angle closer to
$\phi \sim \pi$.

\begin{figure}[htpb]\centering
  \includegraphics[width=\columnwidth]{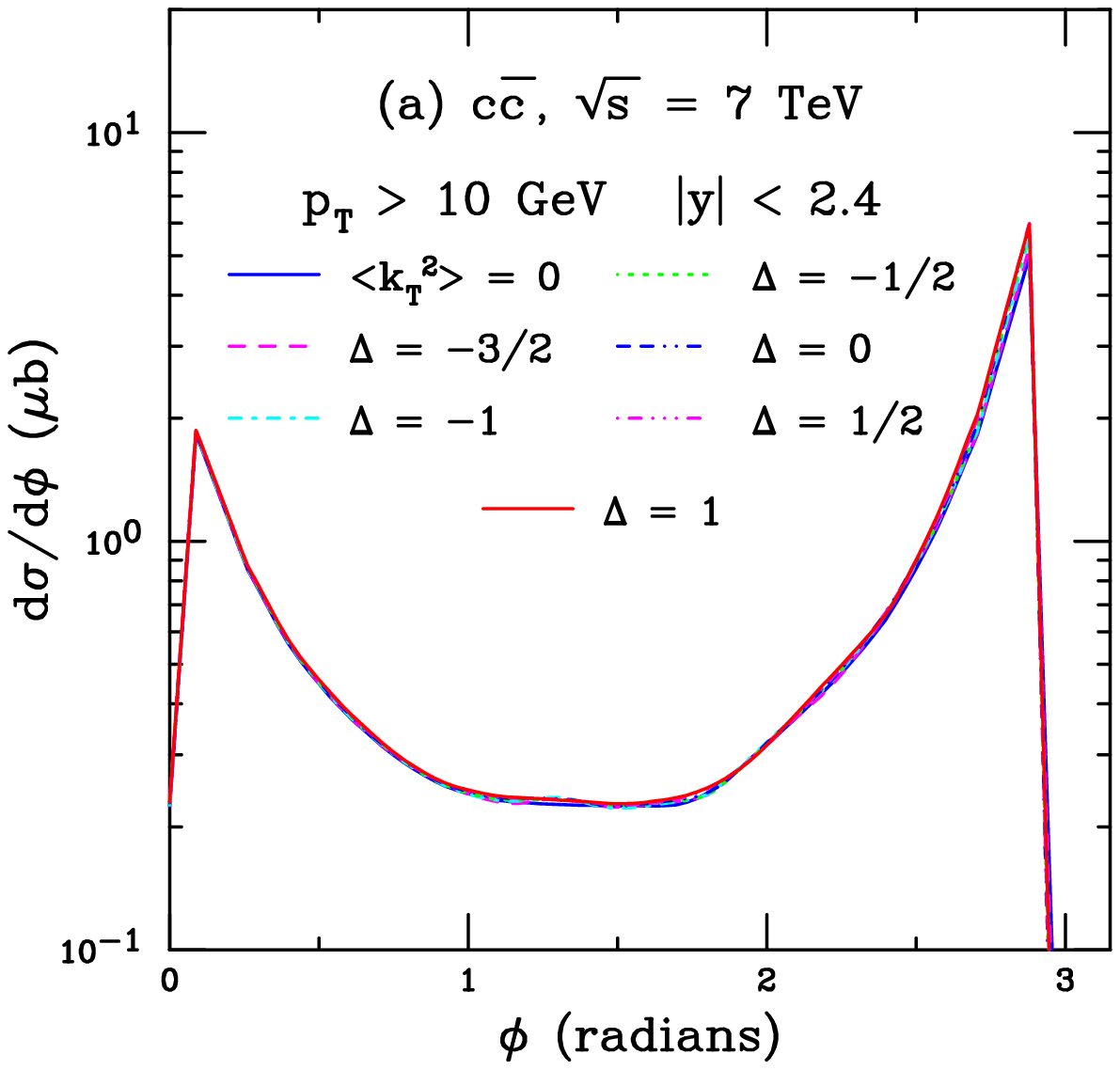}
  \includegraphics[width=\columnwidth]{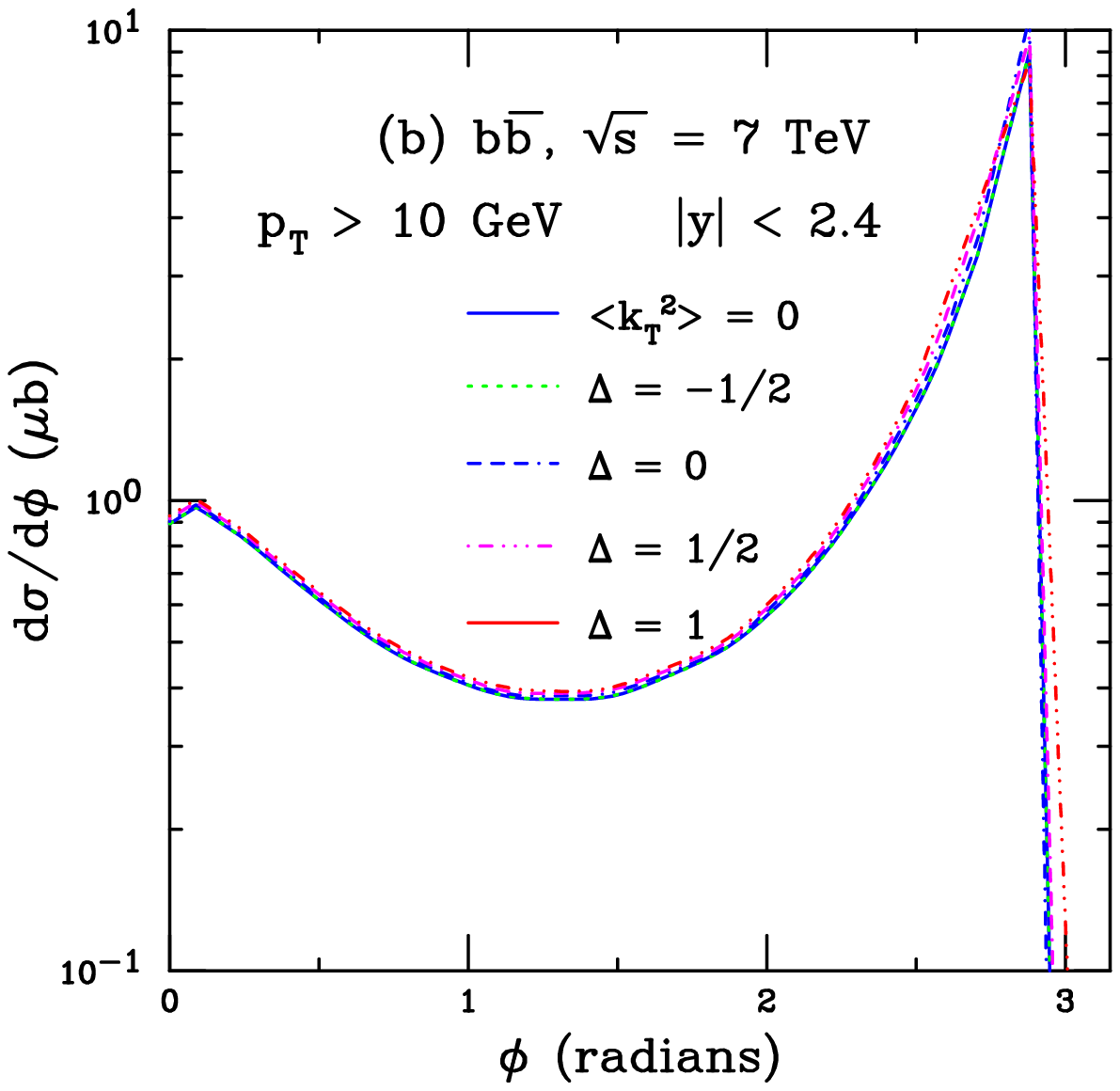}
  \caption[]{(Color online)
    The azimuthal angle distributions for (a) $c \overline c$ and
    (b) $b \overline b$ pairs in the central rapidity range $|y| < 2.4$ with
    $p_T > 10$~GeV.  Calculations are shown with $\langle k_T^2 \rangle = 0$ and
    for values of $\Delta$ from $-3/2$ to 1 for charm and $\Delta = -1/2$ to 1
    for bottom, as in Eq.~(\protect\ref{DeltakT}).
  }
  \label{fig9}
\end{figure}

\begin{figure}[htpb]\centering
  \includegraphics[width=\columnwidth]{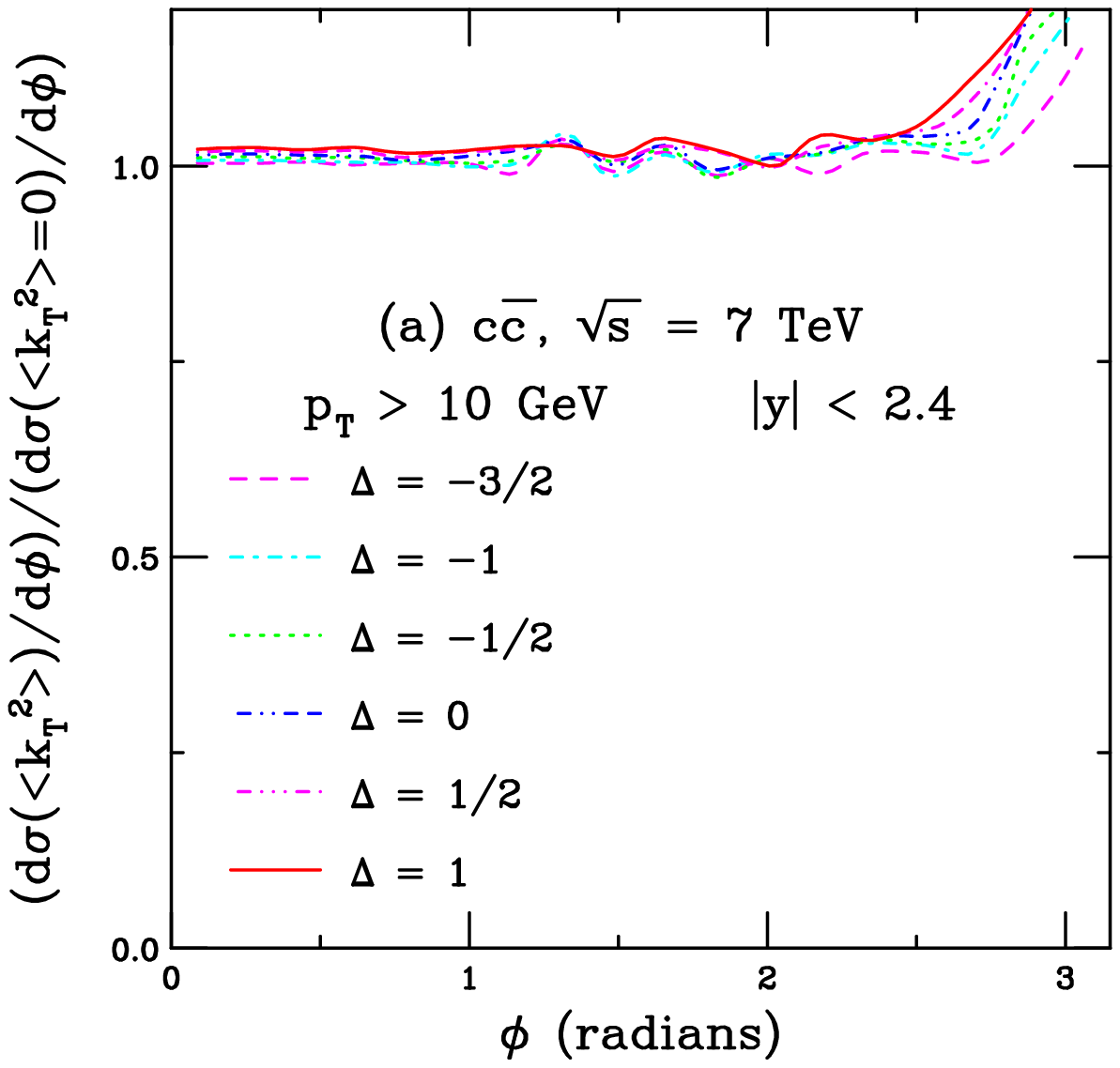}
  \includegraphics[width=\columnwidth]{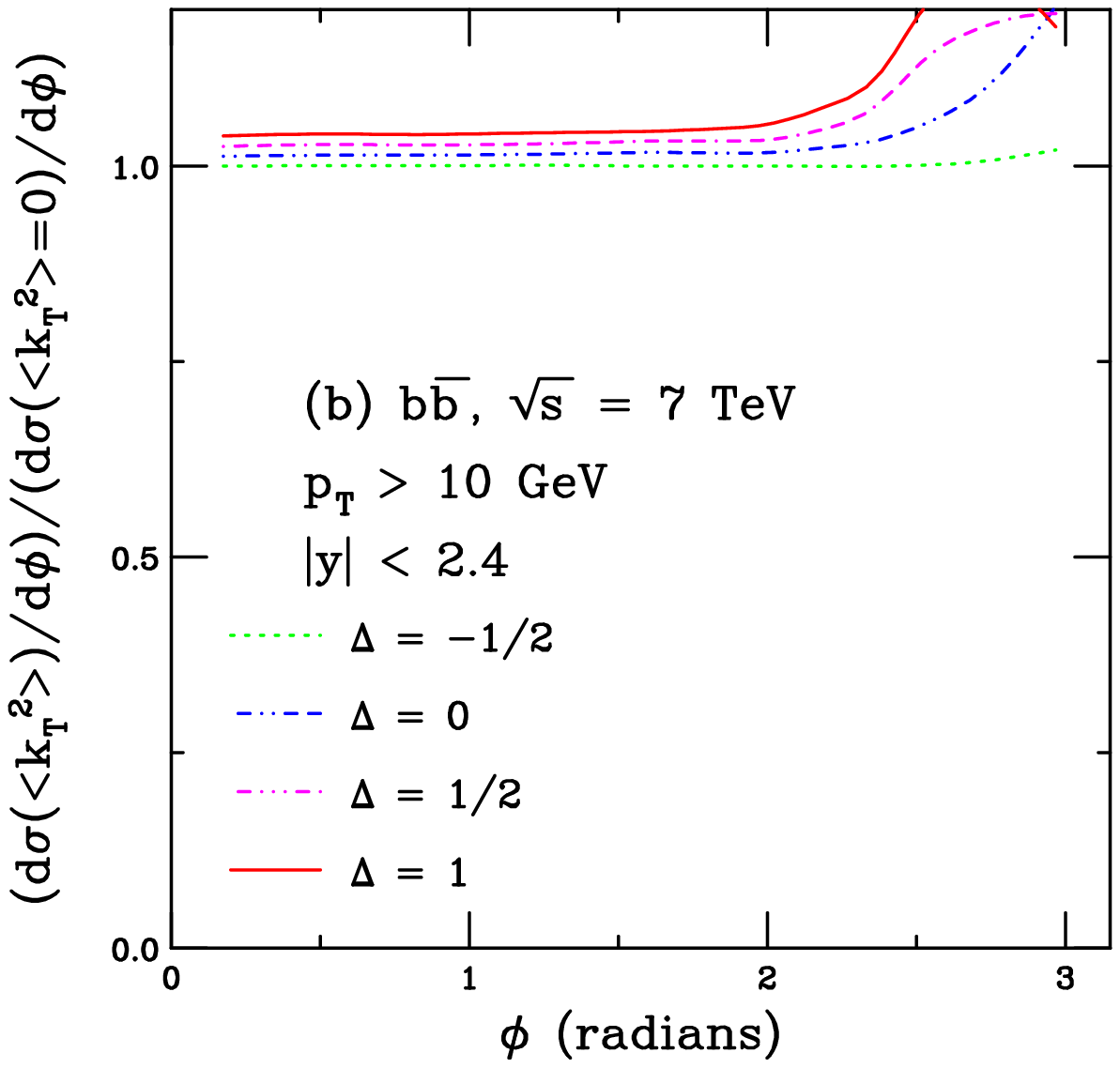}
  \caption[]{(Color online)
    The ratio of azimuthal angle distributions relative to
    that of $\langle k_T^2 \rangle = 0$ for (a) $c \overline c$ and
    (b) $b \overline b$ pairs in the central rapidity range $|y| < 2.4$ with
    $p_T > 10$~GeV.  Calculations are shown
    for values of $\Delta$ from $-3/2$ to 1 for charm and $\Delta = -1/2$ to 1
    for bottom, as in Eq.~(\protect\ref{DeltakT}).
  }
  \label{fig10}
\end{figure}

Figures~\ref{fig9} and \ref{fig10} show $d\sigma/d\phi$ for $p_T > 10$~GeV at
midrapidity and the ratios relative to $d\sigma/d\phi$ with
$\langle k_T^2 \rangle = 0$.  In this case, an enhancement is seen at
$\phi \sim 0$ for both charm and bottom although the enhacement is larger for
charm (note that the scales on the $y$-axes of Fig.~\ref{fig10}
are the same).  The shapes are
consistent with the solid red distributions in Fig.~\ref{fig6}.

It is not possible to distinguish between different values of $\Delta$ for
$c \overline c$, even by examining the ratios in Fig.~\ref{fig10}(a).
Although the ratios increase with $\Delta$ at $\phi \sim 0$, the curves are
almost indistinguishable at low $\phi$.
Larger differences in the $c \overline c$
ratios as $\phi \rightarrow \pi$ can be observed since the ratios at
$\phi > 2.5$ increase at slightly lower $\phi$
for higher values of $\langle k_T^2 \rangle$.  This is because
$\langle k_T^2 \rangle^{1/2} < p_T$ for all $p_T$, resulting in a negligible
effect for charm.  There is also only a small difference between the
$b \overline b$ ratios in Fig.~\ref{fig10}(b).
However, the $b \overline b$ ratios are visibly separated
at $\phi \sim 0$ because $\langle k_T^2 \rangle^{1/2} < p_T$ remains
true for the heavier bottom quarks,
$\langle k_T^2 \rangle^{1/2}/p_T \sim 0.1-0.2$ for bottom, rather than
$\langle k_T^2 \rangle^{1/2}/p_T \sim 0.05-0.1$ for charm.  In addition, the
harder $p_T$ distribution for bottom means that a smaller fraction of the total
$b \overline b$ cross section is contained the region $p_T < 10$~GeV than for
charm where $\sim 98$\% of the charm $p_T$ distribution is contained in the
range $p_T < 10$~GeV.

Finally, it is worth noting that the azimuthal angle distributions and their
ratios at low $p_T$ ($p_T < 10$~GeV) and higher $p_T$ ($p_T > 10$~GeV) at
forward rapidity are effectively identical to those at midrapidity.  Thus,
these results are not shown.  Results for $p_T$ cuts higher than $p_T > 10$~GeV
are not shown because the effect of $k_T$ broadening on higher $p_T$ cuts
would be negligible.  

\section{Comparison to Data}
\label{sec:dataComp}


There are data on heavy quark pair correlations to test these calculations.
There are charm pair correlation data from ALICE \cite{ALICEDDpairs} and
LHCb \cite{LHCbDDpairs} and bottom quark-bottom jet correlations from CMS
\cite{CMSbbpairs}, both from $p+p$ collisions at $\sqrt{s} = 7$~TeV.
In addition, there are $D^0 D^{*-}$ and $D^+ D^{*-}$ data from $p + \overline p$
collisions at $\sqrt{s} = 1.96$~TeV from CDF in Tevatron Run II
\cite{CDFDDpairs}.

\subsection{LHCb $C \overline C$ Data}

LHCb measured $c \overline c$, $cc$, and $(c + \overline c)J/\psi$
correlations in $p+p$
collisions at 7 TeV for $2 < y < 4$ and $3 < p_T < 12$~GeV.  The discussion here
will focus on comparison to some of the $c \overline c$ correlations, in
particular on the $D^0 \overline D^0$, $D^0 D^-$ and $D^+ D^-$ combinations where
fragmentation functions are known.
Thus final states that
include a $D_s$ or $\Lambda_c$ are excluded.
In addition, the final-state $c \overline c$ data measured by LHCb
are more likely to
arise from production of a single $c \overline c$ pair.

\begin{figure}[htpb]\centering
  \includegraphics[width=\columnwidth]{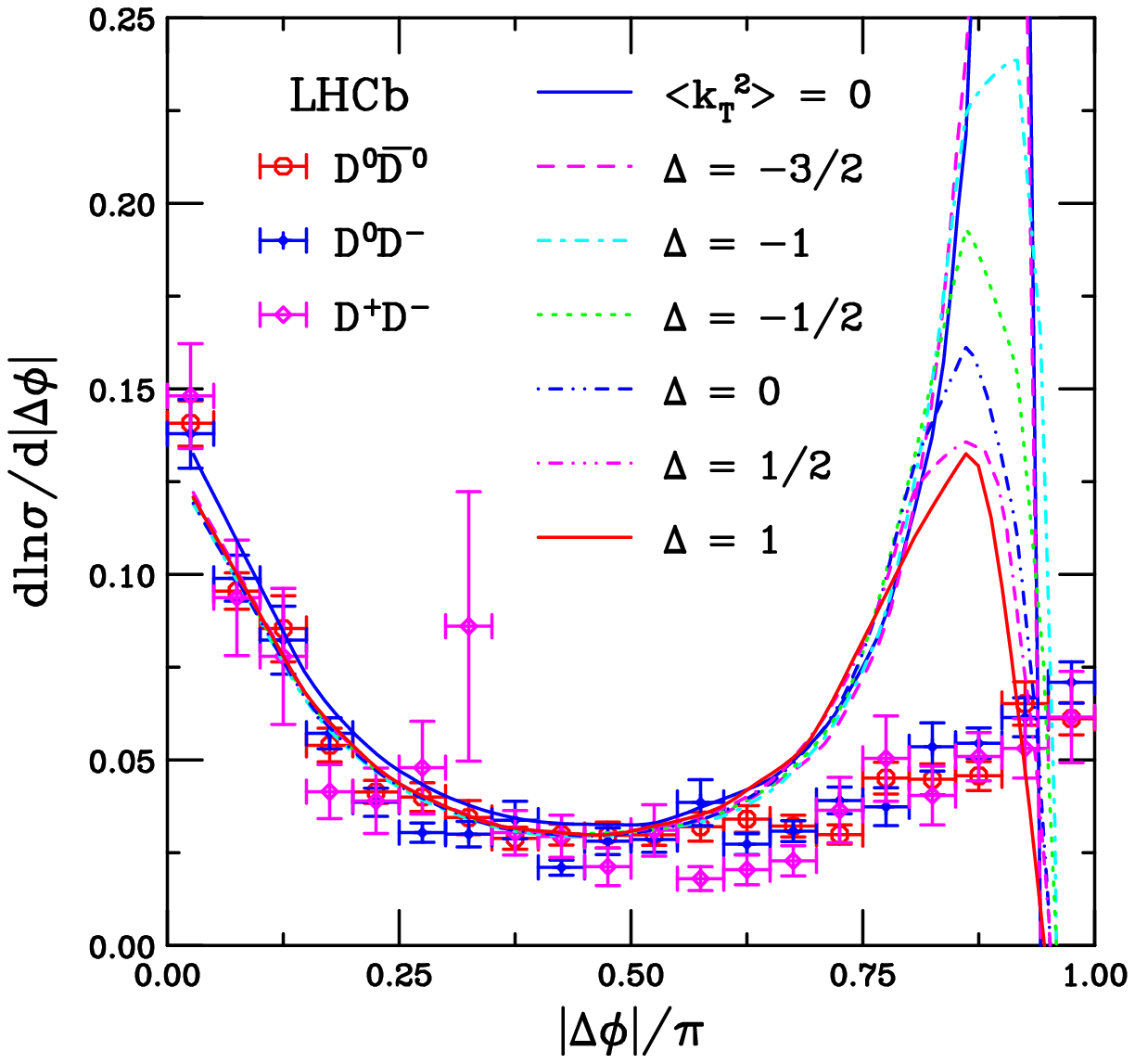}
  \caption[]{(Color online)
    The azimuthal angle distributions for $D^0 \overline D^0$ (red),
    $D^0 D^-$ (blue), and $D^+ D^-$ (magenta) pairs measured in $p+p$ collisions
    at $\sqrt{s} = 7$~TeV by LHCb \protect\cite{LHCbDDpairs}.
    The data are compared
    to calculations in the same acceptance with $\langle k_T^2 \rangle = 0$ and
    for values of $\Delta$ from $-3/2$ to 1 in Eq.~(\protect\ref{DeltakT}).}
  \label{lhcb_azi}
\end{figure}

Figure~\ref{lhcb_azi} shows the azimuthal angle distributions for the same
values of $\Delta$ (from Eq.~(\ref{DeltakT})) employed in the previous section.
The results for $\phi < 2$ are in good agreement with the data and with the
results of Fig.~\ref{fig9}(a) for $p_T > 10$~GeV.  However, as
$\phi \rightarrow \pi$, there is a strong effect on the peak value as a function
of $\Delta$.  The peak in this region gets broader and lower as $\Delta$
increases.  The larger values of $\Delta$ are closer to the data although
the data show only a slight trend upward in this region rather than a second
peak at $\phi \sim \pi$.
The calculations at large $\phi$ differ strongly with those at
$p_T > 10$~GeV in Fig.~\ref{fig9}(a), likely because at $p_T \sim 3$~GeV a
significant contribution from $k_T$ broadening still remains,
see Fig.~\ref{fig1}.  The sharp
peak near $\phi \sim \pi$ is somewhat artificial since, at
$\langle k_T^2 \rangle = 0$, all the peak is essentially contained in a single
$\Delta \phi$ bin. When $k_T$ broadening is included, the peak region is wider
and is shared among several bins although, for $p_T$ values of a few GeV, not
including $p_T < 2\langle p_T \rangle \sim 2m_D$,
it is not completely washed out as it
is for the results with $p_T < 10$~GeV, dominated by $p_T \sim 0$, in
Fig.~\ref{fig7}.  Thus the overall agreement of the calculations with the
LHCb $\phi$ distribution data can be considered quite good.

\begin{figure}[htpb]\centering
  \includegraphics[width=\columnwidth]{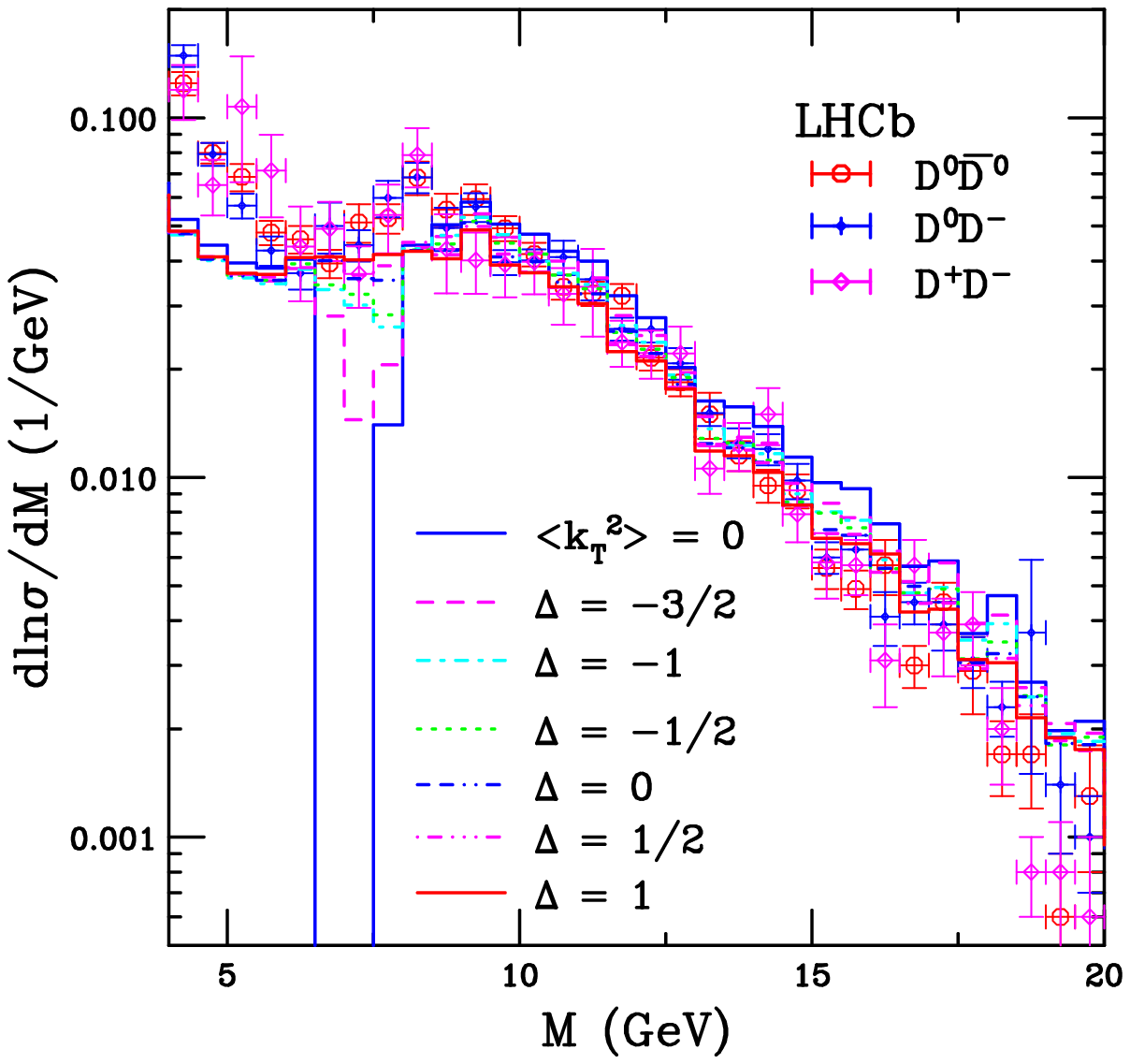}
  \caption[]{(Color online)
    The invariant mass distributions for $D^0 \overline D^0$ (red),
    $D^0 D^-$ (blue), and $D^+ D^-$ (magenta) pairs measured in $p+p$ collisions
    at $\sqrt{s} = 7$~TeV by LHCb \protect\cite{LHCbDDpairs}.
    The data are compared
    to calculations in the same acceptance with $\langle k_T^2 \rangle = 0$ and
    for values of $\Delta$ from $-3/2$ to 1 in Eq.~(\protect\ref{DeltakT}).
  }
  \label{lhcb_mass}
\end{figure}

Calculations of the $D \overline D$ pair invariant mass in the LHCb acceptance
with the data are shown in Fig.~\ref{lhcb_mass}.  The agreement
of the calculations
with the data are very good.  A similar minimum is seen in both for
$M \sim 2\sqrt{m_D^2 + p_T^2}\sim 7$~GeV, using the lower limit of the $p_T$
range for the detected $D$ mesons.  Below
this value of $M$, the data have a higher cross section than the calculations.
The gap in the calculated mass at $M \sim 7$~GeV
for $\langle k_T^2 \rangle = 0$ fills in as $\Delta$
increases.  In addition, at higher masses, the calculated distribution falls off
somewhat faster for larger $\Delta$ so that $\Delta = 1$ agrees best with the
data.

\begin{figure}[htpb]\centering
  \includegraphics[width=\columnwidth]{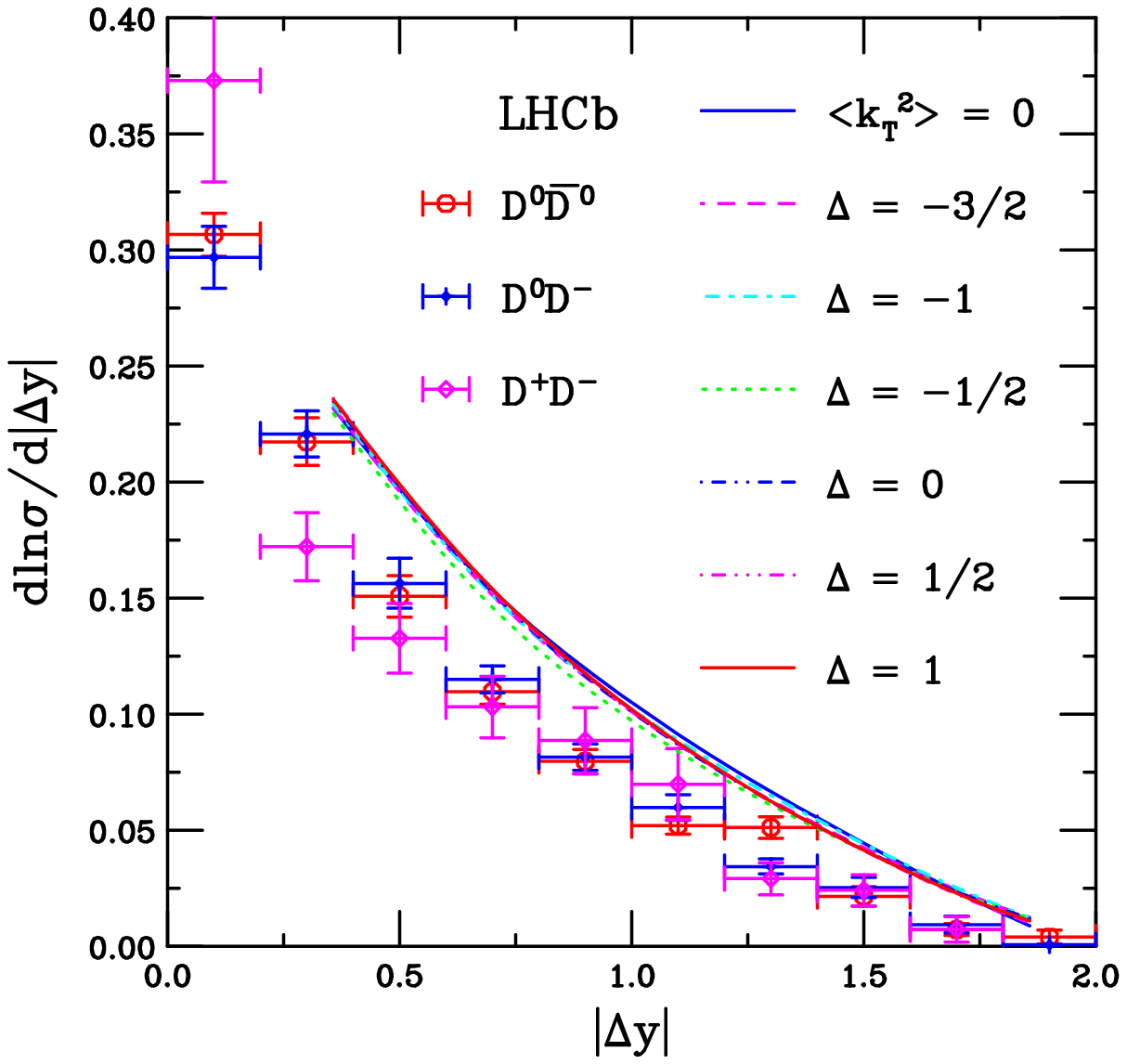}
  \caption[]{(Color online)
    The $|\Delta y|$ distributions for $D^0 \overline D^0$ (red),
    $D^0 D^-$ (blue), and $D^+ D^-$ (magenta) pairs measured in $p+p$ collisions
    at $\sqrt{s} = 7$~TeV by LHCb \protect\cite{LHCbDDpairs}.
    The data are compared
    to calculations in the same acceptance with $\langle k_T^2 \rangle = 0$ and
    for values of $\Delta$ from $-3/2$ to 1 in Eq.~(\protect\ref{DeltakT}).
  }
  \label{lhcb_dely}
\end{figure}

Finally, Fig.~\ref{lhcb_dely} compares the data with the calcualated
$|\Delta y|$ distribution.  In this case, the calculated results are independent
of the $k_T$ broadening.  The data fall off somewhat
faster than the calculations in all cases.  

Overall, the calculated results are in good agreement with the
assumption that the $D$ and $\overline D$ mesons both arise from the
production of a single $c \overline c$ pair.
Discrepancies between the calculations and the data may appear because the
calculations are based on exclusive $Q \overline Q$ production and include only
real corrections at next-to-leading order in addition to vertex corrections.
No soft radiation from parton showers,
as in event generators, which might smooth out the
$\Delta \phi$ distribution at $\phi \sim \pi$, are included in these
calculations, only fragmentation.
In addition, HVQMNR is a negative weight Monte Carlo so that
additional weight is placed on the last $\phi$ bin, $\phi = \pi$.  A positive
weight next-to-leading order exclusive Monte Carlo, such as POWHEG, can be
coupled to PYTHIA or HERWIG and include soft initial or final-state radiation.
If the multiplicities of the $p+p$ events are sufficiently large, heavy quark
scattering with produced particles may also weaken the azimuthal correlation.
See Ref.~\cite{Szczurek} for a comparison of these data to calculations in the
$k_T$-factorization approach.

Note also that, while most of the events are likely to arise from production
of a single $c \overline c$
pair, the $cc$ and $(c + \overline c) J/\psi$ yields
constitute about 10\% of
the total charm hadron pair yields \cite{LHCbDDpairs}.
Because only the $c \overline c$ pairs can be produced in a single
hard scattering,  these other events are more
consistent with production through double parton scattering or independent
production of two $c \overline c$ pairs.  The azimuthal
and rapidity distributions for the $cc$ and $c J/\psi$
events are consistent with isotropic
production \cite{LHCbDDpairs}, as might be expected from two hard scatterings.

In addition the ALICE Collaboration made an analysis of azimuthal
correlations between reconstructed $D$ mesons and a light hadron trigger
particle in $p+p$ collisions at 7~TeV and $p+$Pb collisions at
5.02~TeV \cite{ALICEDDpairs}.
The light hadrons were primary particles, emitted from the collision 
points.  These particles include those from heavy flavor decays, such as
the unreconstructed partner $D$ meson.  
The data were binned according to the transverse
momentum of both the $D$ meson and the light hadron.  The minimum light hadron 
$p_T$ was soft, $p_T > 0.3$~GeV.  These data were further subdivided into
two $p_T$ ranges, $0.3 < p_T < 1$~GeV and $p_T > 1$~GeV.  The $D$ meson $p_T$
was considerably higher: $3 < p_T < 5$~GeV, $5 < p_T < 8$~GeV, and
$8 < p_T < 16$~GeV.  To improve statistics, the ``$D$ meson'' is an average
over the $D^0$, $D^+$ and $D^{*+}$.  The ALICE measurements
cover the central region, $|y| < 0.5$ for
the $D$ and $| \Delta \eta | < 1$ for the light hadron.  The general behavior
is, however, the same as the LHCb $D^0 \overline D^0$ pairs, a peak at 
$\Delta \phi = 0$ and a smaller enhancement at $\Delta \phi = \pi$.  The
peak at $\Delta \phi = 0$ increases with increasing trigger particle $p_T$ and
also with increasing $D$ meson $p_T$ consistent with the trends of these
calculations.  The data were compared to simulations
with various PYTHIA tunes and also POWHEG+PYTHIA.
All simulations reproduced the trends of the data \cite{ALICEDDpairs},
consistent with what one might expect from the results shown here.  New ALICE
data on $D$-hadron correlations in $p+p$ collisions at 13~TeV were presented
recently \cite{BTrzeciak}.  In these higher energy collisions, the same models
describe these data as well.

\subsection{CDF $D \overline D$ Data}
\label{cdf_data}

The CDF Collaboration studied charm hadron correlations in $p + \overline p$
collisions at $\sqrt{s} = 1.96$~TeV \cite{CDFDDpairs}.
CDF combined their single inclusive
charm hadron data to look for fully reconstructed charm quark pairs.  Only
$D^{*\pm}$ were considered as candidates for the second charm hadron, in the
decay chain
$D^{*\pm} \rightarrow (D^0/\overline D^0) \pi^\pm \rightarrow (K\pi)\pi^\pm$, so
that the mass difference between the $D^{*\pm}$ and the first charm hadron,
$D^0/\overline D^0$ or $D^\pm$, $\Delta m = m(K\pi\pi) - m(K\pi)$, can suppress
the combinatorial background.
Several thousand $D^0 D^{*-}$ and $D^+ D^{*-}$ pairs were constructed in this way.

Their goal was to try to better understand the underlying
$Q \overline Q$ production process \cite{CDFDDpairs}.
As discussed in Sec.~\ref{sec:generators}, commonly employed high energy event
generators such as PYTHIA implement prompt $Q \overline Q$ production by
three different leading-order porcesses: `pair creation',
as in the LO diagrams in
Fig.~\ref{lodias}; `flavor excitation', and `gluon splitting',
corresponding to the
NLO processes shown in Fig.~\ref{nlodias} (b) and (c) respectively.

The weighting of these different contributions in PYTHIA
can be tuned to match the data to determine which processes are most important
to heavy flavor production in the phase space covered by an experiment.
According to this interpretation, the final-state $Q \overline Q$ configurations
are treated separately by virtue of their topology with $\phi = \pi$,
corresponding to pair creation, and $\phi = 0$, corresponding to gluon
splitting.  This apporach stands in contrast to the NLO calculations shown in
this paper where all NLO diagrams are summed coherently with weighting according
to color and spin with interference terms.
As described in Sec.~\ref{sec:generators}, at NLO, flavor excitation and gluon
splitting are both part of production by $gg$ interactions in the initial state.
They are not separate production processes and are not treated as such.

\begin{figure}[htpb]\centering
  \includegraphics[width=\columnwidth]{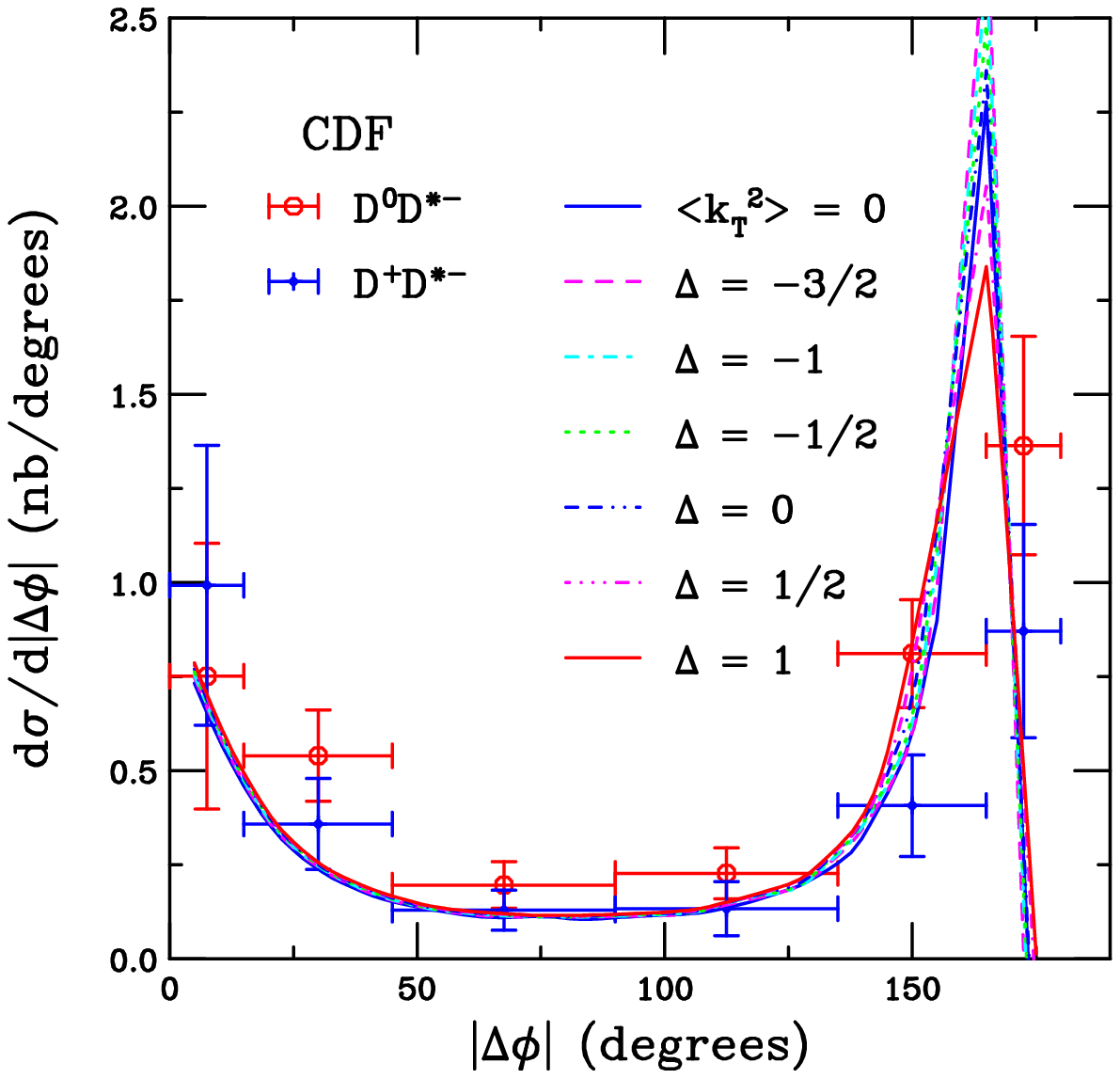}
  \caption[]{(Color online)
    The azimuthal angle distributions for $D^0 D^{*-}$ (red) and
    $D^+ D^{*-}$ (blue) pairs measured in $p+ \overline p$ collisions
    at $\sqrt{s} = 1.96$~TeV by CDF \protect\cite{CDFDDpairs}.
    The data are compared
    to calculations in the same acceptance with $\langle k_T^2 \rangle = 0$ and
    for values of $\Delta$ from $-3/2$ to 1 in Eq.~(\protect\ref{DeltakT}).}
  \label{cdf_azi}
\end{figure}

The data on each final state, in the rapidity interval $|y| < 1$ and $p_T$
ranges $5.5 < p_T^{D^0,D^{*-}} < 20$~GeV, $7 < p_T^{D^+} < 20$~GeV, are shown in
Fig.~\ref{cdf_azi}.  In the calculations, the $y$ interval was the same but the
lower limit on the $D$ meson $p_T$ was taken to be 5.5~GeV in all cases.  The
calculations agree rather well with the data.  The results are independent of
$\Delta$ except when the pairs are completely back-to-back
($|\Delta \phi| = \pi$), as might be expected for the given $p_T$ range.  In
Ref.~\cite{CDFDDpairs}, the comparison with the separate LO `production
mechanisms' in PYTHIA showed that, at least for the PYTHIA tune used by the
CDF Collaboration, the 'gluon splitting' or collinear
contribution ($\Delta \phi = 0$) to production was underestimated.
In the calculation, with the NLO contributions added according to initial-state
parton channel ($gg$, $(q/\overline q)g$ and $q \overline q$), there is no
significant discrepancy.

\subsection{CMS Bottom Quark-Bottom Jet Data}

The CMS Collaboration measured $b \overline b$ correlations through the use of
their jet trigger \cite{CMSbbpairs}.  The analysis employed the single jet
trigger with calorimeter energy above trigger thresholds of 15, 30 and 45~GeV.
The energy scale at which these three triggers are greater than 99\% efficient
corresponds to transverse momentum of the leading jet, $p_T^{\rm jet}$,
of 56, 84 and 120~GeV
respectively.  The leading jet is associated with one of the $b$ quarks.
The triggered events
are required to have one reconstructed jet with the minimum $p_T$ defined as
above
and a reconstructed primary vertex.  The leading jet is requred to be within
the pseudorapidity interval $|\eta^{\rm jet}| < 3$.  The jet along with
identified $B$ hadrons with $p_T^B > 15$~GeV and in the interval $|\eta^B|<2$,
is considered to originate from a single $b \overline b$ pair.  Thus the events
also have two reconstructed secondary vertices.  The angular correlations are
calculated using the $B$ hadron and $b$-jet flight directions.

The CMS results were presented as a function of the azimuthal angle difference,
$\Delta \phi$, and a separation variable, $\Delta R$, that includes the
difference in polar angles between the $B$ hadron and jet, given in terms of
the difference in pseudorapidity, $\Delta \eta$,
$\Delta R = \sqrt{(\Delta \eta)^2 + (\Delta \phi)^2}$.  The CMS
predictions for the
behavior of $\Delta R$ are based on the notion that gluon splitting dominates
at low $\Delta R$ while flavor creation is most important at large $\Delta R$.
This is similar to what one expects from event generators for $\Delta \phi$,
as discussed in Secs.~\ref{sec:generators} and \ref{cdf_data}.

\begin{figure}[htpb]\centering
  \includegraphics[width=\columnwidth]{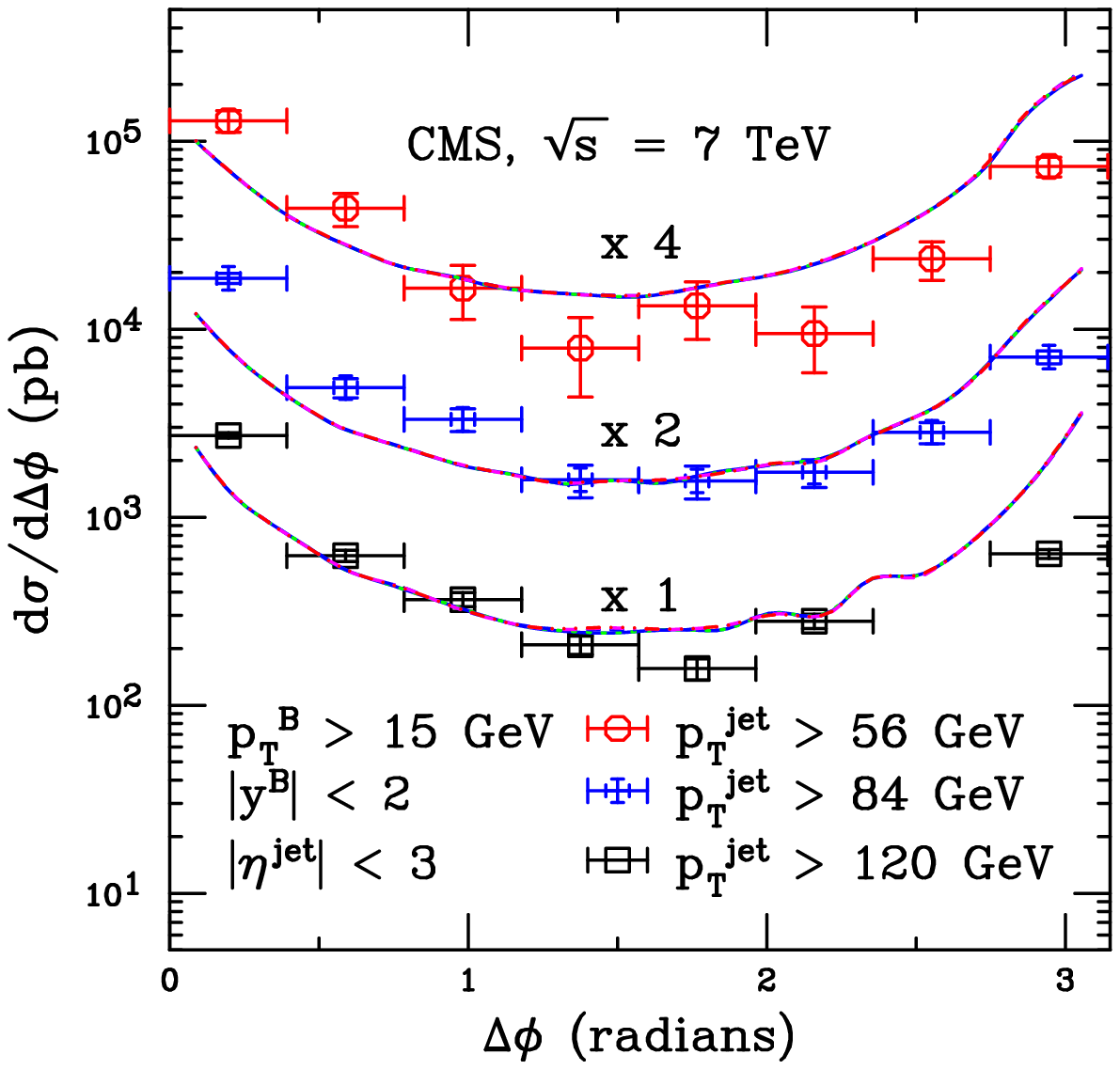}
  \caption[]{(Color online)
    The azimuthal angle distributions for $B \overline B$ pairs measured in
    $p+ p$ collisions
    at $\sqrt{s} = 7$~TeV by CMS \protect\cite{CMSbbpairs}.  The $p_T^{\rm jet}$
    lower limits of 56 GeV (red), 84 GeV (blue) and 120 GeV (black) are
    scaled by factors of 4, 2 and 1 respectively.
    The data are compared
    to calculations in the same kinematics for $\langle k_T^2 \rangle = 0$
    (blue curves) and
    for values of $\Delta$ from $-1/2$ to 1 in Eq.~(\protect\ref{DeltakT}).
    }
  \label{cms_azi}
\end{figure}

The calculations shown here are based on the assumption that a
$b \overline b$ pair is produced with one final state $B$ or $\overline B$
in the kinematic region $p_T^B > 15$~GeV and rapidity $|y^B|<2$.
The jet is assumed
to originate from the second $B$ hadron of the pair with the minimum $p_T$ of
the $B$ equal to the $p_T$ of the jet, greater than 56, 84 and 120~GeV
respectively, while $\eta^{\rm jet}$ is assumed to be
equivalent to the rapidity of the $b$ quark.  Since the pseudorapidity
difference is replaced by a rapidity difference in the calculation of $\Delta R$
and the equivalence of the $b$ quark and the jet $p_T$ might not be
exact, one might expect that the calculation of the $\Delta \phi$ distribution
might be better reproduced than that of $\Delta R$.

\begin{figure}[htpb]\centering
  \includegraphics[width=\columnwidth]{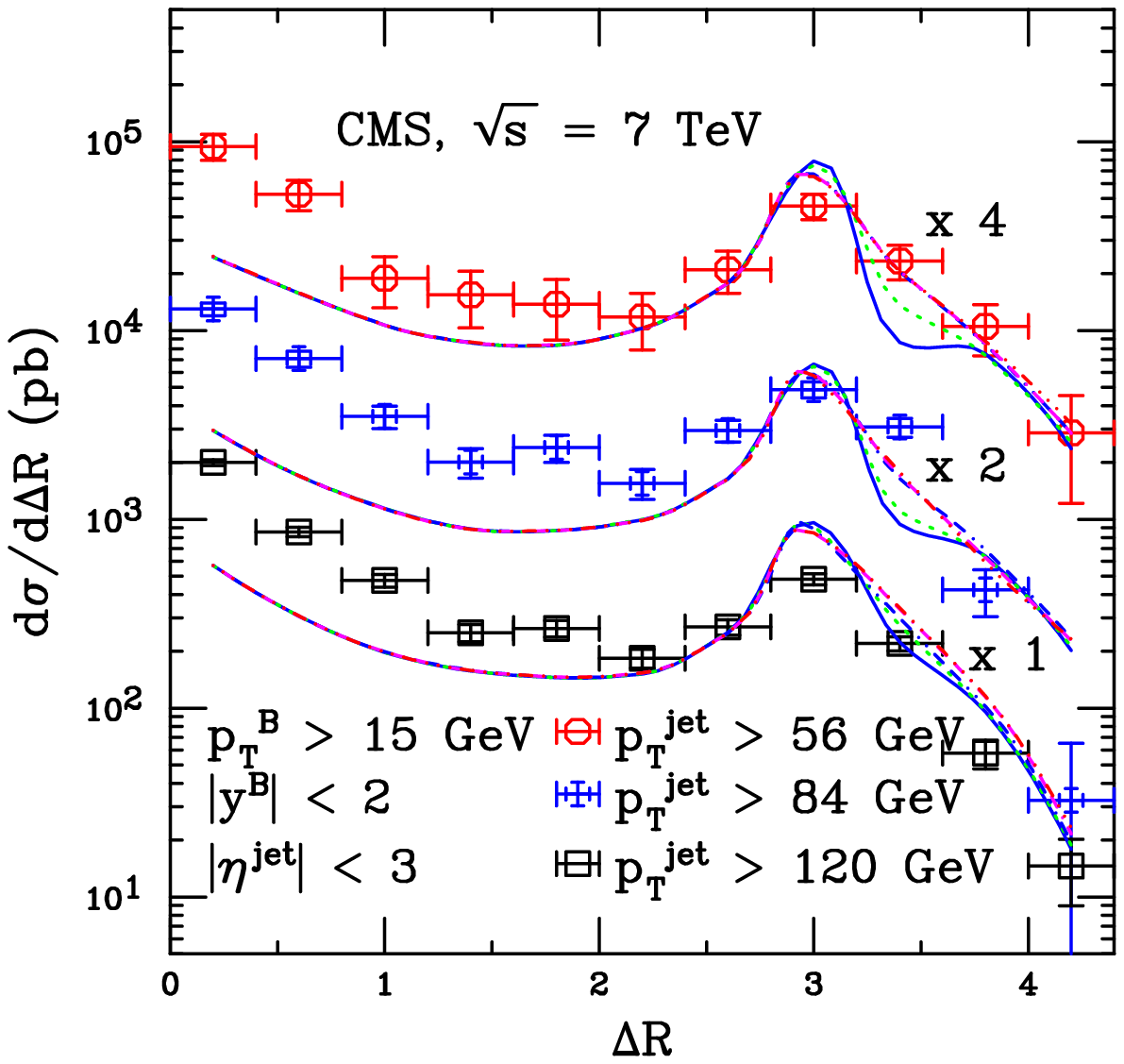}
  \caption[]{(Color online)
    The $\Delta R$ distributions for $B \overline B$ pairs measured in
    $p+ p$ collisions
    at $\sqrt{s} = 7$~TeV by CMS \protect\cite{CMSbbpairs}.  The $p_T^{\rm jet}$
    lower limits of 56 GeV (red), 84 GeV (blue) and 120 GeV (black) are
    scaled by factors of 4, 2 and 1 respectively.
    The data are compared
    to calculations in the same kinematics for $\langle k_T^2 \rangle = 0$
    (blue curves) and
    for values of $\Delta$ from $-1/2$ to 1 in Eq.~(\protect\ref{DeltakT}),
    the green dotted to solid red curves respectively.}
  \label{cms_delR}
\end{figure}

The comparisons of the calculations with the data are shown in
Figs.~\ref{cms_azi} and \ref{cms_delR}.  While the calculations are done with
all values of $\Delta$ in Eq.~(\ref{DeltakT}), no difference between them is
visible except at large $\Delta R$.  As in Ref.~\cite{CMSbbpairs}, the data
and calculations for the different minimum jet $p_T$ are scaled by factors of
4, 2 and 1 for $p_T^{\rm jet}$ greater than 56, 84 and 120~GeV respectively to
more clearly separate the results.

The comparison to the $\Delta \phi$
distributions is very good, as may be expected.  The comparison to $\Delta R$
is not as good at low $\Delta R$ where the CMS Collaboration suggests gluon
splitting is dominant.  However, given the agreement with the low $\Delta \phi$
results, where this mechanism is also expected to be dominant in event
generators, the underestimate may be due to the calculation of $\Delta R$ with
$\Delta y$ instead of $\Delta \eta$.  Note that the large $\Delta R$ results are
best reproduced with $\Delta = 1$ in Eq.~(\ref{DeltakT}).

\section{Cold Nuclear Matter: $p+$Pb Collisions at 5.02 TeV}
\label{sec:pPb}

Finally, the interaction of $c \overline c$ pairs in cold nuclear matter, such
as $p+$Pb collisions at the LHC, are discussed.  It has been suggested
\cite{gossiaux} that energy loss by heavy quarks in heavy-ion collisions could
change the azimuthal correlations.  First, it must be determined how the
charm distributions and their
azimuthal separation are influenced by the presence of
cold nuclear matter.
For example, the effect of cold matter energy loss, which could
be manifested as additional $k_T$ broadening by multiple scattering in the
nucleus, could result in changes in the $p_T$ distributions of heavy quarks
in $p+$Pb relative $p+p$ collisions.  This would be in addition to modification
of the parton densities in the nucleus, referred to as shadowing.

Here the results of shadowing alone on the single charm meson $p_T$
distributions and $c \overline c$
pair azimuthal
distributions are compared to shadowing and $k_T$ broadening, both
with $\langle k_T^2 \rangle \sim 1.5$~GeV$^2$ alone, as in $p+p$ collisions,
and with an additional $k_T$ kick due to the presence of the nuclear medium
in $p+$Pb collisions.

The calculations of the nuclear modification factor $R_{p{\rm Pb}}(p_T)$ are
compared to data from ALICE \cite{ALICEDmesons}.  ALICE measured prompt
production of $D^0$, $D^+$, $D^{*+}$, and $D_s^-$ as well as their charge
conjugates at central rapidity, $-0.96 < y_{\rm cms} < 0.04$ in $p+$Pb collisions
at $\sqrt{s} = 5.02$~TeV.  The yields were compared to those in $p+p$
collisions to form $R_{p{\rm Pb}}(p_T)$.  Note that, for the analysis in
Ref.~\cite{ALICEDmesons}, the
$p+p$ baseline at 5.02~TeV was interpolated between the 2.76~TeV and 7~TeV
data since it was completed before $p+p$ data were available
at 5.02~TeV.

In Ref.~\cite{ALICEDmesons}, the average
of the $D^0$, $D^+$ and $D^{*+}$  were compared to several calculations,
including next-to-leading order calculations with HVQMNR including only
shadowing effects with no momentum braodening nor fragmentation.
The uncertainty band gives
suppression below $p_T \sim 5$~GeV, rising to be equivalent with
unity at higher $p_T$.  See Ref.~\cite{ALICEDmesons} for details of the other
comparisons which included a leading order calculation of shadowing due to
power corrections, $k_T$ broadening and cold matter energy loss \cite{Vitev},
and a calculation assuming color glass condensate in the initial state
\cite{Fuji}.

\begin{figure}[htpb]\centering
  \includegraphics[width=\columnwidth]{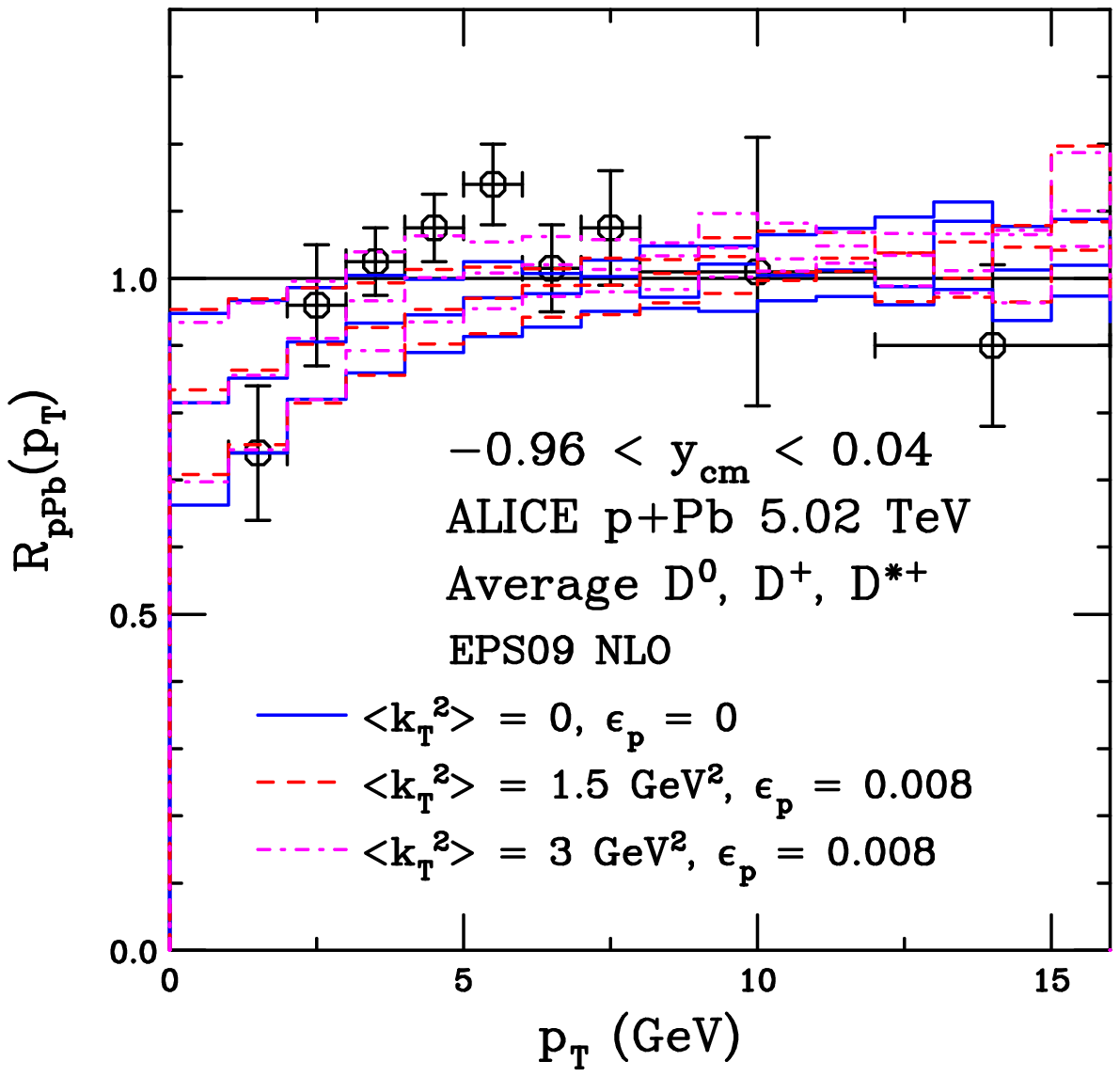}
  \caption[]{(Color online)
    The nuclear modification factor $R_{p{\rm Pb}}(p_T)$ in 5.02 TeV $p+$Pb
    collisions for $-0.96 < y_{\rm cm} < 0.04$.  The ALICE data for the average
    of their $D^0$, $D^+$ and $D^{*+}$ measurements \protect\cite{ALICEDmesons}
    at central rapidity are shown.  Results are shown for: no
    $k_T$ broadening and no fragmentation (solid blue); the standard result from
    Sec.~\protect\ref{sec:azi_dist}, 
    $\langle k_T^2 \rangle = 1.5$~GeV$^2$ and $\epsilon_P = 0.008$ (red dashed);
    and $\langle k_T^2 \rangle = 3$~GeV$^2$ and $\epsilon_P = 0.008$
    (magenta dot-dashed).  The EPS09 NLO uncertainty band, along with the
    central value, is shown.  Note that the additional $k_T$ broadening is
    only applied to 
    $p+$Pb collisions and not to $p+p$ collisions in the magenta curves.}
  \label{shad_pt}
\end{figure}

To go beyond $p+p$ collisions, in the NLO calculations,
the proton parton densities must be replaced by those of
the nucleus.  
If $A$ is a nucleus, the nuclear parton densities, 
$f_i^A(x_2,\mu^2)$, can be assumed to factorize into
the nucleon parton density, $f_i^p(x_2,\mu_F^2)$, independent of $A$;
and a shadowing ratio, $S_i(A,x_2,\mu_F^2)$ that
parameterizes the modifications of the nucleon parton densities in the nucleus.
Here $x_2$ is the fraction of the nucleon momentum carried by the interacting
parton in the nucleus.  While the shadowing effect may also depend on the impact
parameter, $b$, of the parton from the proton with the lead nucleus, only
minimum bias results, independent of impact parameter, are shown here.

The calculations in Fig.~\ref{shad_pt} employ the EPS09 NLO \cite{EPS09}
parameterization for $S_i$, assuming collinear factorization.  The EPS09 sets
include
15 parameters, giving an error set of 30 additional parameterizations created
by varying each parameter within one standard deviation.  
The EPS09 uncertainty
band is obtained by calculating the deviations from the central value
for the 15 parameter variations on either side of the central set and adding
them in quadrature.  There is a more recent shadowing parameterization by
Eskola {\it et al.}, EPPS16 \cite{EPPS16}, that includes some of the LHC
data from the first $p+$Pb run at 5.02~TeV in the global analysis of the
nuclear parton densities.  The central
result for the gluon distribution is very similar to that of EPS09 so that the
central result with EPPS16 should be very similar to that shown here.
However, it employs 5 additional parameters, thus resulting in a larger
uncertainty band than the one in Fig.~\ref{shad_pt}.

The calculations given as a function of $p_T$ in Fig.~\ref{shad_pt}
are in the same rapidity interval as the ALICE data.
Three bands are shown.  The first, with $\langle k_T^2 \rangle = 0$ and
$\epsilon_P = 0$, in blue, corresponds to the EPS09 calculation in
Ref.~\cite{ALICEDmesons} albeit the curves in that paper likely employed a
larger quark mass, $m\sim 1.5$~GeV, which would result in a reduced shadowing
effect at $p_T \sim 0$ relative to the calculation shown here.  The red
histogram, employing the preferred parameters found for $D$ mesons in this
paper, and
assuming the same intrinsic $k_T$ in $p+$Pb and $p+p$ collisions,
gives a nearly identical result to the blue histogram, calculated without
either effect.

However, one might expect that a higher intrinsic $k_T$
is required in a nuclear medium relative to that in $p+p$ due to multiple
scattering in the nucleus, known as the Cronin effect \cite{Cronin}.  If one
assumes that $\langle k_T^2 \rangle$ is doubled in the cold nuclear medium, then
the magenta curves are obtained.  The larger $\langle k_T^2 \rangle$ results in
an increase of the band at intermediate $p_T$, with the upper limit of the
band increasing above unity at $p_T \sim 5$~GeV.  The statistics on the
averaged $D$ meson data are not sufficiently significant to distinguish between
the results: all are consistent with the data.

\begin{figure}[htpb]\centering
  \includegraphics[width=\columnwidth]{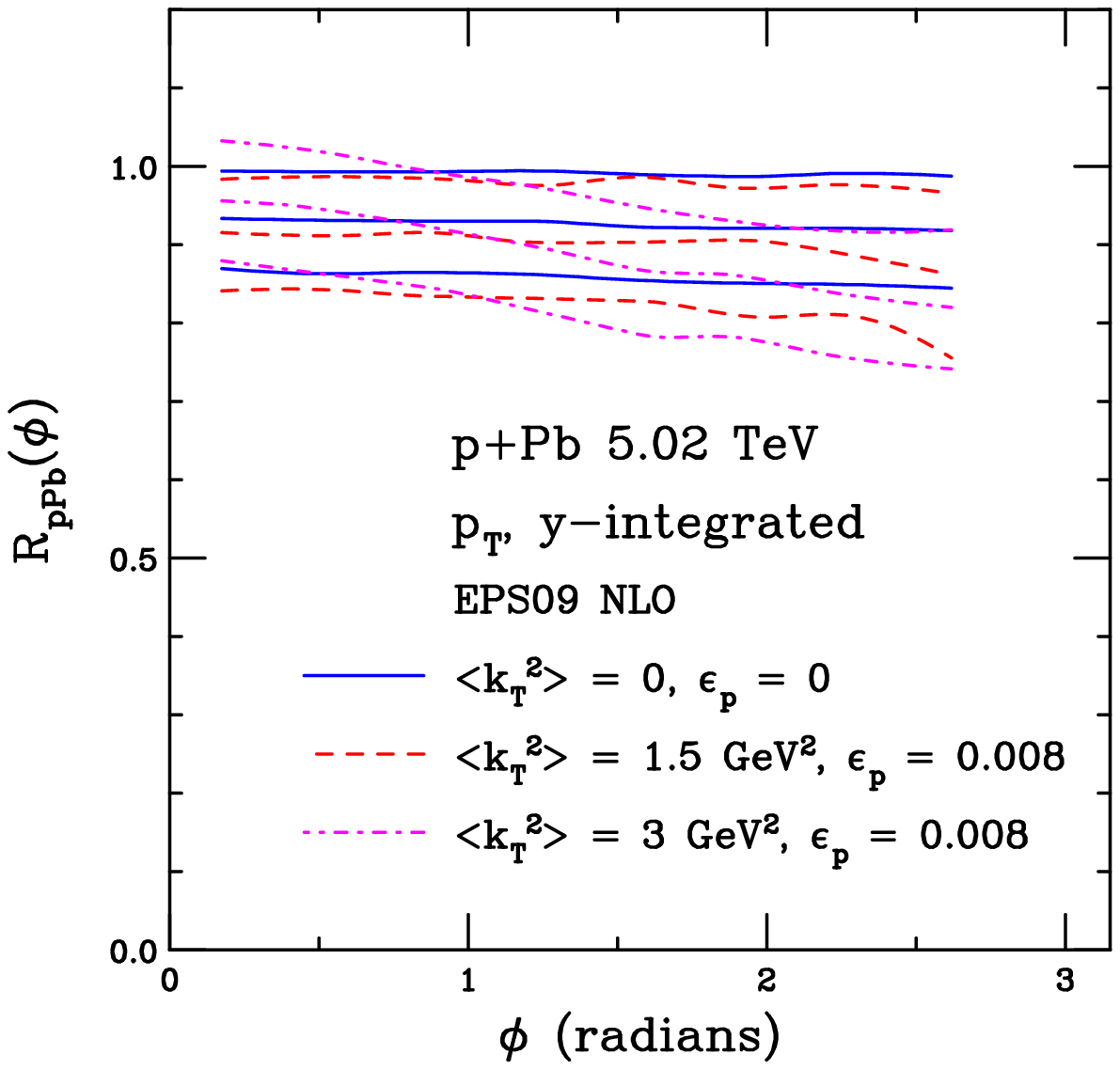}
  \caption[]{(Color online)
    The nuclear modification factor $R_{p{\rm Pb}}(\phi)$ in 5.02 TeV $p+$Pb
    collisions.  Results are shown for: no
    $k_T$ broadening and no fragmentation (solid blue); the standard result from
    Sec.~\protect\ref{sec:azi_dist}, 
    $\langle k_T^2 \rangle = 1.5$~GeV$^2$ and $\epsilon_P = 0.008$ (red dashed);
    and $\langle k_T^2 \rangle = 3$~GeV$^2$ and $\epsilon_P = 0.008$
    (magenta dot-dashed).  The EPS09 NLO uncertainty band is shown along
    with the central value.
    No $p_T$ or rapidity cuts have been imposed.  Note that the additional
    $k_T$ broadening in the magenta curves is only applied to 
    $p+$Pb collisions and not to $p+p$ collisions.}
  \label{shad_azi}
\end{figure}

The results for the nuclear modification factor of the azimuthal angular
correlations between the heavy quarks are shown in Fig.~\ref{shad_azi} for the
same three sets of calculations as in Fig.~\ref{shad_pt}.  In this
case there are no cuts made on $p_T$ or rapidity.

When no intrinsic $k_T$ is included, $R_{p{\rm Pb}}(\phi)$ is independent of
$\phi$.  There is a slight $\phi$ dependence introduced when the $p+p$
value of $\langle k_T^2 \rangle$ is employed.  It is, however, a small deviation
relative to the calculation with no $k_T$ effect.  There is a more significant
effect, as expected, when the intrinsic $k_T$ effect is doubled in $p+$Pb
relative to $p+p$ collisions.  Instead of being independent of $\phi$,
$R_{p{\rm Pb}}(\phi)$ 
now decreases as $\phi$ increases, a result of the increased
$\langle k_T^2 \rangle$, as in Fig.~\ref{fig4}(a).

In heavy-ion collisions, in particular at heavy-ion colliders, multiple
$c \overline c$ pairs are produced in a single event.  Therefore,
for an analysis of
heavy-flavor azimuthal correlations in hot matter it is necessary to find the
pair vertex to ensure that the $c$ and $\overline c$ originate from the same
hard scattering.

\section{Summary}

The similarity of the results for $p_T > 10$~GeV, even for
$\langle k_T^2 \rangle = 0$, shows that the enhancement at $\phi \sim 0$, with
the $Q \overline Q$ pair aligned opposite a hard parton, is independent of
$\langle k_T^2 \rangle$ and arises instead from the $Q \overline Q$ production
mechanism at high $p_T$.  Thus the high $p_T$ behavior of $d\sigma/d\phi$ is
indicative of the contribution of next-to-leading order production while the
low $p_T$ behavior of $d\sigma/d\phi$ is extremely sensitive to the chosen
$\langle k_T^2 \rangle$ and essentially independent of fragmentation.

It appears that, so far, the open heavy flavor results presented by the LHC
collaborations, both single inclusive production and charm pair correlations,
are in agreement with calculations based on collinear factorization with single
hard scatterings.  The exception, the $cc$ and $c J/\psi$ events at LHCb, are
consistent with double parton scattering.

Thus, for $p_T$ cuts on the order of a few GeV, the calculations of the
azimuthal angle distributions are rather
insensitive to fragmentation and $k_T$ broadening which affect the correlations
at low $p_T$.  Thus, hot nuclear matter effects on these correlations should be
rather robust for $p_T \geq 3-5$~GeV.

\section*{Acknowledgments}
I would like to thank A.~Mischke for encouragement and T.~Dahms, M.~Durham
and L.~Vermunt for discussions.
This work was performed under the auspices of the 
U.S. Department of Energy by Lawrence Livermore National Laboratory under 
Contract DE-AC52-07NA27344 and supported by the U.S. Department of Energy, 
Office of Science, Office of Nuclear Physics (Nuclear Theory) under contract 
number DE-SC-0004014.

\end{document}